\documentclass[twoside,english]{emulateapj}

\pdfoutput=1

\makeatletter
\usepackage{amsmath}
\makeatother

\usepackage{babel}
\begin{document}

\title{APPLICATION OF GAS DYNAMICAL FRICTION FOR PLANETESIMALS: II. EVOLUTION
OF BINARY PLANETESIMALS}

\author{Evgeni Grishin \& Hagai B. Perets}

\affil{Physics Department, Technion - Israel Institute of Technology, Haifa,
Israel 3200003}
\begin{abstract}
One of first the stages of planet formation is the growth of small
planetesimals and their accumulation into large planetesimals and
planetary embryos. This early stage occurs much before the dispersal
of most of the gas from the protoplanetary disk. At this stage gas-planetesimal
interactions play a key role in the dynamical evolution of \emph{single}
intermediate-mass planetesimals ($m_{p}\sim10^{21}-10^{25}g$) \emph{through
gas dynamical friction} (GDF). A significant fraction of all Solar
system planetesimals (asteroids and Kuiper-belt objects) are known
to be binary planetesimals (BPs). Here, we explore the effects of
GDF on the evolution of \emph{binary} planetesimals embedded in a
gaseous disk using an N-body code with a fiducial external force accounting
for GDF. We find that GDF can induce binary mergers on timescales
shorter than the disk lifetime for masses above $m_{p}\gtrsim10^{22}g$
at 1AU, independent of the binary initial separation and eccentricity.
Such mergers can affect the structure of merger-formed planetesimals,
and the GDF-induced binary inspiral can play a role in the evolution
of the planetesimal disk. In addition, binaries on eccentric orbits
around the star may evolve in the supersonic regime, where the torque
reverses and the binary expands, which would enhance the cross section
for planetesimal encounters with the binary. Highly inclined binaries
with small mass ratios, evolve due to the combined effects of Kozai-Lidov
cycles with GDF which lead to chaotic evolution. Prograde binaries
go through semi-regular Kozai-Lidov evolution, while retrograde binaries
frequently flip their inclination and $\sim50\%$ of them are destroyed.
\end{abstract}

\section{INTRODUCTION}

Planets form in protoplanetary disks around young stars. Once km sized
planetesimals have been formed, their evolution is determined by three
basic dynamical processes: Viscous stirring, dynamical friction and
coagulation or disruption through collisions (see \citealp{2004ARA&A..42..549G}
and references therein for details). These dynamical processes do
not include the effects from planetesimal-gas interaction in the disk,
which can be important during the early stages of planet formation
when gas is abundant (the first few Myr, with possible suggestions
for longer timescales (\citealp{2014ApJ...793L..34P}).

During the evolutionary phases of planet formation, small dust grains
successively grow into large planetary embryos. The evolution of small
size planetesimals is dominated by gas drag. Gas drag keeps the planetesimal
disk thin, the relative velocities low, and assists in coagulation
of small bodies \citep{ormel+10,2002Icar..155..436O,2011ApJ...727L...3P}.
Larger planetary embryos (i.e. $m_{p}\ge0.1M_{\oplus}$) are also
affected by gas interactions, and can migrate through type I or type
II migration (see \citealp{2006RPPh...69..119P} and references therein).

In our previous paper (\citealt{2015ApJ...811...54G}; hereafter;
paper I), we have studied the planetesimal-disk interactions of\emph{
intermediate mass planetesimals} (IMP), in the mass range $10^{21}\le m_{p}\le10^{25}g$,
using the gas dynamical friction (GDF) approach, and found it to be
efficient in changing the orbital evolution of the IMPs, and significantly
more important than aerodynamical gas drag in this mass range. 

The interaction between a gravitating planetesimal and the nebular gas was first studied by \citet{1988Icar...75..552O}. In their study, the effective drag coefficient increases rapidly with increasing gravitational energy. The first detailed analytic study of   GDF was  by \citet{1999ApJ...513..252O} for objects
in homogenous environment moving on straight line. \citet{2007ApJ...665..432K}
have generalized \citeauthor{1999ApJ...513..252O}'s work for circular
orbits. GDF for binary perturbers was first introduced by \citet{2008ApJ...679L..33K},
and \citet{2014ApJ...794..167S} who studied the evolution of a binary
traveling in a non-stationary gaseous medium, generalizing further
previous works. Here, we focus on the effects of GDF on binary planetesimals
(BPs) which had not been explored before, and complement previous
work on the evolution of lower mass BPs due to aerodynamical gas drag
\citep{2011ApJ...733...56P}. 

Paper I dealt with strictly single planetesimals. However, observations
indicate that a large fraction (at least $20\%$) of all Kuiper Belt
and Trans Neptunian Objects (KBOs, TNOs) are gravitationally bound
binaries \citep{2007hst..prop11113N}. Such binaries could play an
important role in planet formation and significantly accelerate planetesimal
growth (\citealp{2011ApJ...727L...3P,2006AREPS..34...47R} and references
therein). \citet{2002Natur.420..643G} showed that massive BPs could
be formed through dynamical friction due to small planetesimals; a
transient binary becomes bound after losing its energy during an encounter
due to dynamical processes, and is then hardened by collisionless
dynamical friction due to small bodies. They find that $\sim5\%$
of Kuiper Belt Objects could form bound binaries through such mechanism.
\citet{2010AJ....140..785N} suggested that BPs can from through a
gravitational collapse, and \citet{2011PASJ...63.1331K,2014PASJ...66..123K}
explored the dynamical formation of BPs. 

In paper I we found that GDF damps inclinations and eccentricities
of planetesimal orbits around the Sun efficiently even if the mass
is as low as $\sim10^{23}g$, given that the random velocity is not
too high ($e,I\sim\text{few}\times H_{0}$, where $e$ and $I$ are
the orbital eccentricity and inclination respectively, $H_{0}=h/a$
is the aspect ratio, $h$ is the disk scale height). In addition,
large planetary embryos of mass $\sim10^{25}g$ migrate within a few
Myrs. One of the implications of IMP interaction with the gas is the
cooling of the planetesimal disk, and the potential overabundance
of IMPs in the inner region, that could potentially assist the formation
of Super-Earths (see discussion in paper I).

Motivated by recent developments both in GDF theory for binaries and
planet formation, we extend the study of paper I on intermediate mass
\textit{\emph{binary planetesimals (BPs)}}, and show that GDF can
change the binary mutual orbit, either leading to binary inspiral
and even merger or to the expansion of the binary separation, depending
on the properties of the binary orbit around the Sun. 

The paper is organized as follows: In section \textbf{\ref{sec:GAS-PLANETESIMAL-INTERACTION}}
we review the gas-planetesimal interactions. In section \textbf{\ref{sec:FORMULATION-OF-THE}}
we discuss the limitations and applicability of our study. In section
\textbf{\ref{sec:diff acc}} we derive the differential accelerations
of BPs due to GDF. In section \textbf{\ref{sec:EVOLUTION OF ORBITAL ELEMENTS}}
we use the differential accelerations and derive analytic expressions
for the evolutionary timescales of binary orbital elements. In section
\textbf{\ref{sec:NUMERICAL SET UP}} we present the numerical set
up and the initial conditions of the tested scenarios. We present
the numerical results in section \textbf{\ref{sec:RESULTS}}. In section
\textbf{\ref{sec:DISCUSSION}} we discuss the significance of our
results. Finally, we summarize the paper in section \textbf{\ref{sec:Summary}}.

\section{GAS PLANETESIMAL INTERACTIONS}
\label{sec:GAS-PLANETESIMAL-INTERACTION}

The standard models of evolution of planetesimal disks account for
various dynamical processes, including both physical processes due
to planetesimal-planetesimal interactions such as viscous stirring,
dynamical friction and physical collisions followed by coagulation,
as well as gas-planetesimal interactions through aerodynamics gas-drag
on small planetesimals, and planetary migration through tidal torques
acting on large planetary embryos and planets. 

Although gas-planetesimal interaction has been studied in the context
of small planetesimals, massive planetesimals have been largely ignored
as they are decoupled from the gas, and their size and velocity dispersion
is dominated by gravitational interactions. In paper I we have studied
IMPs in the mass range $10^{21}-10^{25}g$ and explored their interactions
with the gas through GDF, which dominate over aerodynamics gas drag
in this mass range. In the following we briefly review the GDF approach
and its comparison to aerodynamic gas drag. More details can be found
in paper I.

The general aerodynamic drag force imposed on a planetesimal of radius
$R$ and relative velocity $v_{rel}$ moving through a gaseous medium
of density $\rho_{g}$ is 
\begin{equation}
\boldsymbol{F}_{D}=-\frac{1}{2}C_{D}\pi R^{2}\rho_{g}v_{rel}^{2}\hat{\boldsymbol{v}}_{rel},\label{eq:drag}
\end{equation}
where $\hat{\boldsymbol{v}}_{rel}$ is the unit vector in the direction
of the relative velocity and $C_{D}$ is the drag coefficient that
for spherical body depend on the Reynolds number, i.e. $C_{D}=C_{D}(\mathcal{R}e)$.
Empirical formula can be used for $C_{D}(\mathcal{R}e)$, fitted for
the range $\log_{10}\mathcal{R}e\in[-3,5]$ (see.\citealp{2003JEE...129.222}
and paper I). 

For the GDF force, consider a perturber with mass $m_{p}$ moving
on a straight line with the same ambient gaseous medium. The perturber
generates a wake, which in turn affects the perturber. Using linear
perturbation theory, \citet{1999ApJ...513..252O} calculated the drag
force felt by the perturber. The GDF force is given by 
\begin{equation}
\boldsymbol{F}_{GDF}=-\frac{4\pi G^{2}m_{p}^{2}\rho_{g}}{v_{rel}^{3}}\boldsymbol{v}_{rel}\mathcal{I}(\mathcal{M}),\label{eq:gdf}
\end{equation}
where  $\mathcal{M}\equiv v_{rel}/c_{s}$ is the Mach number, $c_s$ is the speed of sound and $\mathcal{I}(\mathcal{M})$
is a dimensionless factor given by 
\begin{equation}
\mathcal{I}(\mathcal{M})=\left\{ \begin{array}{cc}
\frac{1}{2}\ln\left(\frac{1+\mathcal{M}}{1-\mathcal{M}}\right)-\mathcal{M} & \mathcal{M}<1\\
\frac{1}{2}\ln(1-\mathcal{M}^{-2})+\ln\left(\frac{v_{rel}t}{R_{min}}\right) & \mathcal{M}>1;\ v_{rel}t>R_{min}
\end{array}\right.,\label{eq:I of m}
\end{equation}
where $\ln v_{rel}t/R_{min}$ is the Coulomb logarithm. The force
is non-vanishing in the subsonic regime, while in the supersonic regime,
a minimal radius $R_{min}$ is introduced to avoid divergence of the
gravitational potential (usually taken to be the physical size of
the perturber, or the accretion radius $Gm_{p}/v_{rel}^{2}$). The
exact value of $R_{min}$ is not well determined; but it can be fitted
through comparison of eqn. (\ref{eq:I of m}) with hydrodynamical
simulations, to find a best fitting value for $R_{min}$ \citep{1999ApJ...522L..35S}.
The maximal scale of the system is usually the disk scale height $h$,
so typical values of the Coulomb logarithm are near $\sim10$. 

We note that the results of \citet{2007ApJ...665..432K} are qualitatively
similar to \citeauthor{1999ApJ...513..252O}'s straight line trajectory
(see fig. 8 of \citealt{2007ApJ...665..432K}).

Another useful formula for the subsonic regime is the expansion in
powers of Mach number 

\begin{equation}
\mathcal{I}(\mathcal{M})=\frac{1}{3}\mathcal{M}^{3}+\frac{1}{5}\mathcal{M}^{5}+\mathcal{O}(\mathcal{M}^{7}).\label{dfdf-1}
\end{equation}
If only the first term is considered, the GDF force scales linearly
with $\boldsymbol{v}_{rel},$ $\boldsymbol{F}_{GDF}\propto-\boldsymbol{v}_{rel}$,
and thus we will occasionally use the term '\emph{linear regime}'
for the subsonic regime.

In paper I we found the critical planetesimal size $R_{\star}$ that
GDF equals to aerodynamic gas drag. It is given by 
\begin{eqnarray}
R_{\star} & = & 0.29\left[\frac{C_{D}(\mathcal{R}e)}{\mathcal{I}(\mathcal{M})}\right]^{1/4}\frac{v_{rel}}{\sqrt{G\rho_{m}}},\label{eq:critical size}
\end{eqnarray}
where $v_{rel}\approx\sqrt{H_{0}^{4}+e^{2}}v_{K}$ , $v_{K}=\sqrt{GM_{\star}/a}$
is the Keplerian velocity, $a$ is the semi-major axis, and $e$ is
the orbital eccentricity. 

In paper I we used a simple flared disk model \citep{1997ApJ...490..368C}
and found that typical masses where GDF dominates over aerodynamical
gas-drag are $\gtrsim10^{21}g$. The parametric dependence of type
I migration and GDF torque is the same, and GDF is comparable to type
I \textit{one sided }torque. The tidal torque is a superposition of
inner and outer torques, and is smaller by a factor of $\sim h/a$
due to the asymmetry of the inner and outer regions of the disk (see
paper I).

\section{FORMULATION OF THE PROBLEM, ASSUMPTIONS
AND APPLICABILITY}
\label{sec:FORMULATION-OF-THE}

The effect of GDF for binary perburbers was studied in detail in \citet{2008ApJ...679L..33K}.
When considering the wakes of both of the bodies, the azimuthal component
of GDF, $F_{\varphi}/\mathcal{M}^{2}$ was reduced due to the companion's
wake. For low Mach numbers, $F_{\varphi}/\mathcal{M}^{2}$ was negligible,
and for $\mathcal{M\sim}1$ it was $\sim30\%$ less. However, the
latter is true for static gas. Consider a binary system that the center of mass (CM) of the binary revolves around the central object (e.g. Earth-Moon system that revolves around the Sun. The inner system is considered as the inner binary (e.g. Earth-Moon system), while the combination of the central object with the CM of the inner binary is considered as the outer binary (e.g. Sun - CM of Earth-Moon system).  If the  outer binary moves with relative velocity $\boldsymbol{v}_{out}$
with respect to the gas, the morphology of the wake is different.
\citet{2014ApJ...794..167S} have shown that the braking torque that
causes the binary to lose angular momentum, is $\Gamma=\alpha a_{bin}F_{DGF,\varphi}^{(1)}$
where $F_{DGF,\varphi}^{(1)}$ is the GDF force applied on the first
body, $a_{bin}$ is the binary separation,  and $\alpha$ is an order of unity parameter that depends on
the minimal radius $R_{min}$; the primary Mach number is $\mathcal{M}_{cm}=v_{out}/c_{s}$;
and the binary Mach number is $\mathcal{M}_{bin}=v_{bin}/c_{s}$, where $v_{bin}$ is the typical binary velocity given by 
\begin{equation}
v_{bin}=\sqrt{\frac{Gm_{bin}}{fr_{H}}}=v_{K}f^{-1/2}Q^{1/3}.\label{eq:vbin}
\end{equation}

In principle, the function $\alpha(R_{min},\mathcal{M}_{cm},\mathcal{M}_{bin})$
can be calculated numerically for each case. It has been calculated
by \citet{2014ApJ...794..167S} for some cases but the full parameter
space has not been explored. Nevertheless, it is of order unity and
$<1$ for most cases. 

We do not account for the companion's wake in our simulation. By doing
that, we overestimate GDF force by at most $2-3$ times, according
to both \citet{2008ApJ...679L..33K} and \citet{2014ApJ...794..167S}.
Moreover, for typical planetesimal masses and for eccentric primary
orbit, $v_{out}\gg v_{bin}.$ In this regime $\alpha\to1$ and the
force is correctly estimated.

In paper I we studied the effects of GDF on single planetesimals.
Here we extend our study to BPs. 

Let us consider a binary with separation of the order of the binary
mutual Hill radius, $r_{H}$, i.e. $a_{bin}=fr_{H}=fa_{out}(m_{bin}/(m_{bin}+M_{\star}))^{1/3}\approx a_{out}fQ^{1/3}$
where $Q\equiv m_{bin}/M_{\star}$ is the star-binary mass ratio,
$m_{bin}=m_{b}+m_{s}$ is the total binary mass, $a_{out}$ is the
semi-major axis of the center of mass (CM) of the binary to the central
star and $0<f<1$ is some fraction. Usually the limit of stable prograde
orbits is near $f\approx0.5$, while the limit of retrograde orbits
is twice as large \citep{1991Icar...92..118H}. Note that $r_{H}$
is an additional length scale in the system, which is important for
the applicability of GDF to BPs.

The density wave propagating in $\hat{\boldsymbol{z}}$ direction
reaches the disk scale height after $\sim1$ orbital period and GDF
is gradually suppressed. In order to tackle the problem, \citet{2011ApJ...737...37M}
solved the equations of motion in a slab geometry. They introduced
an averaged potential at the scale height of the disk. Using Fourier
analysis, they have shown that the dominant contributions are from
perturbations with length scale as the disk scale height $l\sim h$.
While it is marginally applicable for single planetesimals (e.g. one
needs to invoke the Kolmogorov description of turbulence of the gaseous
disk; see discussion in paper I), here the contributions to the perturbation
on the mutual binary orbit are much less than the scale height, since
the relevant scale height is the binary Hill radius which is smaller
than the scale height of the disk. \citet{1999ApJ...513..252O}'s
3D analysis remains valid as long as $r_{H}<h$, or $Q^{1/3}<(h/a)^{3}$.
With $h/a\sim0.022$ in our case, the restriction is $Q\lesssim10^{-5}$.
Hence, only for masses of $\approx4M_{\oplus}$ and above do 3D modeld
break down. In our simulations the highest masses of planetesimals
are $m_{p}\lesssim2\cdot10^{25}g$, so we are well within the thick
disk regime.

To conclude, \citet{1999ApJ...513..252O}'s 3D formalism has difficulties
describing GDF for single planetesimals, but it fits well for describing
the mutual binary orbits for planetesimals of mass $m_{bin}\lesssim4M_{\oplus}$.
As long as the the binary mass is much lower than Earth mass, we can
consider the binary to be embedded in a spherical gaseous halo.

\section{DIFFERENTIAL ACCELERATIONS}
\label{sec:diff acc}

Similarly to the effects of aerodynamic gas drag on a binary planetesimal
\citep{2011ApJ...733...56P} differential force due to the interaction
with the gas can shear apart the binary. In this section we calculate
the differential acceleration. In appendix \ref{sec:binary stability}
we define the \emph{GDF shearing radius} and show that it is much
larger than the Hill radius of the binary, thus concluding that BPs
are stable to shearing effects by GDF. 

Consider, for simplicity, a circular and co-planar binary with masses
$m_{b}\ge m_{s}$ and total binary mass $m_{bin}=m_{b}+m_{s}$. The magnitudes of the binary velocities of each body are $v_{b}=(m_{s}/m_{bin})v_{bin}$
and $v_{s}=(m_{b}/m_{bin})v_{bin}$, where the binary velocity is given in eq. \ref{eq:vbin}. The relative velocities of each
body with respect to the gas are 
\begin{equation}
\boldsymbol{v}_{b,rel}=\boldsymbol{v}_{out}+\boldsymbol{v}_{b}=v_{out}\boldsymbol{\hat{\varphi}}(\nu_{out})+\frac{m_{s}}{m_{bin}}v_{bin}\boldsymbol{\hat{\varphi}_{bin}}(\nu_{bin,b}),\label{eq:vbrel}
\end{equation}
 and 
\begin{equation}
\boldsymbol{v}_{s,rel}=\boldsymbol{v}_{out}+\boldsymbol{v}_{s}=v_{out}\boldsymbol{\hat{\varphi}}(\nu_{out})+\frac{m_{b}}{m_{bin}}v_{bin}\boldsymbol{\hat{\varphi}_{bin}}(\nu{}_{bin,s}),\label{eq:vsrek}
\end{equation}
 where $v_{out}\equiv\varepsilon v_{K}$ is the center of mass velocity,
$\varepsilon\approx3H_{0}^{2}$ is the relative velocity scaling between
the gas and a single planetesimal in a circular orbit (see paper I
for details); $\boldsymbol{\hat{\varphi}}\equiv(-\sin\nu_{out},\cos\nu_{out})$
is the unit vector of the binary's center of mass around the star;$\nu_{out}=\Omega_{out}t$
is the true anomaly of the primary; $\Omega_{out}=\sqrt{GM_{\star}/a_{out}^{3}}$
is the mean motion of the primary around the star;$\boldsymbol{\hat{\varphi}_{bin}}=(-\sin\nu_{bin},\cos\nu_{bin})$
is the unit vector of the binary system; $\nu_{bin}=\Omega_{bin}t$
is the binary true anomaly; and $\Omega_{bin}=\sqrt{Gm_{min}/(fr_{H})^{3}}$
is mean motion of the binary. Note that 
\begin{equation}
\Omega_{bin}=\sqrt{\frac{Gm_{bin}}{f^{3}r_{H}^{3}}}=f^{-3/2}\Omega_{out}\gg\Omega_{out}.\label{eq:sep. of times}
\end{equation}

For circular binary, there is always a phase difference of $\pi$
in the binary's true anomalies, (i.e. $|\nu_{bin,s}-\nu_{bin,b}|=\pi$),
hence in Eqs. (\ref{eq:vbrel}) and (\ref{eq:vsrek}) $\boldsymbol{\hat{\varphi}_{bin}}(\nu_{bin,b})=-\boldsymbol{\hat{\varphi}_{bin}}(\nu_{bin,s})$.
We chose the initial phase such that eqn. (\ref{eq:vbrel}) has a
minus sign in the second term, and omit the dependence on the true
anomaly to simplify notation.

\subsection{Linear force regime}

Consider the subsonic case, where the eccentricity of the center of
mass (i.e. the primary), $e_{out}$ is small or zero. In this case,
the relative velocities involved are small compared to the sound speed
$c_{s}$, and the acceleration of each mass is linearly proportional
to the velocity, i.e. 
\begin{equation}
\boldsymbol{a}_{GDF,i}=C_{sub}m_{i}\boldsymbol{v_{rel}},
\label{agdfaux}
\end{equation}
where $C_{sub}=4\pi G^{2}\rho_{g}/3c_{s}^{3}$. The differential acceleration
is 
\begin{equation}
\label{eq:6-1}
\begin{aligned}
\frac{\boldsymbol{\Delta a_{GDF}}}{C_{sub}}  =  m_{b}\boldsymbol{v}_{b}-m_{s}\boldsymbol{v}_{s}\\
  =  m_{b}(\varepsilon v_{K}\boldsymbol{\hat{\varphi}}-\frac{m_{s}}{m_{bin}}v_{bin}\boldsymbol{\hat{\varphi}_{bin}}) 
 \\
  - m_{s}(\varepsilon v_{K}\boldsymbol{\hat{\varphi}}+\frac{m_{b}}{m_{bin}}v_{bin}\boldsymbol{\hat{\varphi}_{bin}}).
  \end{aligned}
\end{equation}
Collecting terms proportional to either $\boldsymbol{\hat{\varphi}}$
or $\boldsymbol{\hat{\varphi}_{bin}}$ we get 
\begin{equation}
\frac{\boldsymbol{\Delta a_{GDF}}}{C_{sub}}=m_{b}(1-q)\varepsilon v_{K}\boldsymbol{\hat{\varphi}}-2\mu v_{bin}\boldsymbol{\hat{\varphi}_{bin}},\label{eq:linear1}
\end{equation}
where $\mu=m_{b}m_{s}/(m_{b}+m_{s})=m_{s}/(1+q)$ is the reduced mass, and $q=m_s/m_b$ is the binary mass ratio.

\subsection{Supersonic regime}

If the eccentricity of the primary is larger than the aspect ratio,
(i.e.$e_{out}>2H_{0}$), the relative velocity of the primary $v_{out}$
is supersonic. In the supersonic regime, the acceleration of each
body is proportional to $\boldsymbol{v}_{rel}/v_{rel}^{3}$, i.e.
\begin{equation}
\boldsymbol{a}_{GDF,i}=C_{super}^{(i)}m_{i}\boldsymbol{v_{rel}}/v_{rel}^{3},\label{eq:agdf}
\end{equation}
where 
\begin{equation}
C_{super}^{(i)}=4\pi G^{2}\rho_{g}\left(\ln\Lambda+\frac{1}{2}\ln\left(1-1/\mathcal{M}_{i}^{2}\right)\right),\label{eq:Csupersonic}
\end{equation}
 with $\mathcal{M}_{i}$ being the instantaneous Mach number of body
$i=b,s$. The second term in eqn. (\ref{eq:Csupersonic}) is negligible
compared to the Coulomb logarithm and can be omitted. Thus, $C_{super}^{(i)}\approx C_{super}$,
where $C_{super}$ includes only the first term. Using the relative
velocities in eqns. (\ref{eq:vbrel}) and (\ref{eq:vsrek}), the differential
acceleration is 
\begin{equation}
\begin{aligned}
\frac{\boldsymbol{\Delta a}_{GDF}}{C_{super}}  =  m_{b}\frac{\boldsymbol{v}_{rel,b}}{v_{rel,b}^{3}}-m_{s}\frac{\boldsymbol{v}_{rel,s}}{v_{rel,s}^{3}}\\
  =  m_{b}\frac{\varepsilon v_{K}\boldsymbol{\hat{\varphi}}-(m_{s}/m_{bin})v_{bin}\boldsymbol{\hat{\varphi}_{bin}}}{|\varepsilon v_{K}\boldsymbol{\hat{\varphi}}-(m_{s}/m_{bin})v_{bin}\boldsymbol{\hat{\varphi}_{bin}}|^{3}}\\
  -m_{s}\frac{\varepsilon v_{K}\boldsymbol{\hat{\varphi}}+(m_{b}/m_{bin})v_{bin}\boldsymbol{\hat{\varphi}_{bin}}}{|\varepsilon v_{K}\boldsymbol{\hat{\varphi}}+(m_{b}/m_{bin})v_{bin}\boldsymbol{\hat{\varphi}_{bin}}|^{3}}.\label{eq:dasuper1}
\end{aligned}
\end{equation}
In order for the supersonic regime to apply for mass range of planetesimals
we consider, the primary orbit must be sufficiently eccentric or inclined.
For large $e_{out},$ the relative velocity of the primary is dominated
by the eccentricity. In this case $\varepsilon\approx e_{out}.$ It
is convenient to define $\beta=v_{bin}/\varepsilon v_{K}\ll1$ to
get 
\begin{equation}
\begin{aligned}
\frac{\boldsymbol{\Delta a}_{GDF}}{C_{super}}  =  m_{b}\frac{e_{out}v_{K}(\boldsymbol{\hat{\varphi}}-(m_{s}/m_{bin})\beta\boldsymbol{\hat{\varphi}_{bin}})}{e_{out}^{3}v_{K}^{3}|\boldsymbol{\hat{\varphi}}-(m_{s}/m_{bin})\beta\boldsymbol{\hat{\varphi}_{bin}}|^{3}}\\
-m_{s}\frac{e_{out}v_{K}(\boldsymbol{\hat{\varphi}}+(m_{b}/m_{bin})\beta)\boldsymbol{\hat{\varphi}_{bin}}}{e_{out}^{3}v_{K}^{3}|\boldsymbol{\hat{\varphi}}+(m_{b}/m_{bin})\beta\boldsymbol{\hat{\varphi}_{bin}}|^{3}}.\label{eq:dasuper2}
\end{aligned}
\end{equation}
In appendix \ref{sec:Derivation} we show that to first order in $\beta$,
eqn. \ref{eq:dasuper2} can be written as 
\begin{equation}
\frac{\boldsymbol{\Delta a}_{GDF}}{C_{super}}=\frac{m_{b}e_{out}v_{K}(1-q)\boldsymbol{\hat{\varphi}}+3qm_{b}v_{bin}\boldsymbol{\hat{\varphi}}-2\mu v_{bin}\boldsymbol{\hat{\varphi}_{bin}}}{e_{out}^{3}v_{K}^{3}}.\label{eq:dasuper}
\end{equation}

\section{EVOLUTION OF BINARY ORBITAL
ELEMENTS}
\label{sec:EVOLUTION OF ORBITAL ELEMENTS}

In paper I we studied the evolution of single planetesimals by following
the primary orbital semi-major axis $a_{out},$ eccentricity $e_{out}$
and inclination $I_{out}.$ For BPs, we will study the evolution of
the inner binary separation $a_{bin},$ binary eccentricity $e_{bin}$
and binary inclination angle $I_{bin}$, relative to the orbital plane
of the primary around the star. In this section we derive the timescales
for the change of binary separation and comment on changes in the
binary eccentricity and inclination.

In its most general form, the differential acceleration in eqns. (\ref{eq:linear1})
and (\ref{eq:dasuper}) can be decomposed into 
\begin{equation}
\boldsymbol{\Delta a}_{GDF}=\mathcal{A}_{out}\boldsymbol{\hat{\varphi}}+\mathcal{A}_{bin}\boldsymbol{\hat{\varphi}_{bin}},\label{eq:da_sep}
\end{equation}
where $\mathcal{A}_{out}$ and $\mathcal{A}_{bin}$ depend on the
specific regime and mass ratio. The total angular momentum of the
binary per unit mass is $l=a_{bin}v_{bin}$, where $a_{bin}$ is the
binary separation. The change in the absolute angular momentum per
unit mass is given by the specific torque, i.e. $\dot{l}=|\boldsymbol{a}_{bin}\times\boldsymbol{\Delta a}_{GDF}|$.
The average torque over one binary period is 
\begin{equation}
\langle|\boldsymbol{a_{bin}}\times\boldsymbol{\Delta a}_{GDF}|\rangle = \frac{1}{2\pi}\intop_{0}^{2\pi}|\boldsymbol{a_{bin}}\times\boldsymbol{\Delta a}_{GDF}|d\nu_{bin}.\label{eq:average1}
\end{equation}

Let us consider a circular binary orbit, where $\boldsymbol{\hat{a}_{bin}}\perp\boldsymbol{\hat{\varphi}_{bin}}$,
and the averaged torque is 
\begin{equation}
\begin{aligned}
\langle|\boldsymbol{a_{bin}}\times\boldsymbol{\Delta a}_{GDF}|\rangle  =  \frac{a_{bin}}{2\pi}\intop_{0}^{2\pi}|\boldsymbol{\hat{\varphi}_{bin}\cdot}\boldsymbol{\Delta a}_{GDF}|d\nu_{bin}\\
  =  \frac{a_{bin}}{2\pi}\intop_{0}^{2\pi}|\mathcal{A}_{cm}\boldsymbol{\hat{\varphi}\cdot\hat{\varphi}_{bin}}+\mathcal{A}_{bin})|d\nu_{bin}=\mathcal{A}_{bin}a_{bin}.\label{eq:average2}
\end{aligned}
\end{equation}
 The last equality comes from the assumption on separation of times,
given in Eqn. (\ref{eq:sep. of times}). In this approximation, $\langle\boldsymbol{\hat{\varphi}}\cdot\boldsymbol{\hat{\varphi}_{bin}}\rangle\approx0$
since $\boldsymbol{\hat{\varphi}_{bin}}$ oscillates very rapidly
with respect to $\boldsymbol{\hat{\varphi}},$ therefore $\boldsymbol{\hat{\varphi}}$
is constant in the frame of the binary (similar to the approach used by \citealp{2011ApJ...733...56P}). The validity
of the approximation depends on the binary separation (i.e. $\langle\boldsymbol{\hat{\varphi}}\cdot\boldsymbol{\hat{\varphi}_{bin}}\rangle=\langle\cos((\Omega_{bin}-\Omega)t)\rangle=f^{3/2}+\mathcal{O}(f^{3})$
). It is roughly $<3\%$ for $f=0.1$ but could be significant for
larger separations. The inspiral timescale is then given by 
\begin{equation}
\tau_{ins}=\frac{1}{2}\frac{l}{\dot{l}}=\frac{v_{bin}}{\mathcal{A}_{bin}}.\label{eq:inspiral timescale}
\end{equation}
This is the timescale for which the binary loses approximately half
of its initial separation, similarly to the definition of migration
timescale in Paper I. 

An alternative derivation of eqn (\ref{eq:inspiral timescale}) can
be obtained from considering the equations of motion for a disturbing
force. For the binary orbital elements, it is best to describe the
change in orbital elements due to the disturbing force, and consequently,
the disturbing acceleration $\boldsymbol{\Delta A}=\Delta A_{r}\hat{\boldsymbol{r}}+\Delta A_{\varphi}\hat{\boldsymbol{\varphi}}+\Delta A_{z}\hat{\boldsymbol{z}}$
\citep{1999ssd..book.....M}, similarly to the case of single planetesimals
in paper I 
\begin{equation}
\begin{aligned}
 \frac{da_{bin}}{dt}  =  2\frac{a_{bin}^{3/2}}{\sqrt{Gm_{bin}(1-e_{bin}^{2})}}\\ \times [\Delta A_{r}e_{bin}\sin\nu_{bin}+\Delta A_{\varphi}(1+e_{bin}\cos\nu_{bin})]\label{eq:dabindt}
\end{aligned}
\end{equation}
\begin{equation}
\begin{aligned}
\frac{de_{bin}}{dt}  =  \sqrt{\frac{a_{bin}(1-e_{bin}^{2})}{Gm_{bin}}}\\ \times [\Delta A_{r}\sin\nu_{bin}+\Delta A_{\varphi}(\cos\nu_{bin}+\cos E_{bin})]\label{eq:debindt}
\end{aligned}
\end{equation}

\begin{equation}
\frac{dI_{bin}}{dt}  =  \sqrt{\frac{a(1-e_{bin}^{2})}{Gm_{bin}}}\frac{\Delta A_{z}\cos(\omega_{bin}+\nu_{bin})}{1+e_{bin}\cos\nu_{bin}},\label{eq:dibindt}
\end{equation}
where $E_{bin}$ is the eccentric anomaly and $\omega_{bin}$ is the
argument of peri-center of the binary. For circular binary, eqn (\ref{eq:dabindt})
reduces to 
\[
\frac{da_{bin}}{dt}=2\frac{a_{bin}^{3/2}}{\sqrt{Gm_{bin}}}\Delta A_{\varphi},
\]
and the inspiral time, 
\begin{equation}
\tau_{ins}=\frac{a_{bin}}{\dot{a}_{bin}}=a_{bin}\sqrt{\frac{Gm_{bin}}{a_{bin}^{3}}}\frac{1}{2\Delta A_{\varphi}}=\frac{v_{bin}}{2\Delta A_{\varphi}},\label{eq:inspiral2}
\end{equation}
 is the same as in eqn. (\ref{eq:inspiral timescale}), where $\Delta A_{\varphi}$
is recognized as $\mathcal{A}_{bin}$ when averaged over binary orbital
period.

Given the latter, alternative derivation of eqn. (\ref{eq:inspiral2}),
several points naturally rise. First, eqn (\ref{eq:dabindt}) is correct
even if the binary has a small inner eccentricity, similarly to paper
I. In addition, for binary with $f<0.1$, the terms proportional to
trigonometric function of $\nu_{bin}$ are averaged out to zero over
one binary orbit, due to separation of times. Hence, $e_{bin}$ and
$I_{bin}$ are constant at first approximation.

In the following we will calculate the inspiral timescale for each
regime.

\subsection{Linear regime}
\label{sec:LINEAR REGIME}

If the primary is in a nearly circular orbit, the Mach numbers involved
in the GDF force are small, and we are able to apply the approximation
of linear regime. In the linear regime, using Eqns. (\ref{eq:inspiral timescale})
and (\ref{eq:inspiral timescale}) the inspiral timescale is 
\begin{equation}
\tau_{ins}=\frac{1}{4C_{sub}\mu}=\frac{3(1+q)}{16\pi}\frac{c_{s}^{3}}{G^{2}\rho_{g}m_{s}}.\label{eq:inspiral linear}
\end{equation}
For equal mass binaries with $m\sim2\cdot10^{23}g$ we get $\tau_{ins}\sim0.4$Myr. The inspiral timescale is comparable to the protoplanetary disk lifetime of ~$10$ Myr (\citealp{2014ApJ...793L..34P}) for planetesimals with mass of $m\sim10^{22}g$. Thus, planetesimals with larger mass will inevitably merge, while planetesimals with lower mass can survive after the dispersal of the protoplanetary disk. Further in the disk, the gas density decreases as $\rho_g \propto a^{-16/7}$, thus the inspiral timescale is longer, and the critical mass for merger is larger. This fact has implications on the frequency and properties of BPs in the debris disk, and is discussed in the discussion. 

It is worth noting that the merger timescale is obtained by integrating
eqn. (\ref{eq:inspiral timescale}) to obtain $\tau_{merge}=\tau_{ins}\ln(a_{in}/R_{c})$
where $a_{in}$ is the initial separation and $R_{c}$ is the contact
radius of the binary \citep{2011ApJ...733...56P}.

Note that the timescale in eqn. (\ref{eq:inspiral timescale}) is
shorter than the migration timescale $\tau_{a}$ in paper I (eqn.
11) by a factor of $\approx H_{0}^{2}(1+q)$. It is due to the fact
that for single planetesimals, the length scale involved is $a_{out}\sim v_{K}/\Omega\sim H_{0}^{-2}v_{rel}/\Omega$,
while for BPs, it is $a_{bin}\sim v_{bin}/\Omega_{bin}$. In the linear
regime ($\mathcal{M}\ll1$) where $F_{GDF}\propto v_{rel}$, the characteristic
time for the change in $a_{out}$, taking a relative velocity of $\sim v_{rel}$
is$\sim H_{0}^{-2}$ longer than the characteristic time for the change
in $a_{bin}$ due to GDF with relative velocity $\sim v_{bin}.$

It is worth noting that $\mathcal{A}_{out}=0$ for $q=1$. In addition,
if $q\ll1,$ then $\mathcal{A}_{out}\gg\mathcal{A}_{bin},$ so that
even if separation of times (Eqn. \ref{eq:sep. of times}) is satisfied,
the larger body inspirals to the center of mass much faster that its
companion. Thus, for the remaining time of the evolution $v_{b},$
and consequently $\mathcal{A}_{out}$ vanish. Hence the inspiral time
in eqn. (\ref{eq:inspiral timescale}) is valid.

\subsubsection{Deviations from the linear regime}

In paper I we have seen that for single planetesimals, the equations
and times scale with the mass as $\tau\propto m_{p}^{-1}$. The latter
is true for any Mach number, not only in the linear regime. One of
the reasons is that the only relevant length scale is the semi-major
axis $a_{out}$ (and consequently the only relevant velocity scale
is $v_{K},$ which is unaffected by the mass of the planetesimal).
For BPs this is not the case. An additional length scale, the binary
separation $a_{bin}$, exists (and consequently an additional velocity
scale $v_{bin}$), and it depends on the mass of the binary $m_{bin}.$
Changing the mass of the binary will result in a change in $v_{bin}$,
and consequently in the binary's Mach number and it's dynamical behavior
due to GDF. However, for the linear regime, $\tau_{ins}$ given by
eqn. (\ref{eq:inspiral linear}), will not depend neither on $a_{bin}$,
nor on $v_{bin}$, and will scale as $\propto\mu^{-1}$. 

We can calculate the deviation from the linear regime in terms of
the Mach number, taking the next term in eqn. (\ref{dfdf-1}). The
non-linear contribution is $\Delta a_{GDF}=\Delta a_{GDF,lin}(1+3\mathcal{M}^{2}/5)$
which leads to 
\begin{equation}
\tau_{ins}=\tau_{ins,lin}\left(1+\frac{3}{5}\mathcal{M}^{2}\right)^{-1}\approx\tau_{ins,lin}\left(1-\frac{3}{5}\mathcal{M}^{2}\right).\label{eq:deviation}
\end{equation}

\subsection{Supersonic regime}
\label{sec:SUPERSONIC REGIME}
Consider now an equal mass binary on an eccentric orbit, such that
$v_{cm}>c_{s}$ (or $e_{out}\gtrsim2H_{0}$). Generally, the Mach
number changes with time. Let us decompose the Mach number into the
orbital Mach number $\mathcal{M}_{cm}\equiv|\boldsymbol{v}_{out}|/c_{s}$,
and the binary Mach number $\mathcal{M}_{bin}\equiv|\boldsymbol{v}_{bin}|/c_{s}$,
similar to the approach of \citet{2014ApJ...794..167S}. We can use
the impulse approximation, where the torque is maximal when the line
connecting the binary components is parallel to the sun-CM line (i.e.
$\nu_{out}=\nu_{bin}$) and where the binary Mach numbers, $\mathcal{M}_{bin}$
is maximized / minimized. 

\begin{figure}
\begin{centering}
\includegraphics[width=8.5cm]{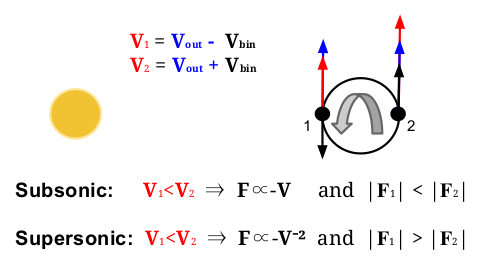}
\par\end{centering}

\caption{\label{fig:Differential-forces-for}A sketch of a system containing
the Sun and a binary planetesimal in orbit around it. Relative velocities
are marked in red, the primary CM velocity in blue, and the binary
velocity in black lines. Differential forces for the subsonic and
the supersonic binaries are shown. Note that $v_{2}$ is always larger
than $v_{1}$. For $\mathcal{M}<1$, the GDF force increases with
velocity, and decreases with velocity for $\mathcal{M}>1$, hence
the total torque is reversed (see details in text). }
\end{figure}

Fig. \ref{fig:Differential-forces-for} illustrates the difference
in the total torque between the subsonic and the supersonic binaries.
For $\mathcal{M}_{1}<\mathcal{M}_{2}<1$, the induced torques are
$T_{1}=a_{bin}F_{1}/2<a_{bin}F_{2}/2=T_{2},$ and the total torque
is positive. On the other hand, for $\mathcal{M}_{2}>\mathcal{M}_{1}>1$,
the change in $\mathcal{I}(\mathcal{M})$ with respect to is $\mathcal{M}$
is negative, (i.e.$\text{ \text{sgn}(}d\mathcal{I}(\mathcal{M})/d\mathcal{M})=-1$),
hence $T_{1}=a_{bin}F_{1}/2>a_{bin}F_{2}/2=T_{2}$ and the total torque
is negative. As long as the supersonic orbit pertains, the torque
is on the opposite direction and it adds angular momentum to the binary,
and the binary gains orbital energy leading to increase in the binary
separation.

The phenomena of increasing $a_{bin}$ in the supersonic CM orbit
can be connected to Fig. (12) of \citet{2014ApJ...794..167S},
where they calculate averaged radial and azimuthal forces for various
values of $\mathcal{M}_{cm}$ and $\mathcal{M}_{bin}$. If both are
subsonic, then $\langle F_{\varphi}\rangle$ is negative and \citet{2014ApJ...794..167S}
quote $\langle F_{\varphi}\rangle=-0.18$ (in normalized code units)
for $\mathcal{M}_{cm}=0.5$ and $\mathcal{M}_{bin}=0.3$, but in our
case $\mathcal{M}_{bin}\ll1$, so $\langle F_{\varphi}\rangle$ is
much lower. For $\mathcal{M}_{cm}=2.0$ and $\mathcal{M}_{bin}=0.5$,
\citet{2014ApJ...794..167S} quote that ``the average of $\langle F_{\varphi}\rangle$
is close to zero''. It is plausible that for our set of parameters
(i.e. $\mathcal{M}_{bin}\ll1$), $\langle F_{\varphi}\rangle$ is
positive. Since for the supersonic case $dI(\mathcal{M})/d\mathcal{M}$
issteeper at the vicinity of the peak, we expect $|\boldsymbol{\Delta a}_{GDF}|$
to grow as the orbit circularizes, and then abruptly to change sign.
Then, for $\mathcal{M}_{cm}\lesssim1+\mathcal{M}_{bin}$ GDF becomes
very efficient in extracting angular momentum, and the binary separation
will sharply decrease.

Since the torque is reversed in the supersonic regime, we will define
the ``break-up timescale'' similarly to the inspiral timescale as
in eqn. (\ref{eq:inspiral linear}). Using eqns. (\ref{eq:dasuper})
and (\ref{eq:inspiral timescale}), the break-up timescale is 
\[
\tau_{break}=\frac{e_{out}^{3}v_{K}^{3}}{4\mu C_{super}}=\frac{e_{out}^{3}v_{K}^{3}(1+q)}{16\pi m_{s}G^{2}\rho_{g}\ln\Lambda}.
\]

Comparison with the inspiral timescale in Eqn, (\ref{eq:inspiral linear})
yields 
\begin{equation}
\frac{\tau_{break}}{\tau_{ins}}=\frac{1}{3\ln\Lambda}\left(\frac{e_{out}}{H_{0}}\right)^{3},\label{eq:tbreak}
\end{equation}
and for $\ln\Lambda\simeq10$ we get $\tau_{break}\approx3.13\cdot10^{3}\tau_{ins}e_{out}^{3}.$ 

For larger eccentricities $e_{out}\gg H_{0}$, the break up timescale
is long and the effect is negligible. The timescales are comparable
for $e_{out}\approx0.068$, and $\tau_{break}\approx0.26\tau_{ins}$
near the trans-sonic limit where $e_{out}\sim2H_{0}$. The binary
break-up is determined by the initial separation. The decay of the
outer eccentricity is $\dot{e}_{out}\sim-e_{out}^{-2}$ (Paper I),
therefore the outer binary spends more time in higher eccentricities
states. Note that the break up timescale is comparable to the eccentricity damping timescale (paper I), thus $e_{out}$ cannot be kept large enough to stay in the supersonic regime, and it is expected that for binaries with small enough separation, the binary expansion inevitably stops before the break-up. The final fate of the binary depends mostly on its initial separation. Determining the rate of supersonic expansion and the critical
separation for break-up will be studied numerically and discussed in the discussion section.
It is worth noting that for $q=1$, $\mathcal{A}_{out}\sim\mathcal{A}_{bin},$
and eqn. (\ref{eq:inspiral timescale}) is valid due to separation
of times (Eqn. \ref{eq:sep. of times}). The case of $q\ll1$ is similar
to the linear regime. Hence the expression for the inspiral time in
eqn. (\ref{eq:inspiral timescale}) is also valid.

\section{NUMERICAL SET UP}
\label{sec:NUMERICAL SET UP}

In order to model the detailed evolution of BPs numerically, we use
GravityLab, an N-body integrator with a shared but variable time step,
using the Hermite 4th order integration scheme documented in \citet{1995ApJ...443L..93H}
. 

\begin{table}
\begin{centering}
\begin{tabular}{|c|c|c|c|c|c|}
\hline 
\#run & $Q$ & $q$ & $e_{out}$ & $f$ & $e_{bin}$\tabularnewline
\hline 
\hline 
1 & $2\cdot10^{-10}$ & $1$ & $0$ & $0.3$ & $0$\tabularnewline
\hline 
2 & $2\cdot10^{-10}$ & $1$ & $0$ & $0.3$ & $0.4$\tabularnewline
\hline 
3 & $2\cdot10^{-10}$ & $1$ & $0$ & $0.1$ & $0$\tabularnewline
\hline 
4 & $2\cdot10^{-10}$ & $1$ & $0$ & $0.1$ & $0.4$\tabularnewline
\hline 
5 & $2\cdot10^{-10}$ & $1$ & $0.1$ & $0.1$ & $0$\tabularnewline
\hline 
6 & $2\cdot10^{-10}$ & $1$ & $0.1$ & $0.1$ & $0.4$\tabularnewline
\hline 
7 & $2\cdot10^{-10}$ & $1$ & $0.3$ & $0.1$ & $0$\tabularnewline
\hline 
8 & $2\cdot10^{-8}$ & $1$ & $0$ & $0.1$ & $0$\tabularnewline
\hline 
9 & $2\cdot10^{-8}$ & $1$ & $0$ & $0.1$ & $0.4$\tabularnewline
\hline 
10 & $2\cdot10^{-8}$ & $1$ & $0.1$ & $0.1$ & $0$\tabularnewline
\hline 
11 & $2\cdot10^{-8}$ & $1$ & $0.1$ & $0.1$ & $0.4$\tabularnewline
\hline 
12 & $10^{-8}$ & $10^{-2}$ & $0$ & $0.1$ & $0$\tabularnewline
\hline 
13 & $10^{-8}$ & $10^{-2}$ & $0.1$ & $0.1$ & $0$\tabularnewline
\hline 
\end{tabular}
\par\end{centering}

\caption{\label{tab:Various-runs-for}Various runs for different initial conditions.
The numbers are in simulation units. In all simulations, the semi-major
axis of the orbit around the Sun , $a_{out}$, is $1$. }
\end{table}

The simulation units are $G=M_{\odot}=a_{\odot}=1$, hence at 1AU,
$v_{K}=1,$ $T=2\pi$. We add a fiducial GDF force that mimics the
GDF force. At each step we calculate the additional external acceleration
and jerk due to GDF. More Details of the modified code are summarized
in the appendix of paper I. The disk properties are the same as in
paper I and we neglect accretion and any other sources of drag other
than GDF.

\begin{table}
\begin{centering}
\begin{tabular}{|c|c|c|c|c|}
\hline 
\#run & $Q$ & $q$ & $e_{out}$ & $I_{bin}$\tabularnewline
\hline 
\hline 
14 & $2\cdot10^{-8}$ & $1$ & $0$ & $0^{\circ}$\tabularnewline
\hline 
15 & $2\cdot10^{-8}$ & $1$ & $0$ & $30^{\circ}$\tabularnewline
\hline 
16 & $2\cdot10^{-8}$ & $1$ & $0$ & $60^{\circ}$\tabularnewline
\hline 
17 & $2\cdot10^{-8}$ & $1$ & $0$ & $120^{\circ}$\tabularnewline
\hline 
18 & $2\cdot10^{-8}$ & $1$ & $0$ & $150^{\circ}$\tabularnewline
\hline 
19 & $2\cdot10^{-8}$ & $1$ & $0$ & $180^{\circ}$\tabularnewline
\hline 
20 & $2\cdot10^{-8}$ & $1$ & $0.1$ & $0^{\circ}$\tabularnewline
\hline 
21 & $2\cdot10^{-8}$ & $1$ & $0.1$ & $30^{\circ}$\tabularnewline
\hline 
22 & $2\cdot10^{-8}$ & $1$ & $0.1$ & $60^{\circ}$\tabularnewline
\hline 
23 & $2\cdot10^{-8}$ & $1$ & $0.1$ & $120^{\circ}$\tabularnewline
\hline 
24 & $2\cdot10^{-8}$ & $1$ & $0.1$ & $150^{\circ}$\tabularnewline
\hline 
25 & $2\cdot10^{-8}$ & $1$ & $0.1$ & $180^{\circ}$\tabularnewline
\hline 
26 & $10^{-8}$ & $10^{-2}$ & $0$ & $0^{\circ}$\tabularnewline
\hline 
27 & $10^{-8}$ & $10^{-2}$ & $0$ & $60^{\circ}$\tabularnewline
\hline 
28 & $10^{-8}$ & $10^{-2}$ & $0$ & $120^{\circ}$\tabularnewline
\hline 
29 & $10^{-8}$ & $10^{-2}$ & $0.1$ & $0^{\circ}$\tabularnewline
\hline 
30 & $10^{-8}$ & $10^{-2}$ & $0.1$ & $60^{\circ}$\tabularnewline
\hline 
31 & $10^{-8}$ & $10^{-2}$ & $0.1$ & $120^{\circ}$\tabularnewline
\hline 
\end{tabular}
\par\end{centering}

\caption{\label{tab:incline}Various runs for different initial conditions.
The numbers are in simulation units. Mach number is not an initial
condition, but rather derived from the relative velocity of the initial
orbit. In all simulations, the semi-major axis to the center of mass
of the binary planetesimal, $a_{out}$ is $1$. }
\end{table}

\begin{figure*}
\begin{centering}
\includegraphics[height=5.5cm]{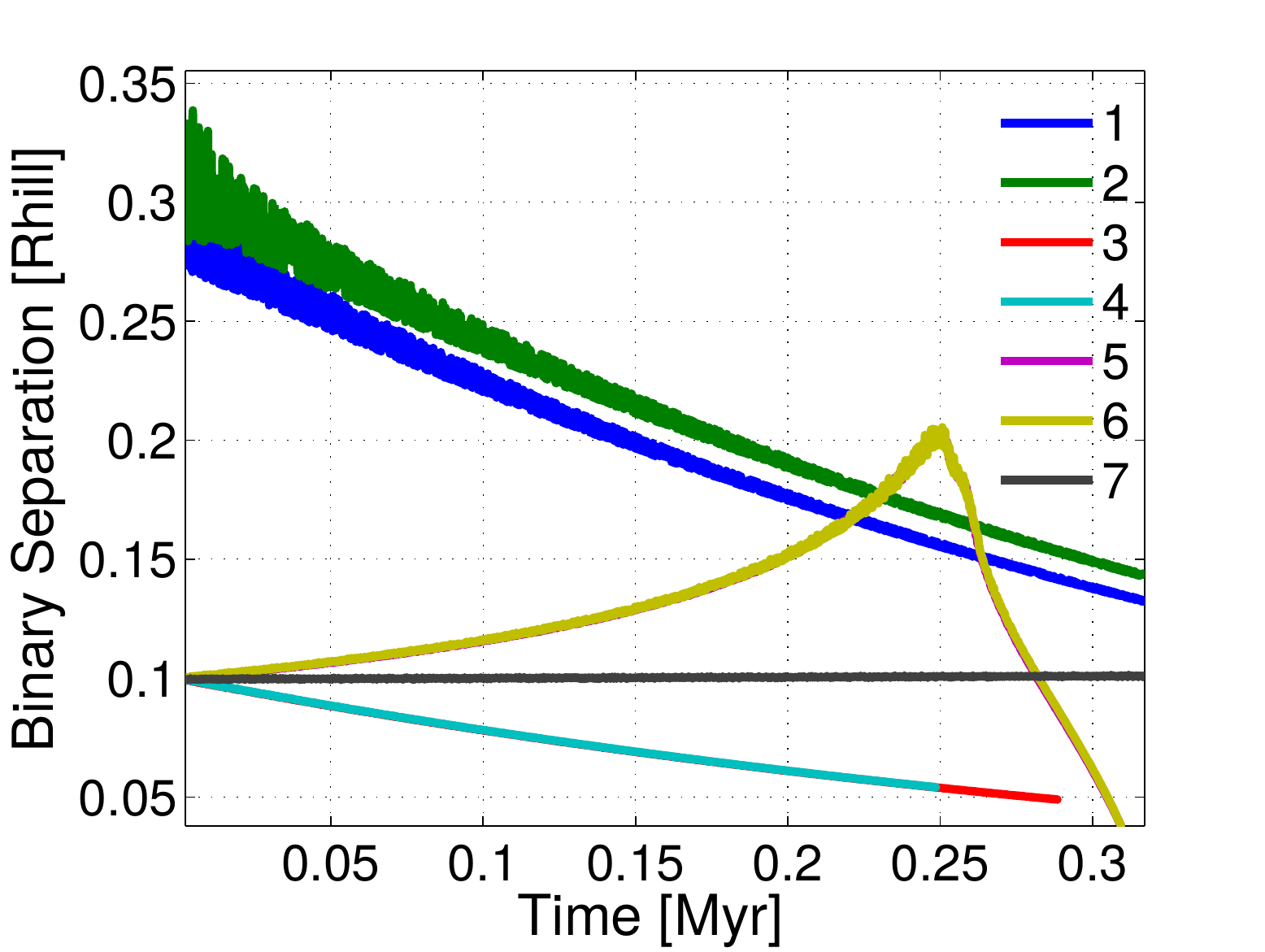}\includegraphics[height=5.5cm]{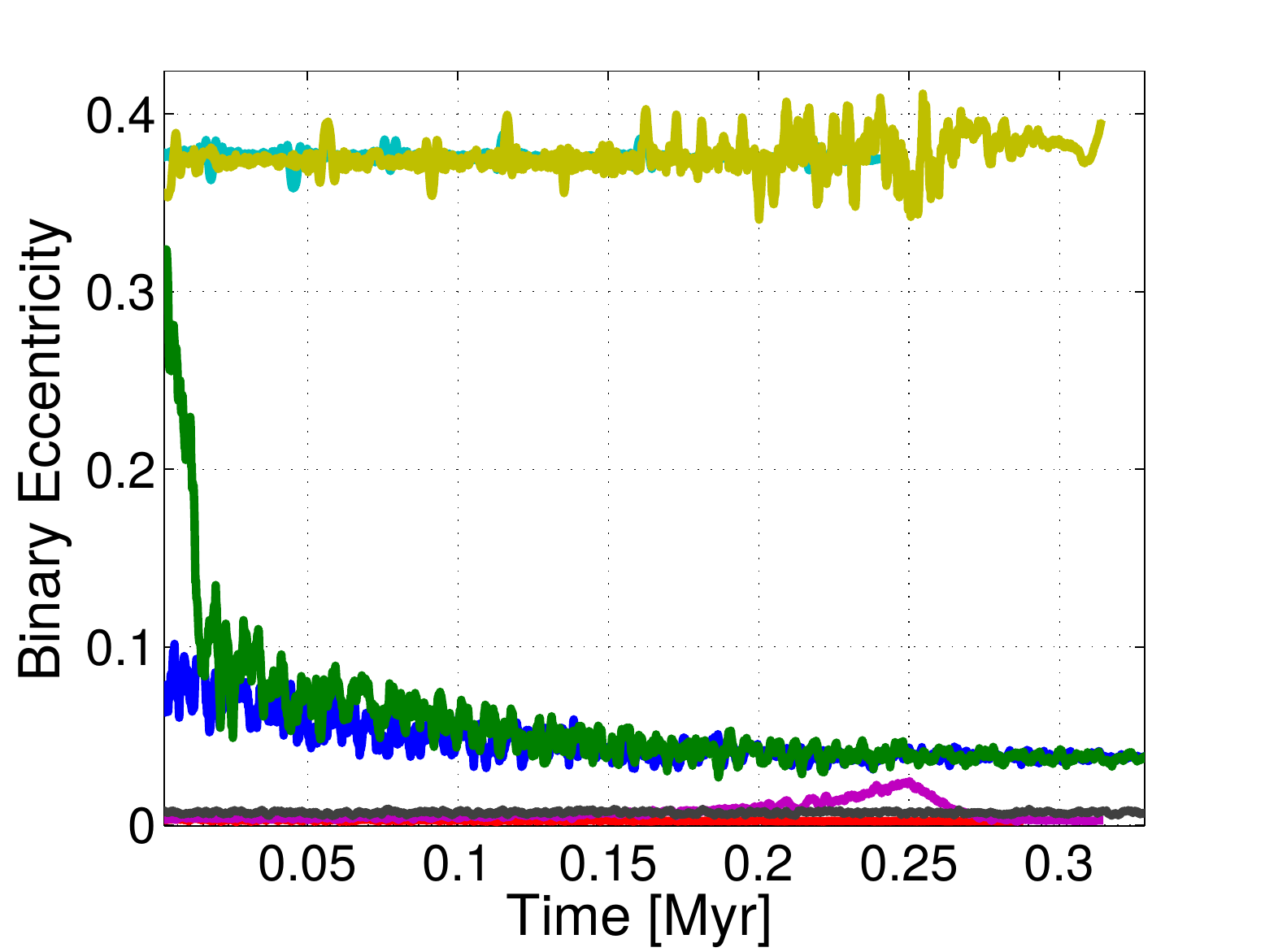}
\par\end{centering}

\begin{centering}
\includegraphics[height=5.5cm]{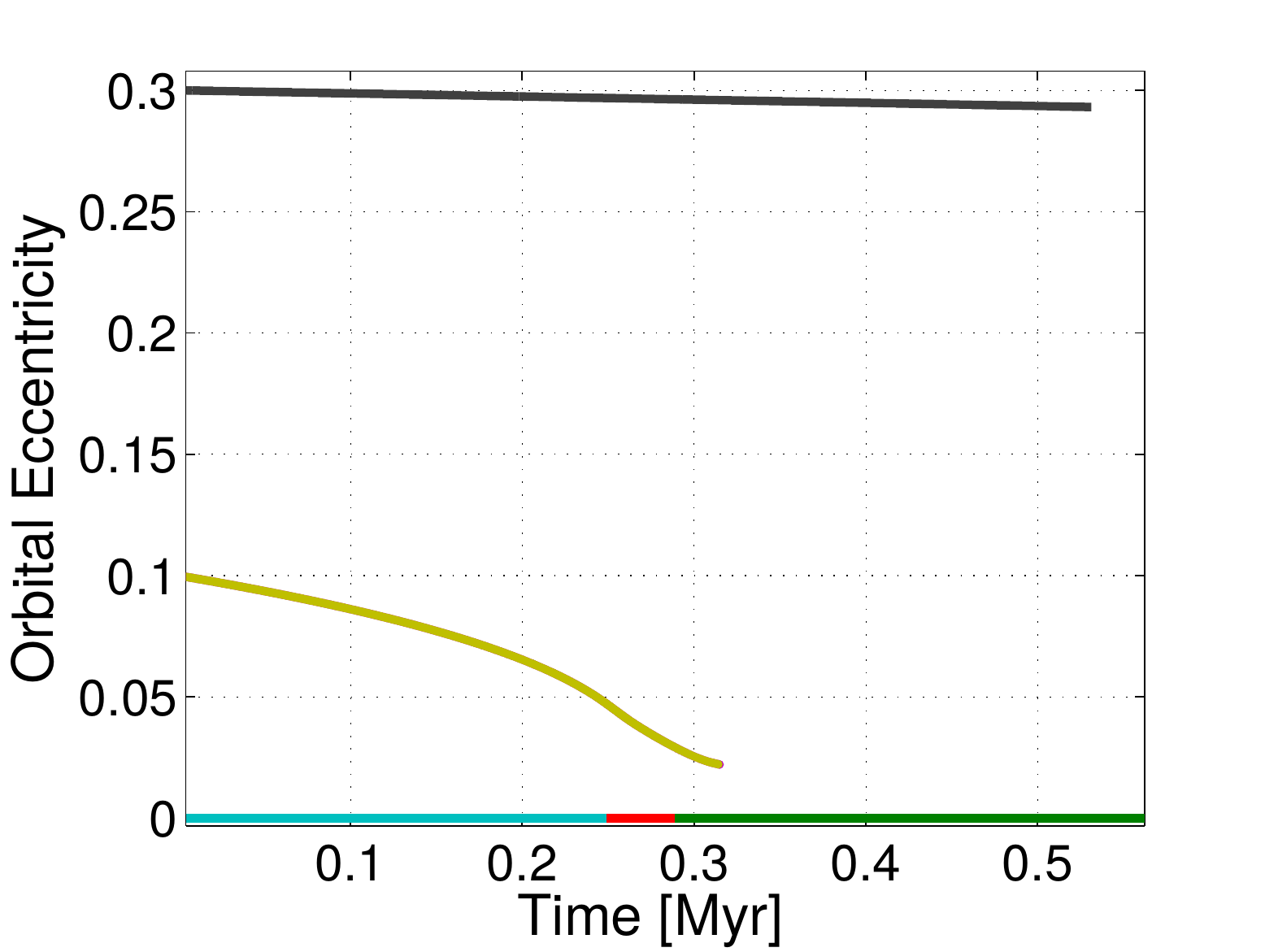}\includegraphics[height=5.5cm]{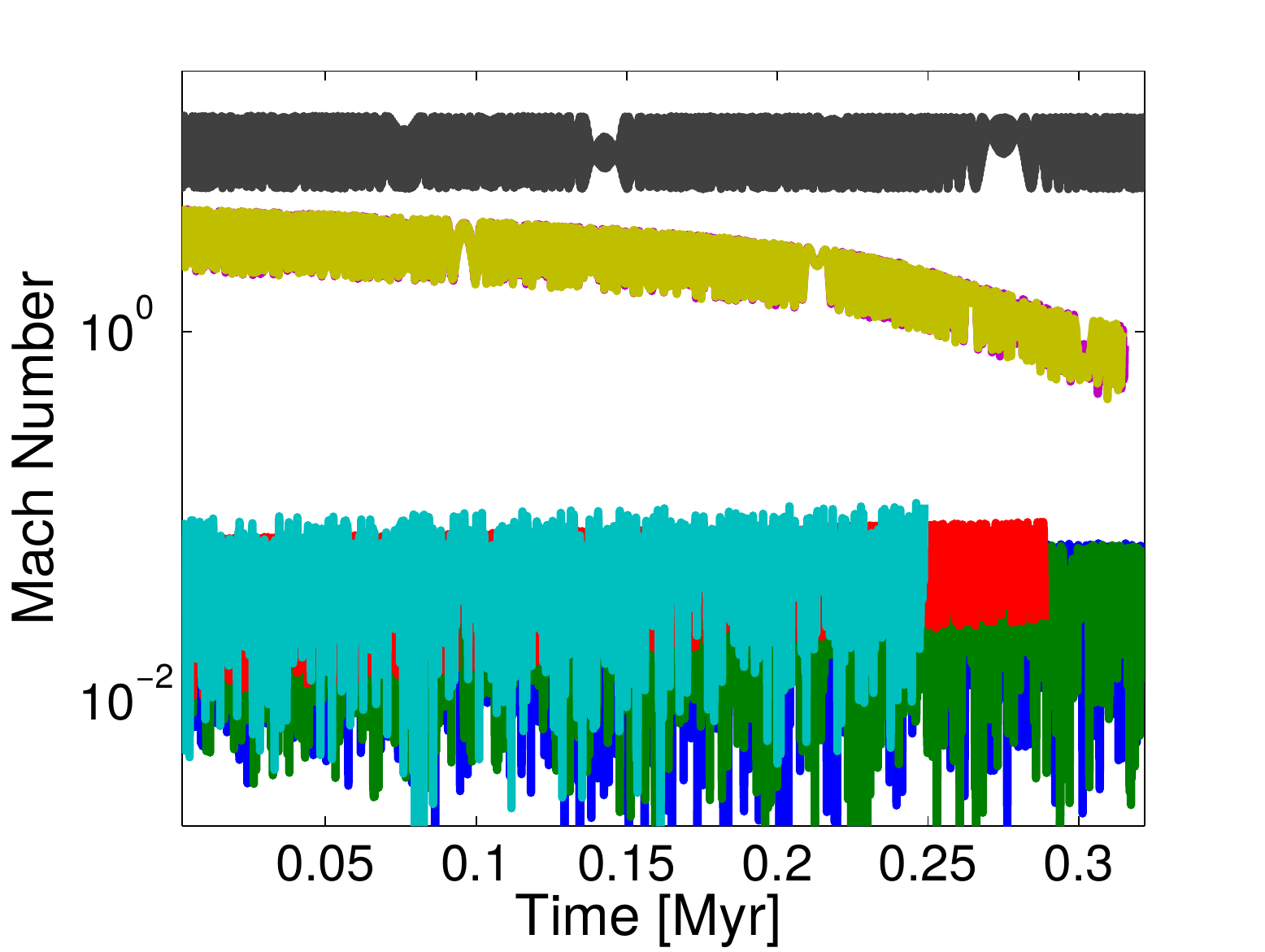}
\par\end{centering}

\caption{\label{fig:Results-for-runs}Evolution of co-planar binaries (runs
1-7). In each run the mass ratios are $Q=2\cdot10^{-10}$ and $q=1.$
Top left: binary separation vs. time. Top right: binary eccentricity
vs. time. For clarity, the data points are averaged such that each
value is an average of 10 nearest data points, reducing the fluctuation
noise. Bottom left: Orbital eccentricity vs. time. Bottom right: Mach
number vs. time.}
\end{figure*}

We run several N-body simulations with a central object and a bound
binary planetesimal with various initial conditions. We first present
the results for co-planar binaries ($I_{bin}=0$); which initial conditions
are summarized in table \ref{tab:Various-runs-for}. We then present
the results for inclined binaries; which initial conditions are summarized
in table \ref{tab:incline}.

\section{RESULTS}
\label{sec:RESULTS}

In the following we present the results of the evolution of BPs due
to GDF. We first explore the parameter space of co-planar BPs (Table
\ref{tab:Various-runs-for}), and then we present the results of inclined
BPs (Table \ref{tab:incline}).

In order to check eqns. \ref{eq:inspiral timescale} and \ref{eq:tbreak},
we ran several simulations with different values of $c_{s}$ and found
that $\tau_{ins}\propto c_{s}^{3}$ and that $\tau_{break}$ is independent
of $c_{s},$ as expected.

\subsection{Co-planar Binaries}
\label{sec:CO PLANAR}

\subsubsection{Equal mass binaries, $q=1$}

\begin{figure*}
\begin{centering}
\includegraphics[height=5.05cm]{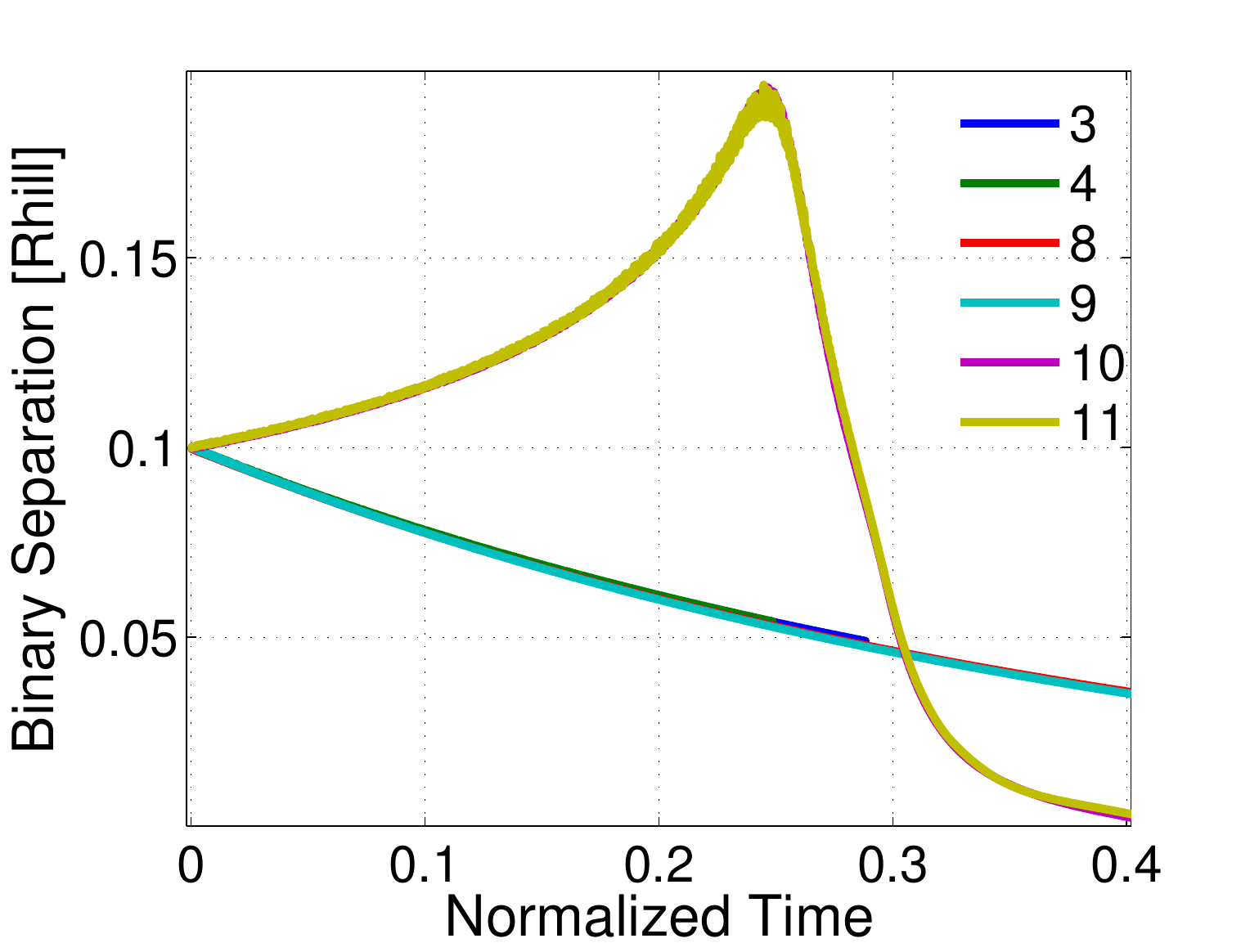}\includegraphics[height=5.05cm]{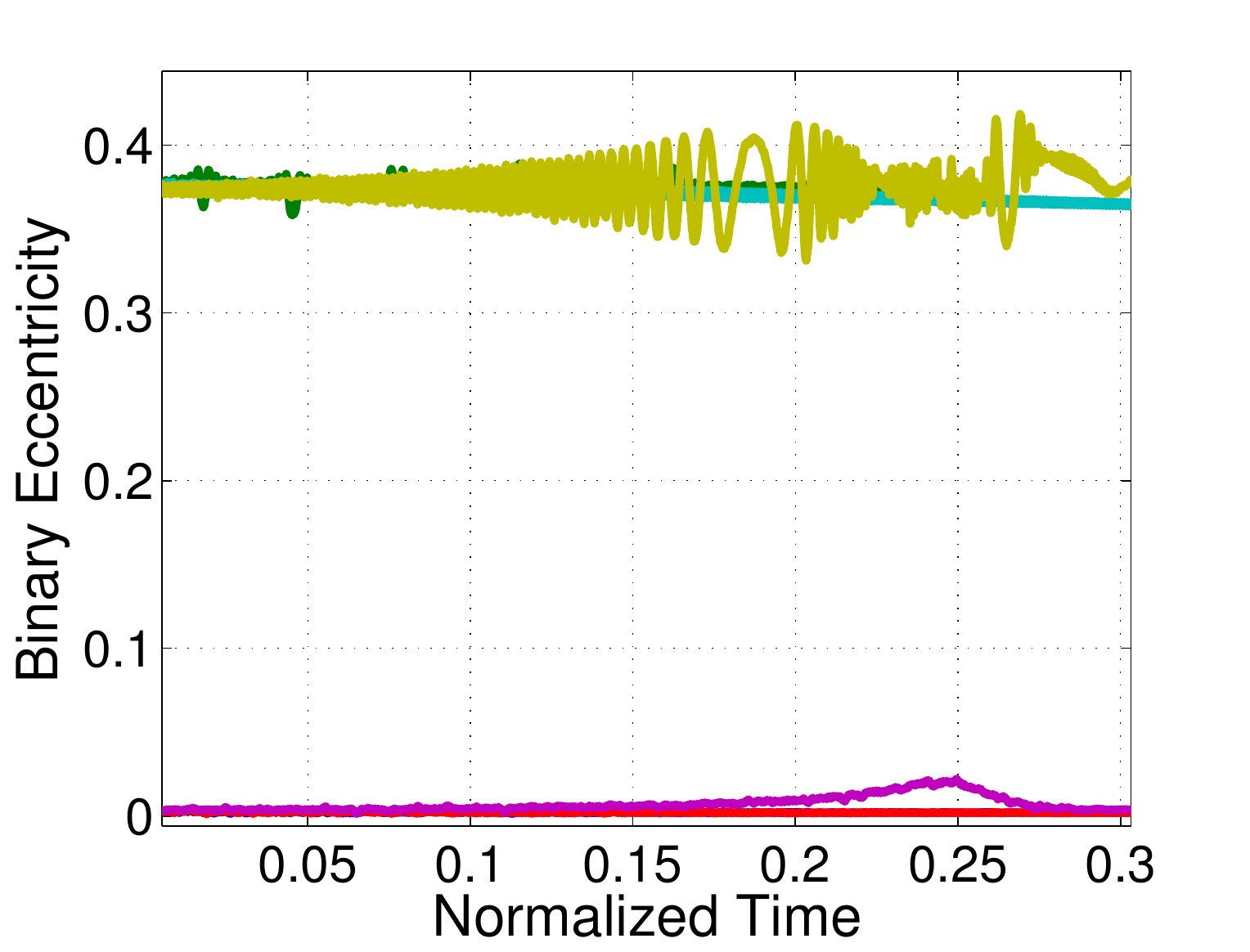}
\par\end{centering}

\begin{centering}
\includegraphics[height=5.05cm]{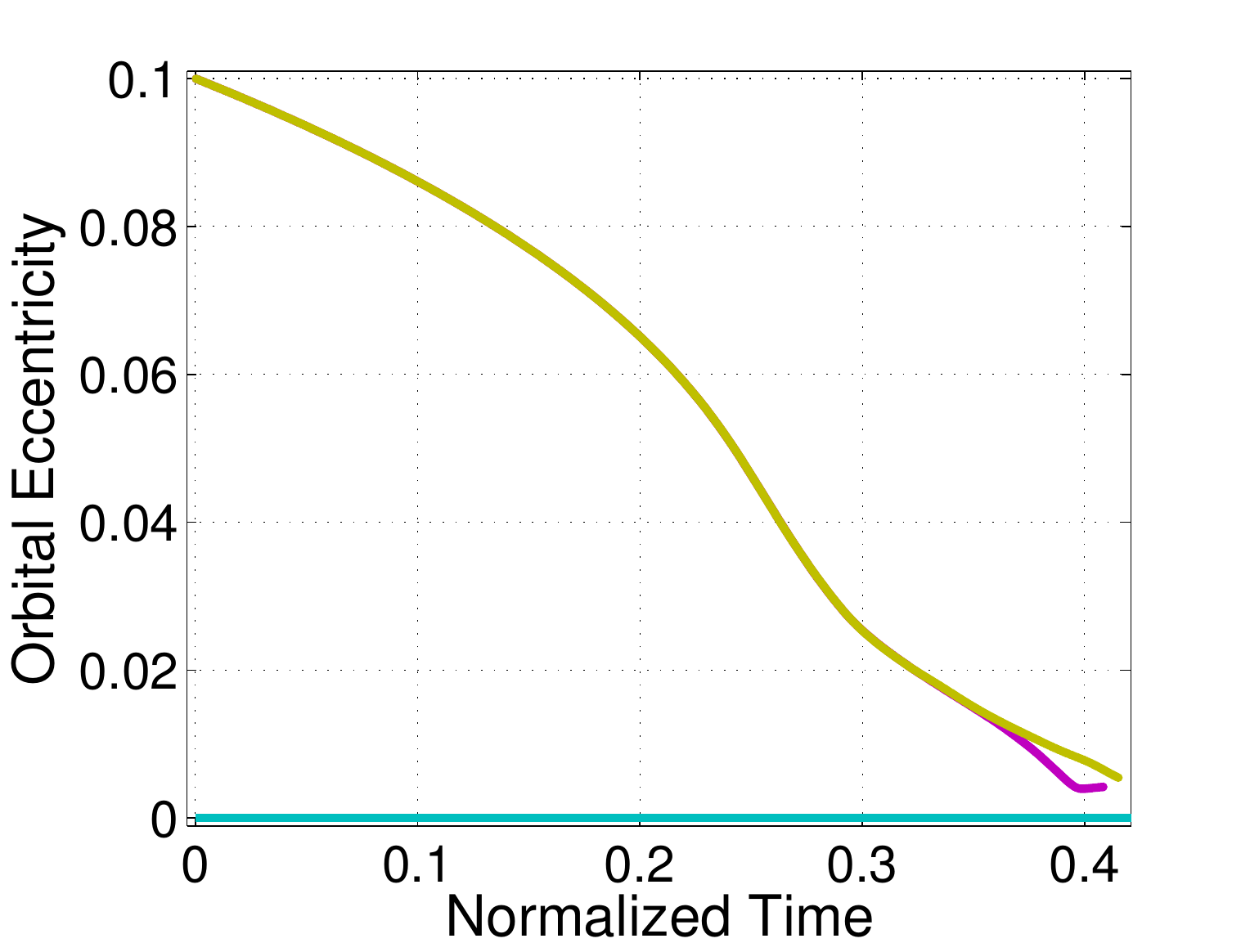}\includegraphics[height=5.05cm]{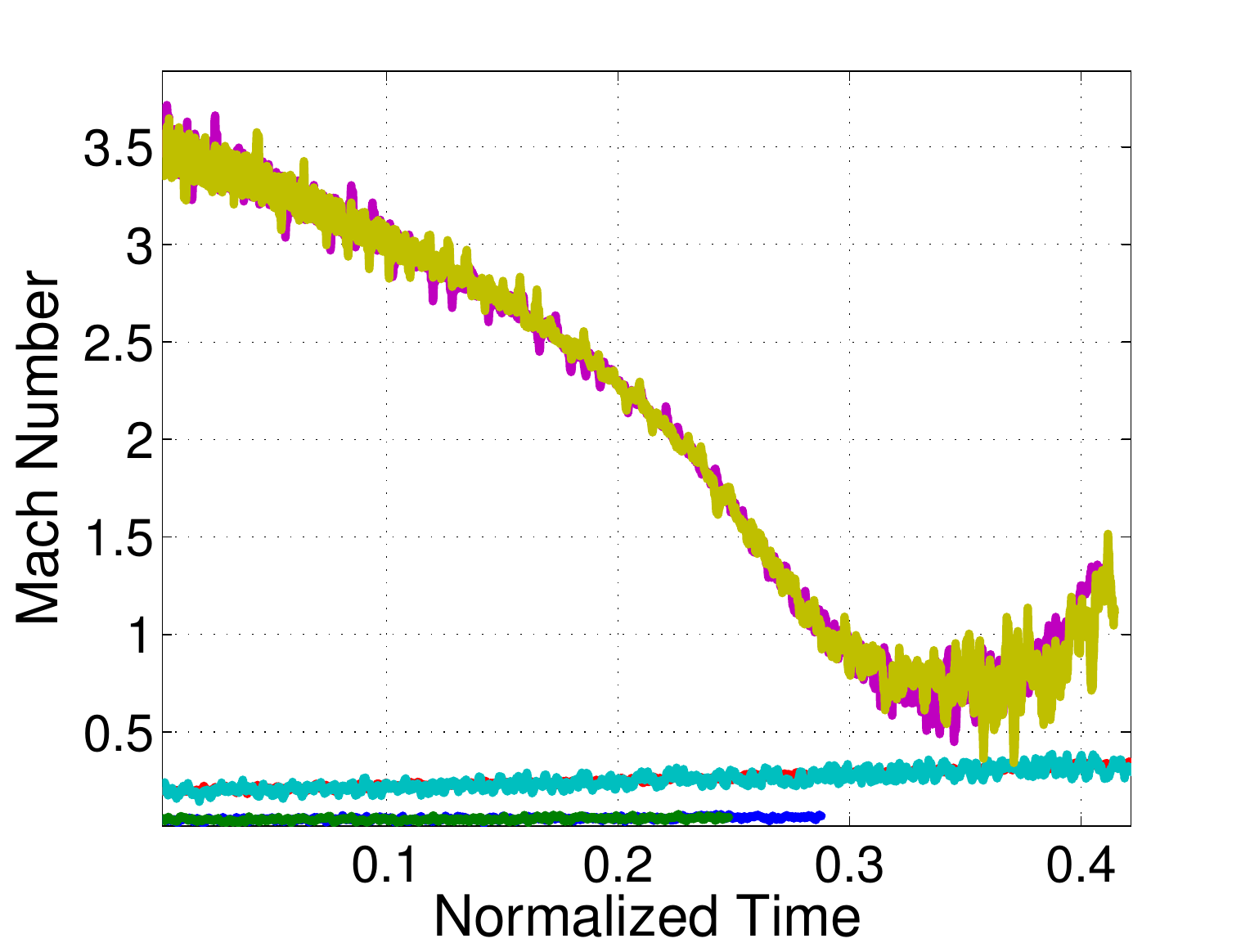}
\par\end{centering}

\caption{\label{fig:fig4}Evolution of co-planar binaries (runs 3-4 and 8-11).
The initial mass ratios are $Q=2\cdot10^{-10}$ and $Q=2\cdot10^{-8}$
respectively, and $q=1$. For runs with $Q=2\cdot10^{-8}$ we multiplied
the time by 100 to check the deviation from the linear regime. Runs
with $Q=2\cdot10^{-10}$ are displayed in their real time (in Myr).
Top left: binary separation vs. time. Top right: binary eccentricity
vs. time. Bottom left: Orbital eccentricity vs. time. Bottom right:
Mach number vs. time. Plotted data points are averaged, as in Fig.
\ref{fig:Results-for-runs}}
\end{figure*}

In figure (\ref{fig:Results-for-runs}), we plot the evolution of
BPs from runs 1-7. On the top left panel, we plot the evolution of
binary separation versus time. Runs 1 and 2 had started with $f=0.3$
while runs 3 and 4 had started with $f=0.1$. We see that they both
inspiral at the same timescales, hence $\tau_{ins}$ does not depend
on the binary separation $f$. Runs 5 and 6 had started with $e_{out}=0.1$
and indeed, the binary is seen to gain angular momentum and expand
as long as the regime is supersonic, and the expansion rate increases
as the Mach number decreases, as expected. The binary eccentricity
$e_{bin}$ is not important, and the plots of runs 5 and 6 are indistinguishable.
Run 7 had started with $e_{out}=0.3$. Due to the large Mach numbers
involved, GDF has little effect on the binary. On the top right panel
we plot the evolution of the binary eccentricity versus time. We see
that for runs with $f=0.3$, $e_{bin}$ drops rapidly, and stays constant
for $f=0.1$. The reason is separation of times (\ref{eq:sep. of times}),
and the RHS terms in eqn. \ref{eq:debindt} oscillates rapidly, such
that $\langle\dot{e}_{bin}\rangle_{n}\approx0$, while for $f=0.3$
the latter is not true. Note that in cases where the separation of
runs 5 and 6 is maximal ($a_{bin}\sim0.2r_{H}),$ the amplitude of
the oscillations of $e_{bin}$ are larger for the same reason. In
order to test the dependence on eccentricity $\tau_{break}\propto e_{out}^{3}$
from Eqn \ref{eq:tbreak}, we have used a power law fit to the data
in run5 and got a power law fit of $\tau_{break}\propto e_{out}^{2.993}$,
consistent with the expectations. In the bottom left panel we plot
the orbital eccentricity versus time. The eccentricity damping has
the same trend for $e_{out}=0.1$, while only a little effect is observed
for for $e_{p}=0.3$. Finally, at the bottom right we plot the Mach
numbers of one of the binary members versus time. We see that the
transition to the subsonic regime for runs 5 and 6 is consistent with
the eccentricity damping in the previous panel.

\begin{figure*}
\begin{centering}
\includegraphics[height=5.05cm]{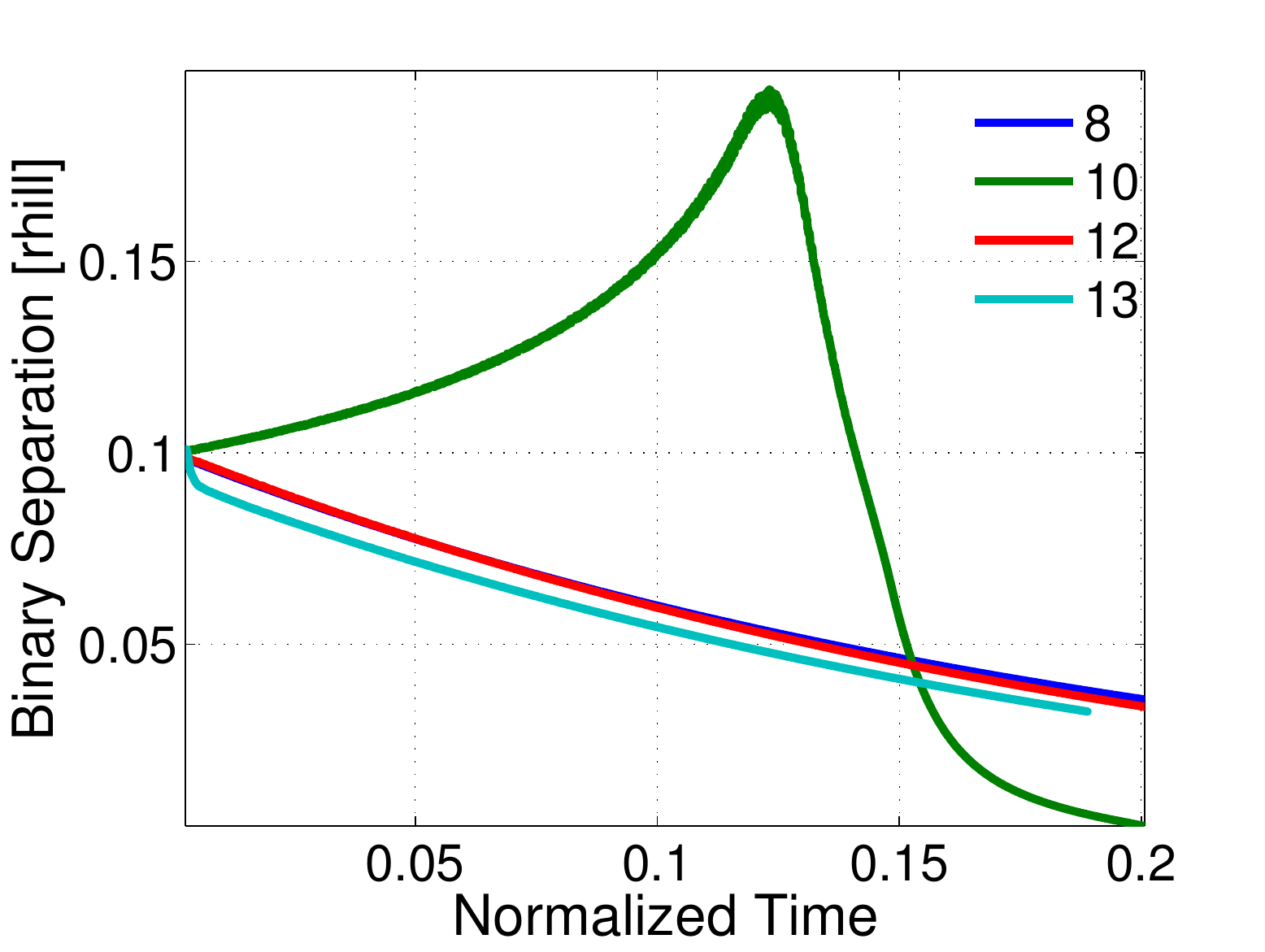}\includegraphics[height=5.05cm]{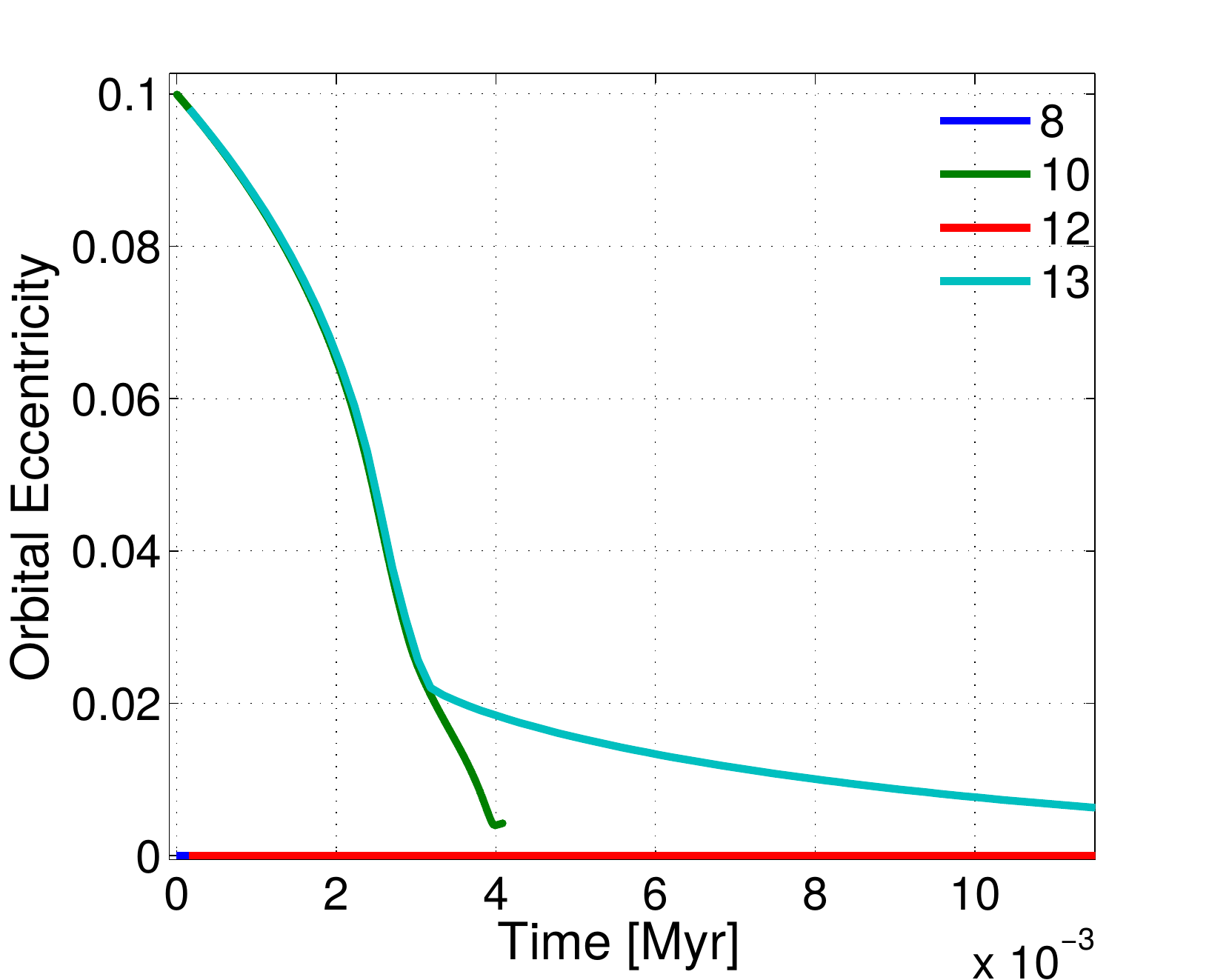}
\par\end{centering}

\begin{centering}
\includegraphics[height=5.05cm]{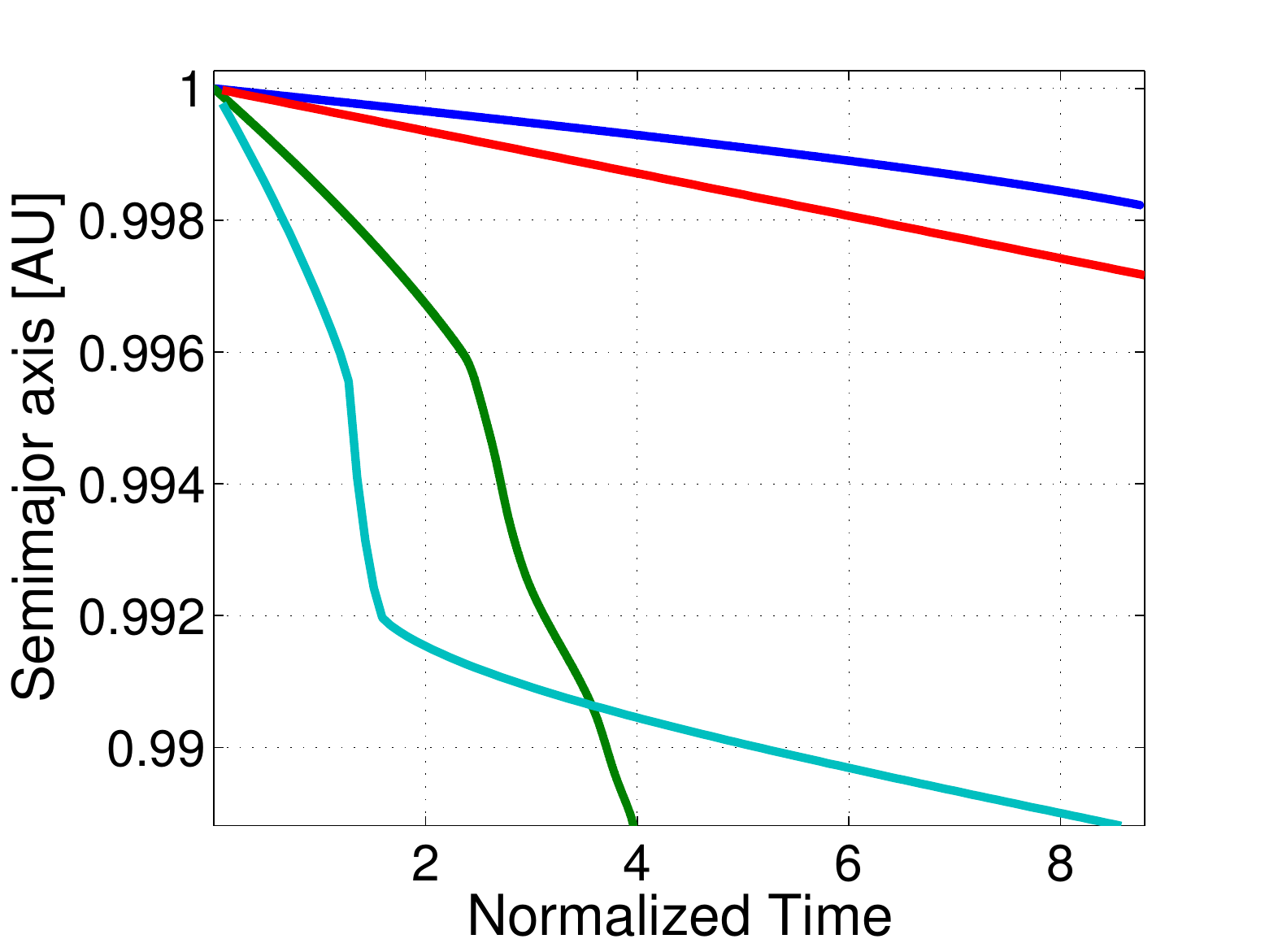}\includegraphics[height=5.05cm]{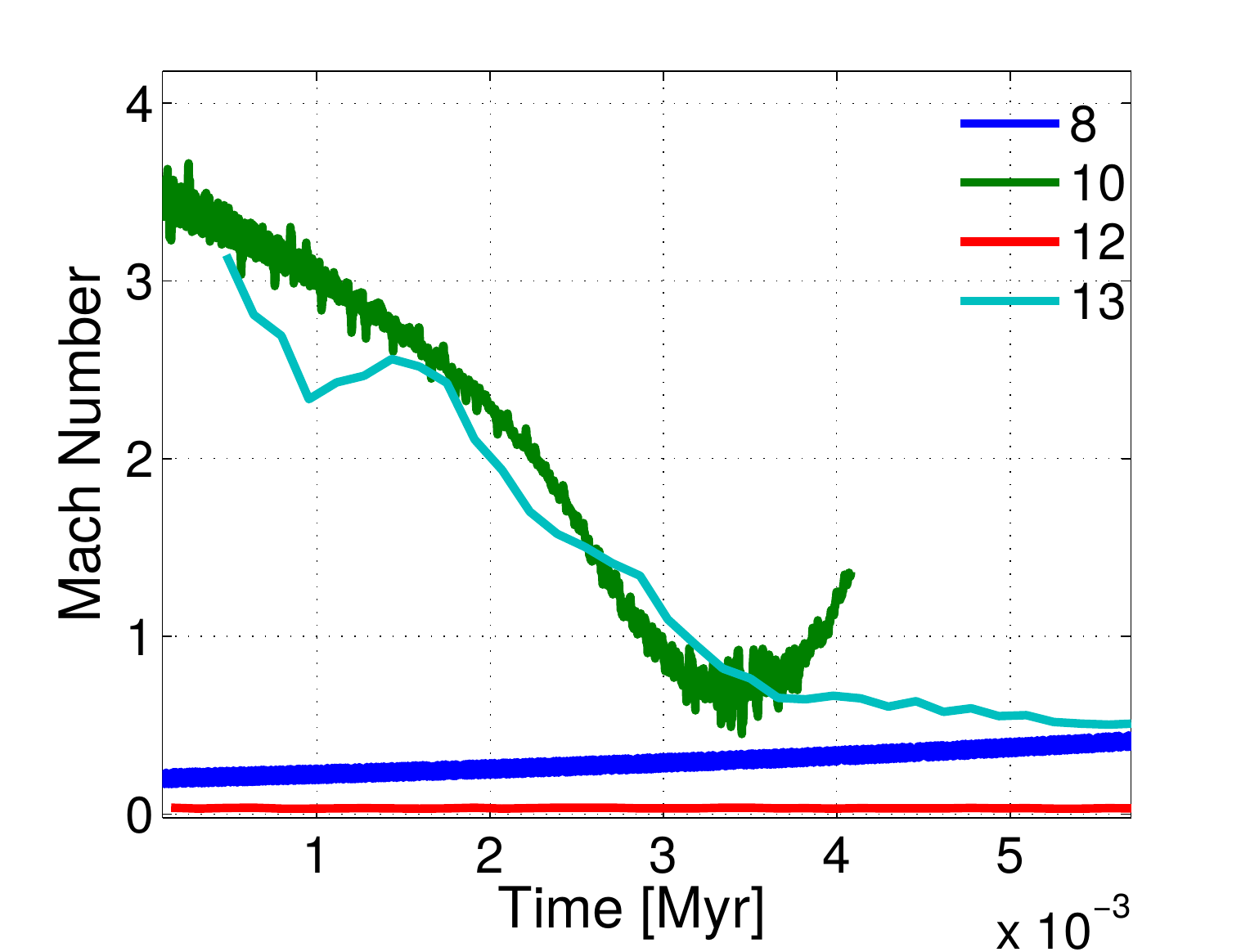}
\par\end{centering}

\caption{\label{fig:5} Evolution of co-planar binaries (runs 8,10 and 12,13).
The initial mass ratios are $Q=2\cdot10^{-10}$ and $Q=10^{-8}$ respectively,
and $q=1$ and $q=10^{-2}$ respectively. Top left: binary separation
vs. time. For runs with $q=1$ we multiplied the time by 50 to compare
different mass ratios. Runs with $q=10^{-2}$ are displayed in their
real time (in Myr). Top right: binary eccentricity vs. time. Bottom
left: Semi-major axis versus time. The time is multiplied by 2 for
runs 12 and 13 since their $Q$ is smaller by a factor of two. Bottom
right: Mach number over time. Plotted data points are averaged, as
in Fig. \ref{fig:Results-for-runs} (but using only 5 nearest ata
points).}
\end{figure*}

\begin{figure*}
\begin{centering}
\includegraphics[height=5.5cm]{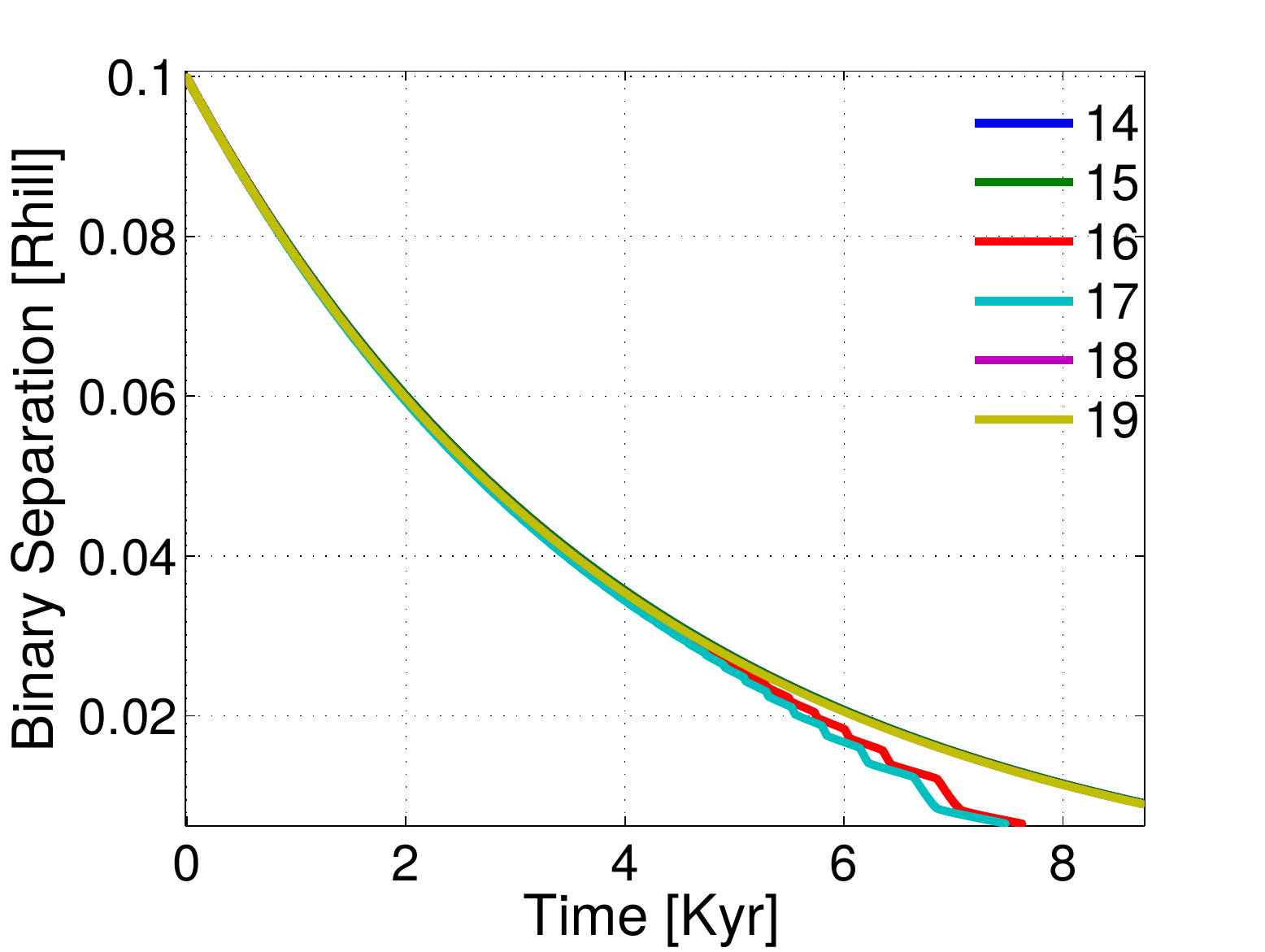}\includegraphics[height=5.5cm]{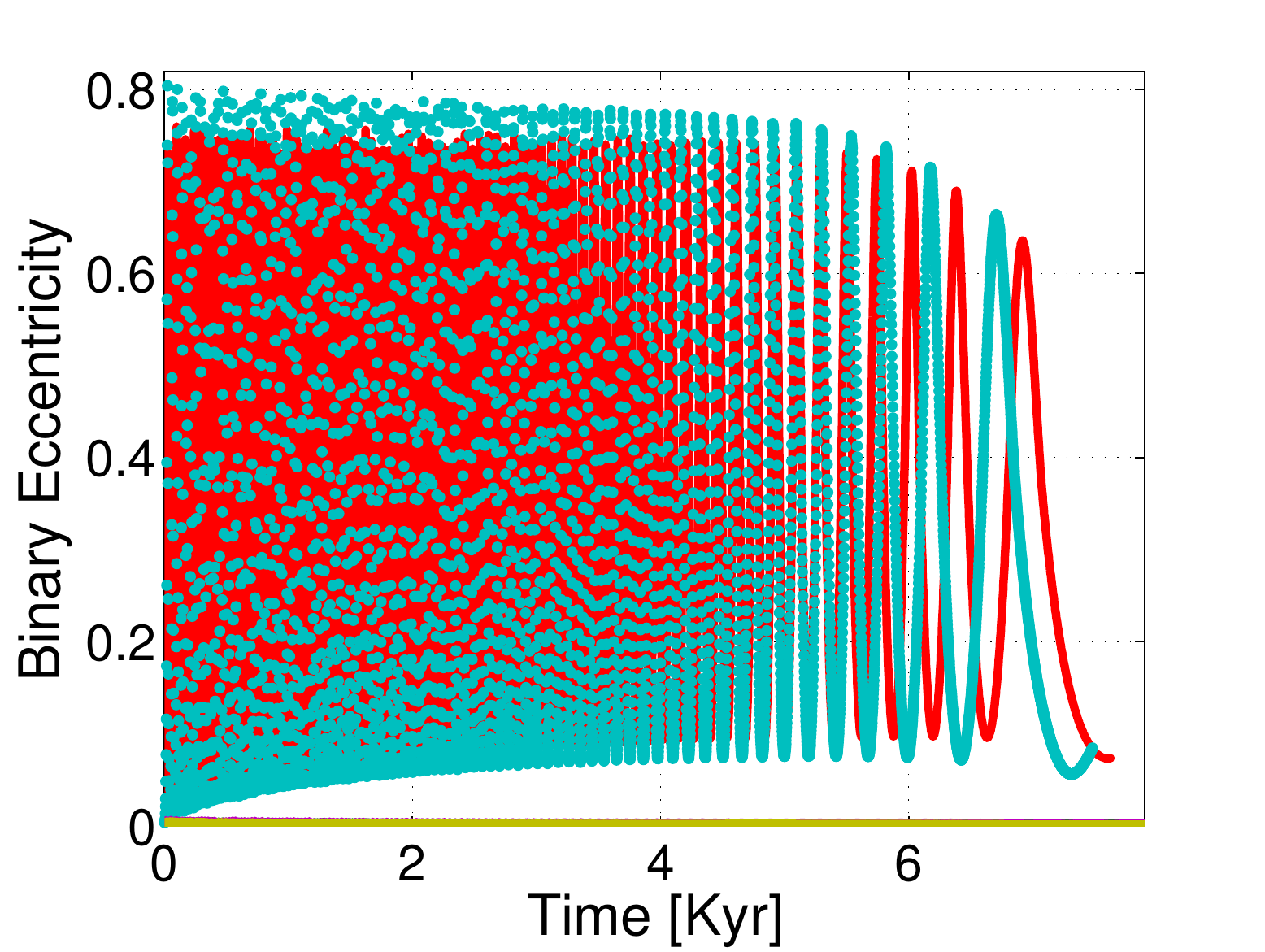}
\par\end{centering}

\begin{centering}
\includegraphics[height=5.5cm]{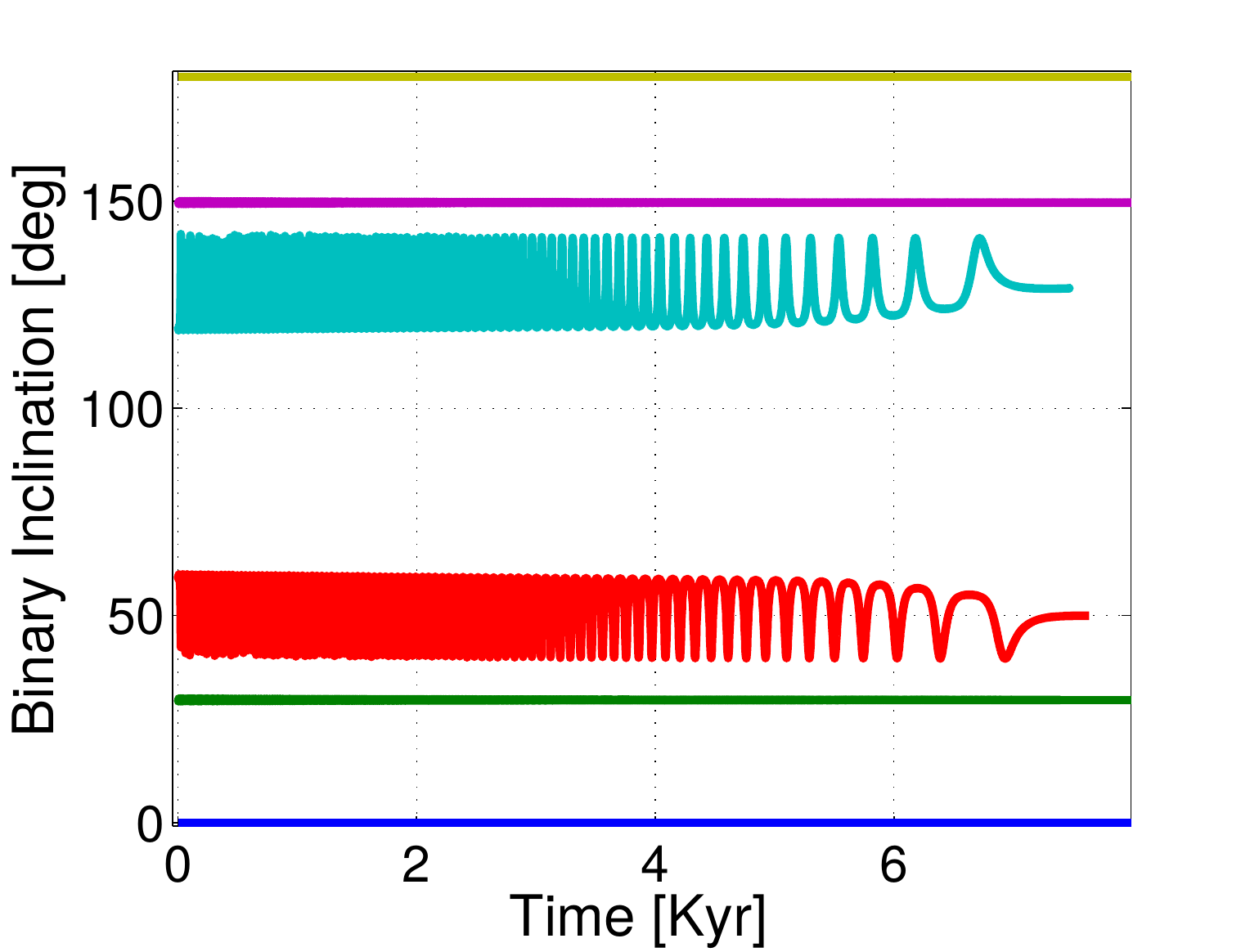}\includegraphics[height=5.5cm]{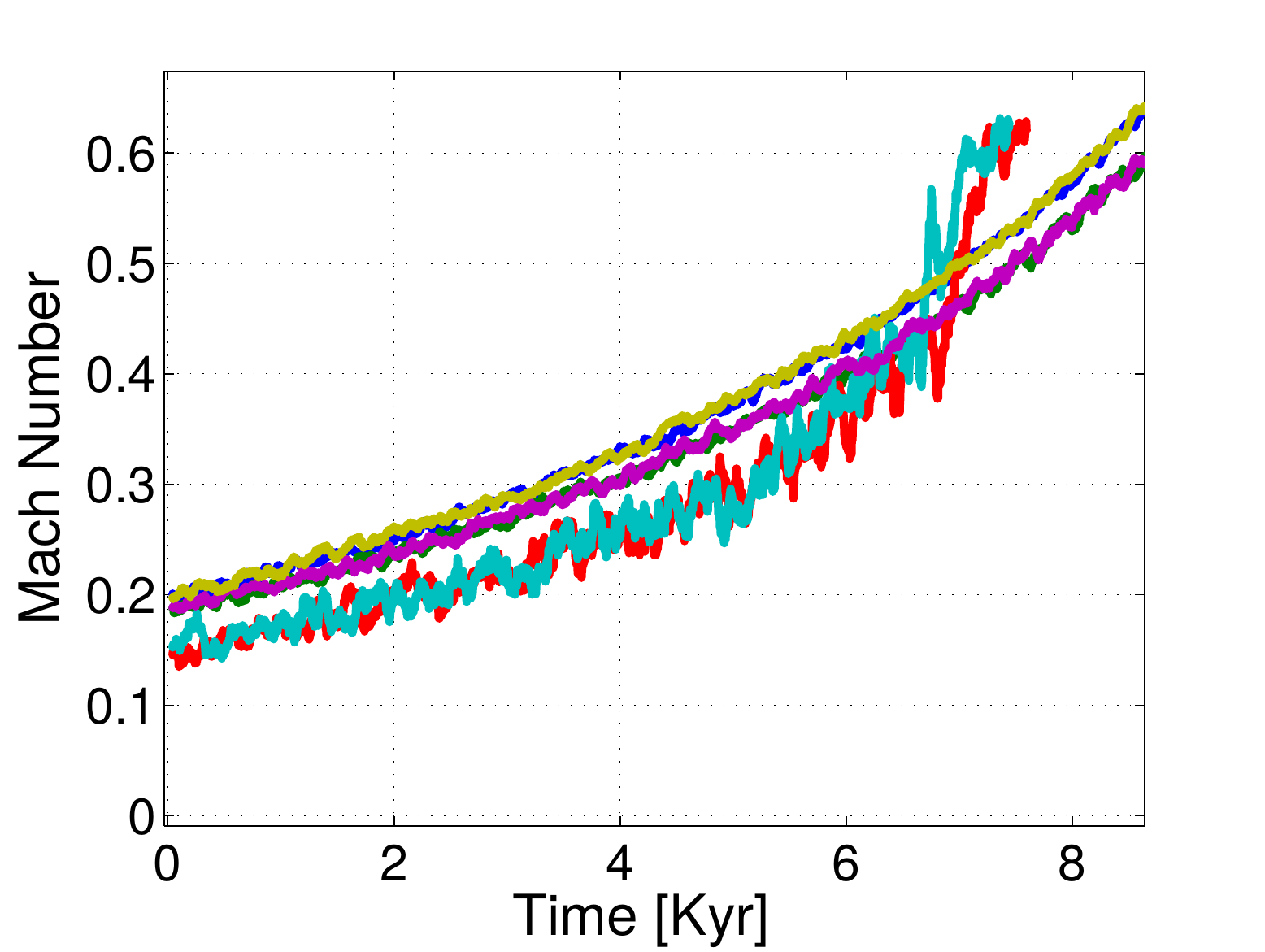}
\par\end{centering}

\caption{\label{inc1}Binary elements for $Q=2\cdot10^{-8}$ and equal mass
binaries with with various binary inclinations. Top left: Evolution
of the binary separation. Top right: Evolution of the binary eccentricity. To avoid overlapping, the data of run 17 is plotted as scatter.  
Bottom left: Evolution of the binary inclination. Bottom right: Evolution
of the Mach number. The data for Mach number have been averaged with
50 nearest data points to reduce fluctuations.}
\end{figure*}

\begin{figure*}
\begin{centering}
\includegraphics[height=4.8cm]{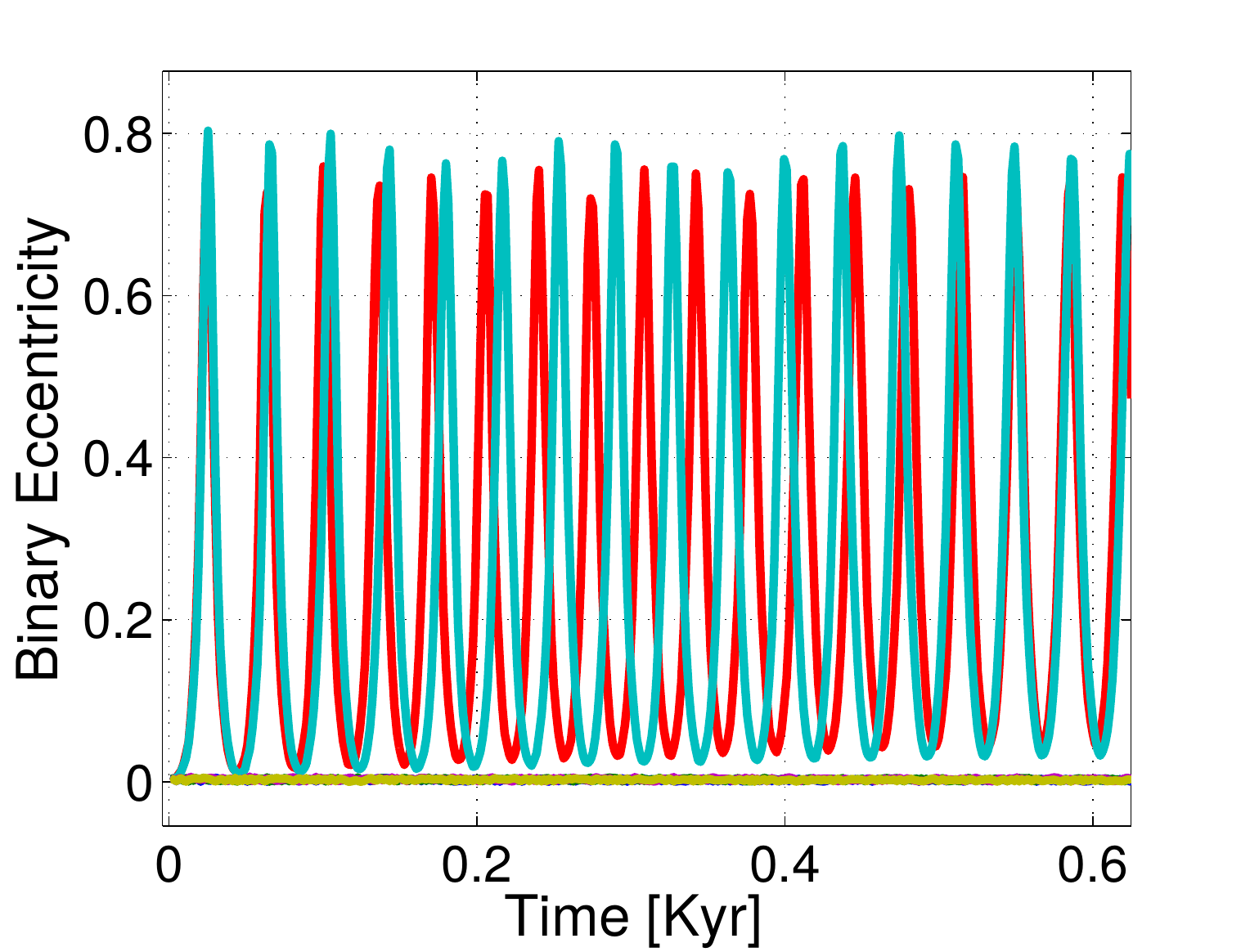}\includegraphics[height=4.8cm]{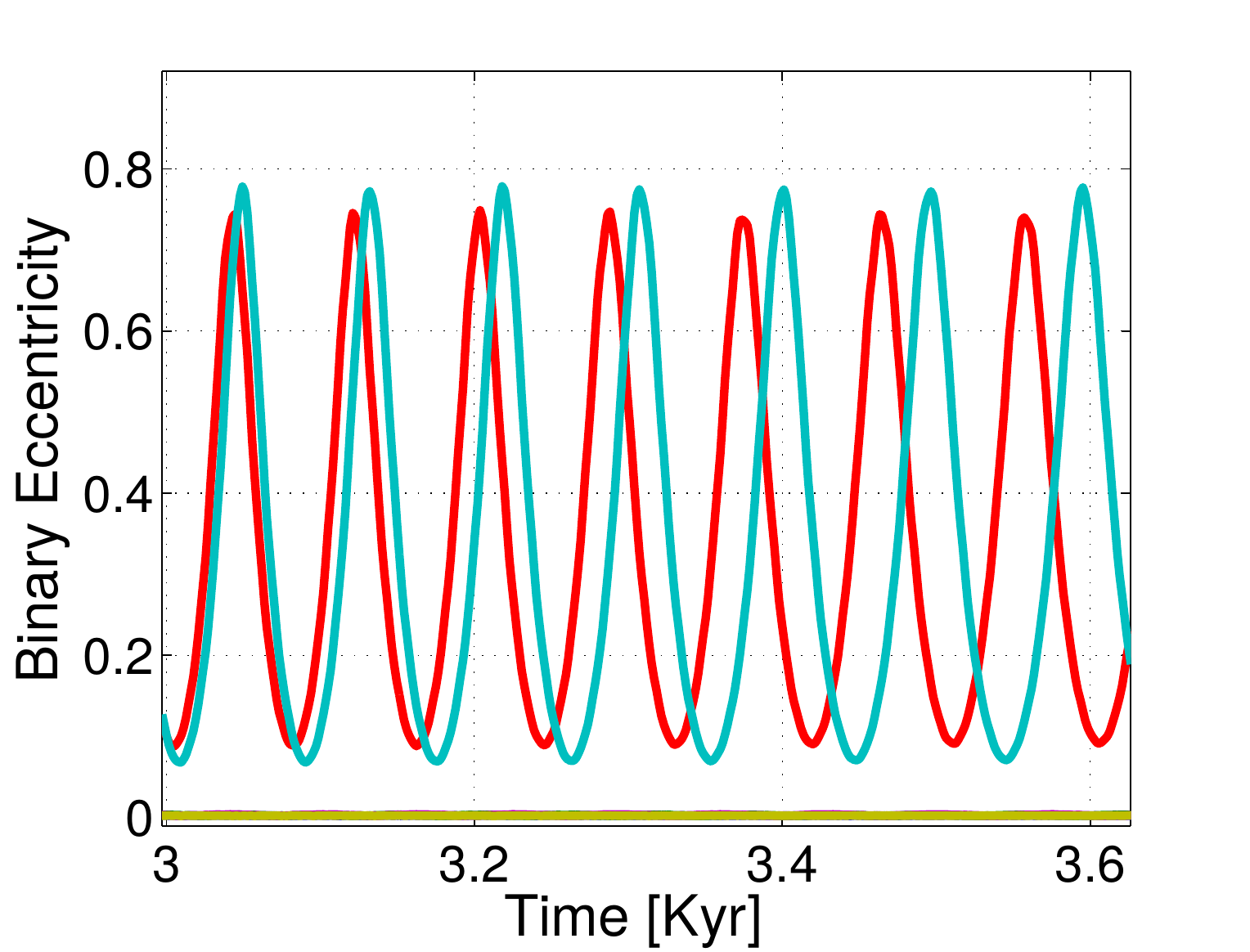}\includegraphics[height=4.8cm, width=6.3cm]{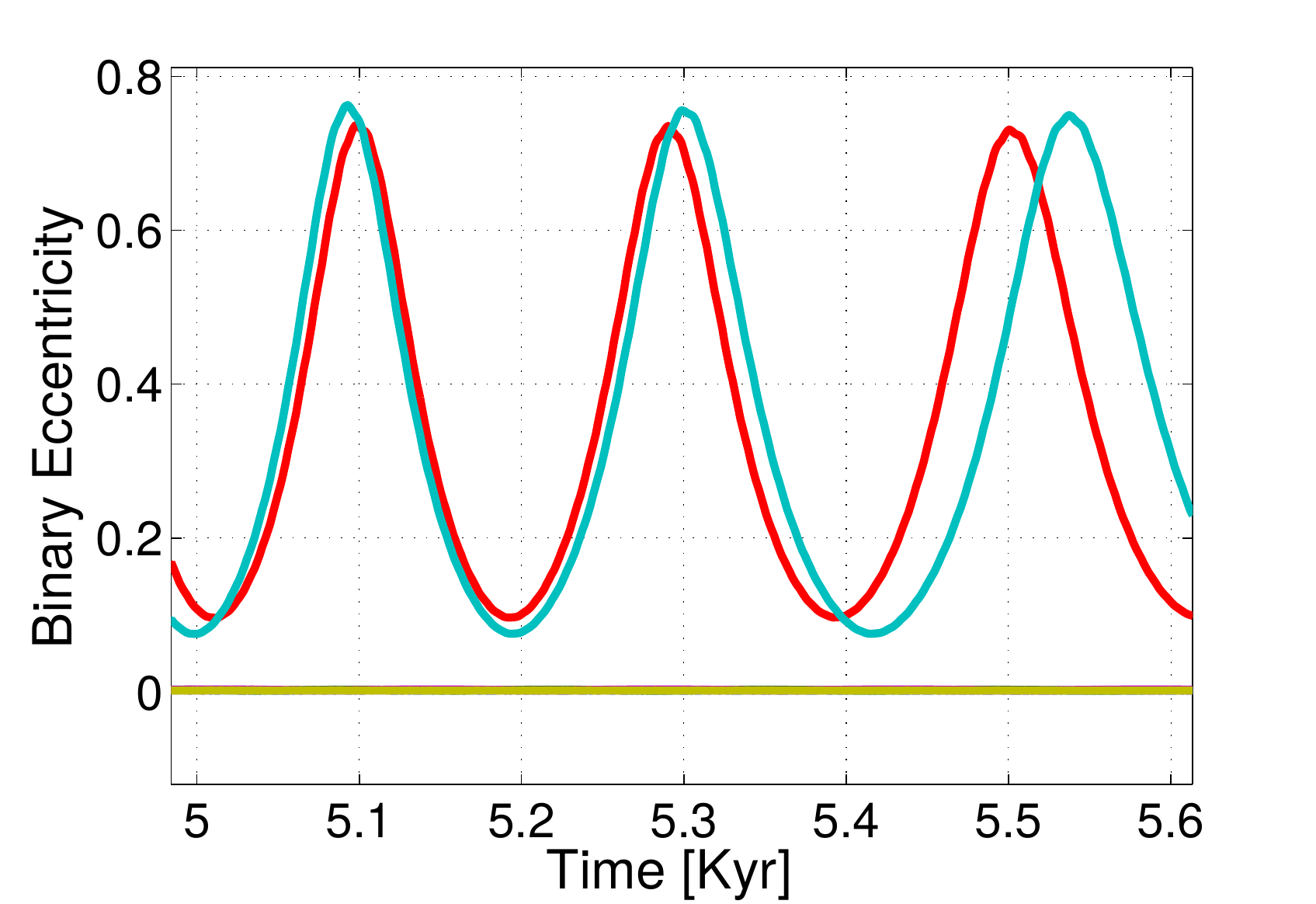}
\par\end{centering}

\begin{centering}
\includegraphics[height=4.8cm]{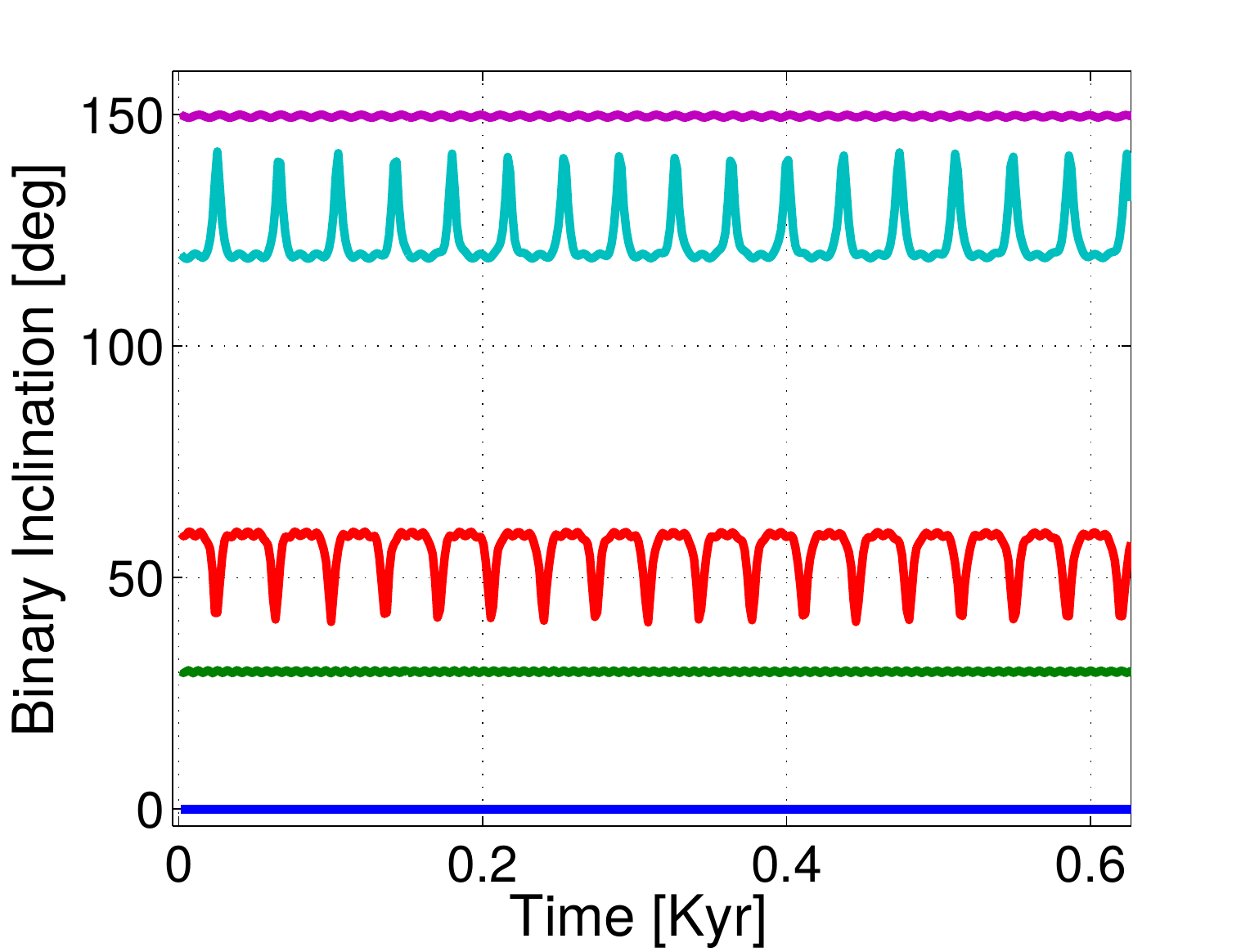}\includegraphics[height=4.8cm]{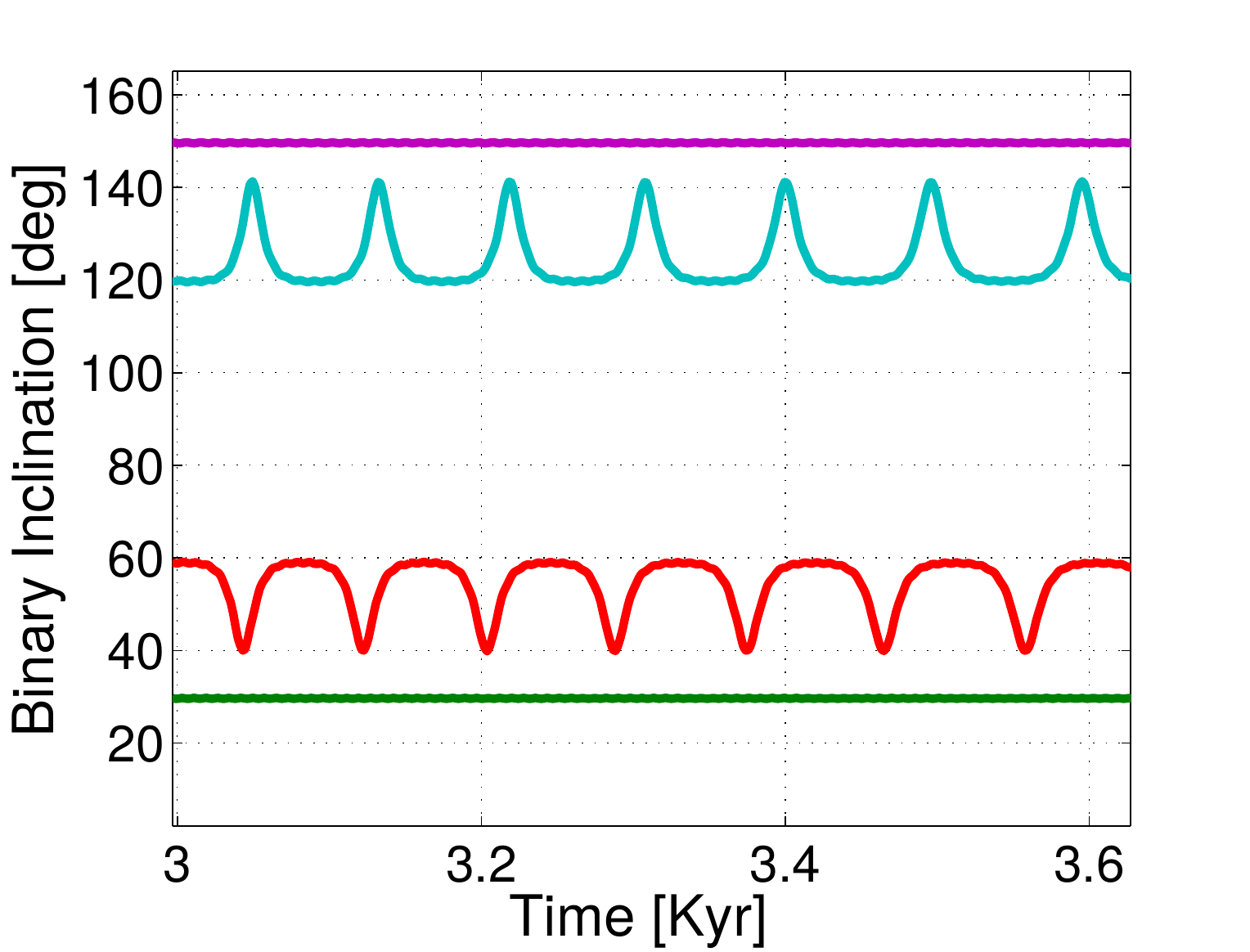}\includegraphics[height=4.8cm]{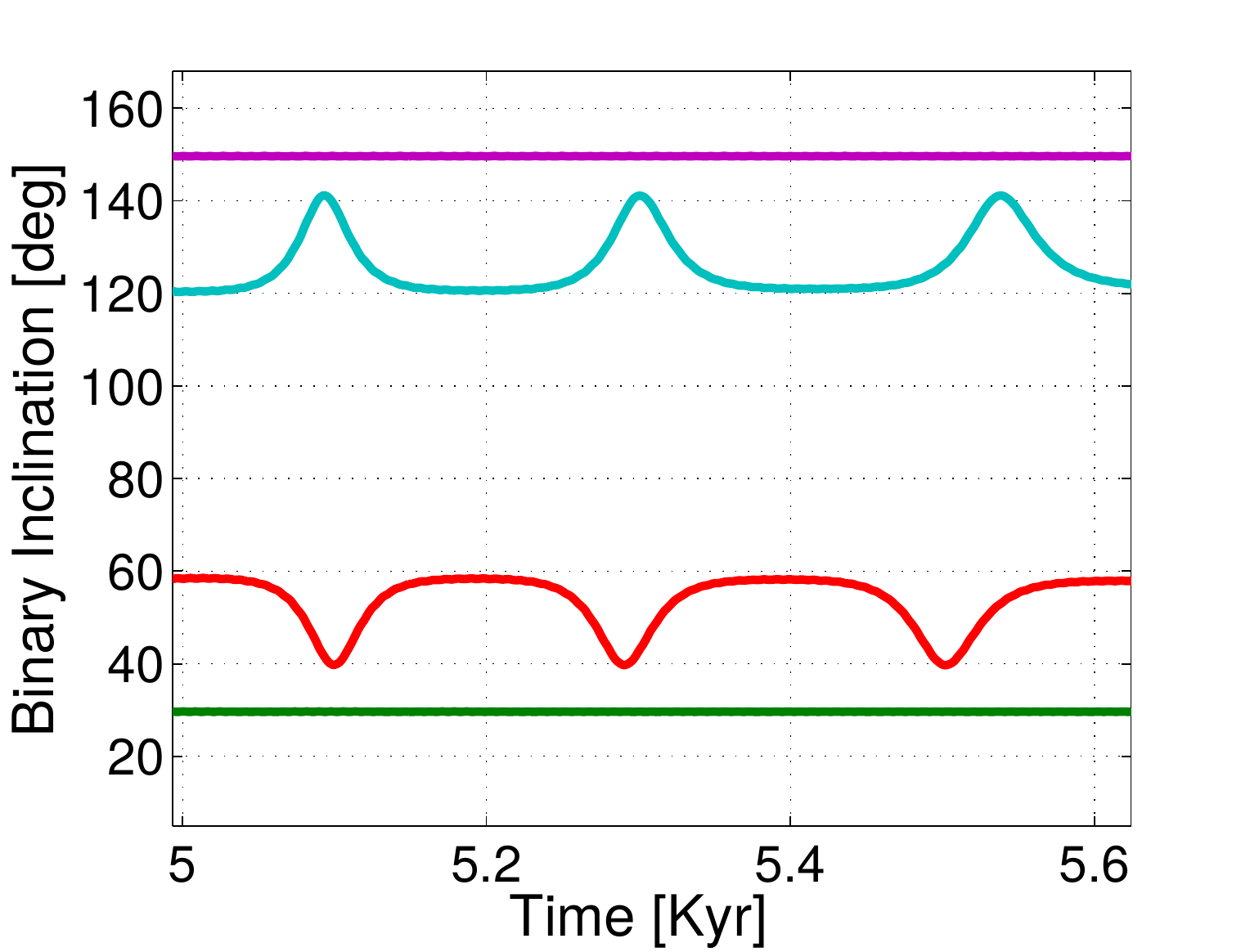}
\par\end{centering}

\caption{\label{inc2}Zoom-in on the binary eccentricity (top) and inclination
(left) evolution. The zoomed-in regions span $\sim0.6$Kyrs with starting
time of $0$, $3$ and $5$ Kyrs, from left to right, respectively. }
\end{figure*}

In figure (\ref{fig:fig4}), we plot the evolution of BPs from runs
3-4 and 8-11. For runs with $Q=2\cdot10^{-8}$ the time has been re-normalized
(see fig. description). The panels are the same as in fig. \ref{fig:Results-for-runs}.
In the top left panel, the trend is the same as in fig. \ref{fig:Results-for-runs}. The same trend is also evident in the eccentricities on the top right
and bottom left panels. The only difference is seen in the top right
panel, where the Mach numbers of the more massive binaries (runs 8
and 9) are larger than in the cases of the less massive binaries (runs
3 and 4). The latter is due to the fact that the binary velocity scales
as $v_{bin}\propto Q^{1/3}$. For runs 10 and 11, the Mach numbers
decrease as $e_{out}$ decreases, and increase again when the binary
rapidly inspirals and $v_{bin}$ increases to sonic velocities before
the binary merger.

\begin{figure}
\begin{centering}
\includegraphics[width=7.5cm]{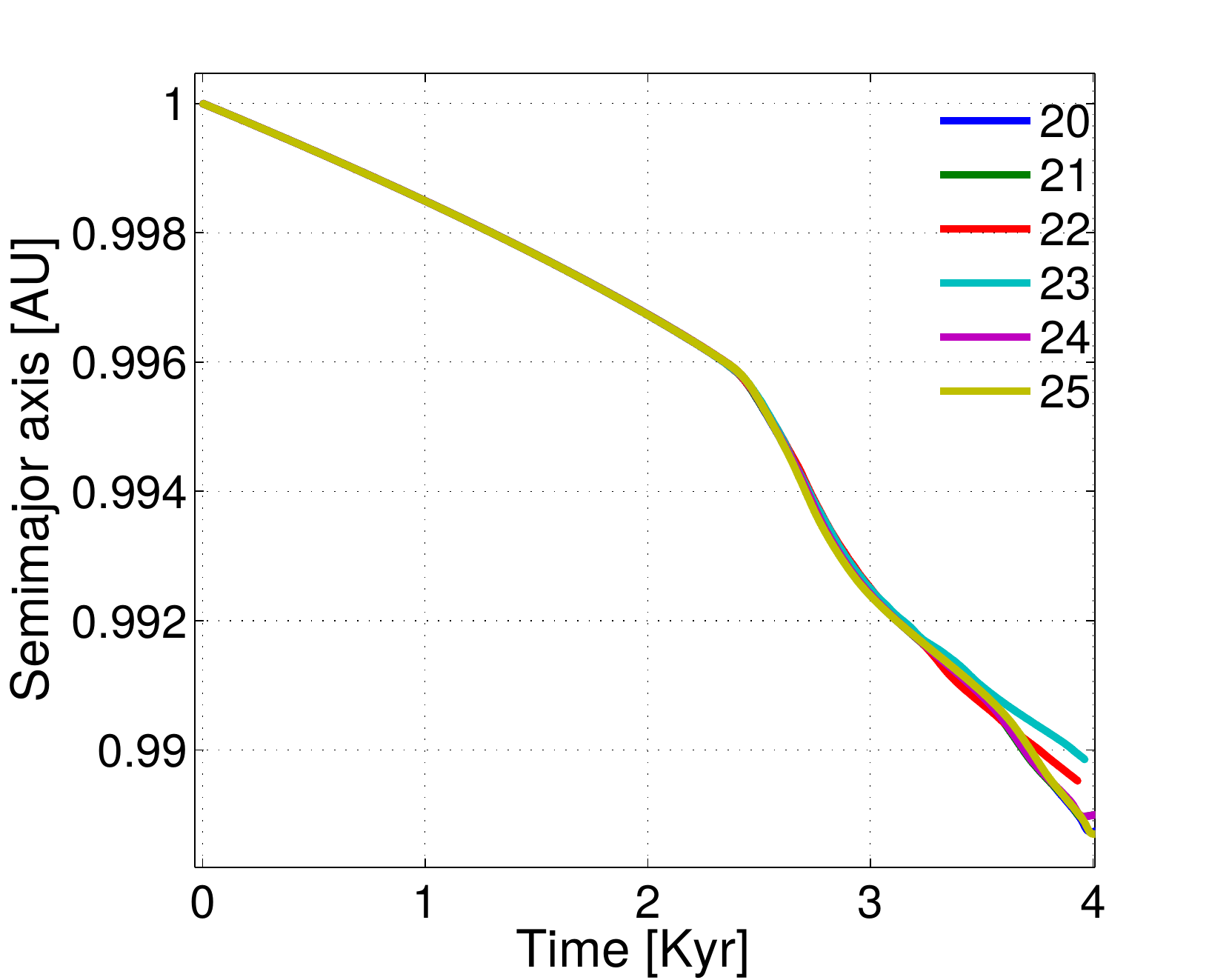}
\includegraphics[width=7.5cm]{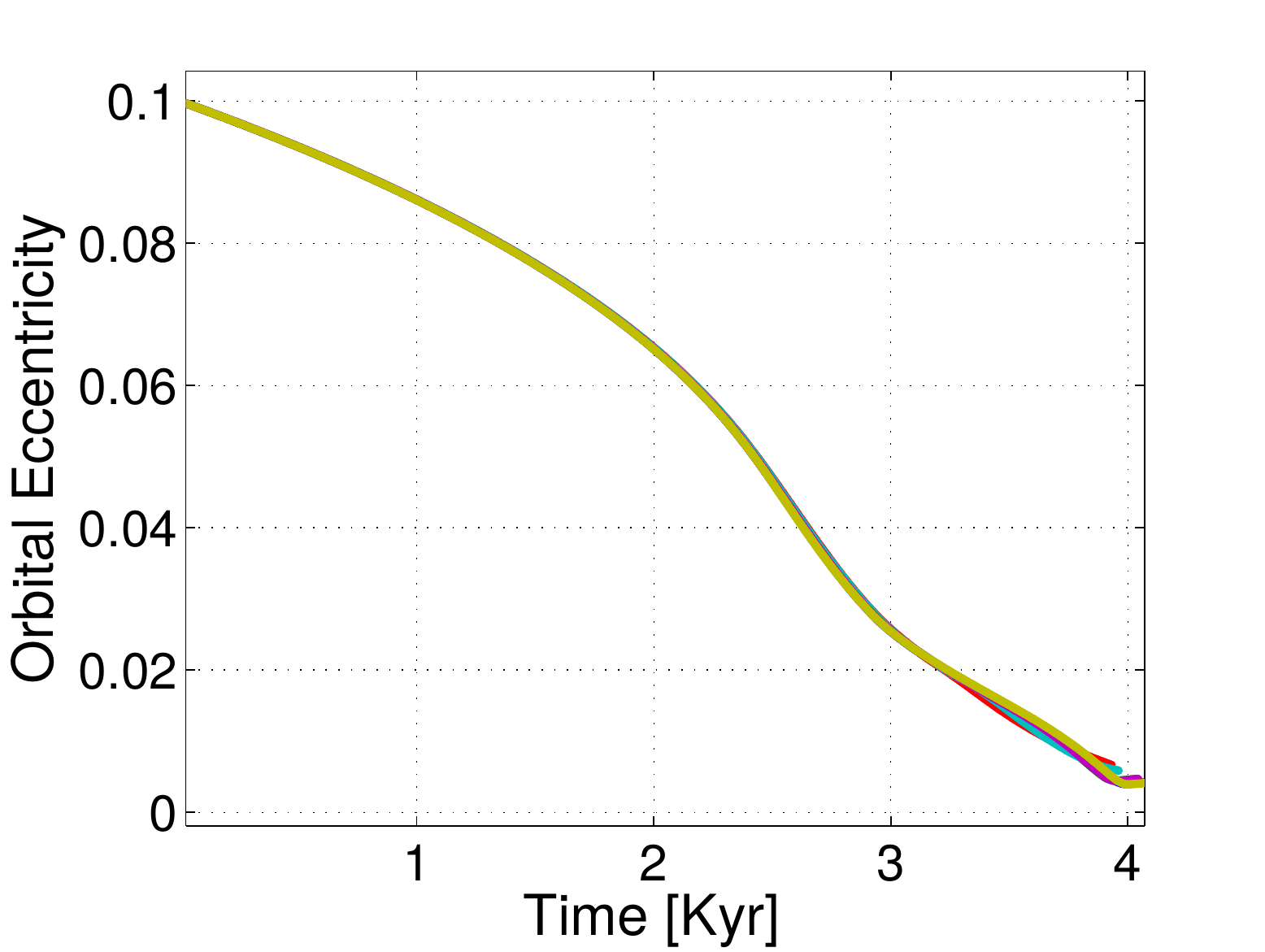}
\par\end{centering}

\caption{\label{primary_inc}Evolution of the orbital semi-major axis (left)
and eccentricity (right) for runs 20-25.}
\end{figure}

\subsubsection{Deviation from linear regime}

In fig. \ref{fig:fig4}, the most visible difference between runs
3 and 8 (that have identical initial conditions except for $Q$) is
the Mach number. We can estimate the deviation from the linear regime
by studying the small differences in the binary separation. We examine
the binary separation of runs 3 and 8 at $t_{f}=2.5$ normalized time
units, where both runs have lost $\sim1/2$ their initial separation.
At $t_{f}$, $f_{3}=0.054$ for run 3, and $f_{8}=0.053$ for run
8. The difference between both runs is $\Delta f/f\approx1.85\%$.
For run 8, the Mach number at $t_{f}$, the Mach number is $\mathcal{M}_{3}=0.09$
for run 3 and $\mathcal{M}_{8}=0.267$, which leads to relative error
of eqn. \ref{eq:deviation} of $\delta\tau_{ins}/t_{f}=(1-(1-0.6\mathcal{M}_{8}^{2})/(1-0.6\mathcal{M}_{3}^{2}))\approx3.8\%$.
We conclude that the deviation is within the estimated error in eqn.
\ref{eq:deviation}.

\subsubsection{Low mass ratio binaries ($q\ll1$)}

In fig. \ref{fig:5} we compare runs 12 and 13 with $q=10^{-2}$ with
previous runs 8 and 10, for which the initial conditions are the same
except for $q$. For hierarchical binaries with $q\ll1$, the general
trend is that the orbital primary elements (with subscript ``out'')
scale with the binary to star mass ratio $Q$, and the binary elements
scale with the reduced mass $\mu$ ( for small $q,$ $\mu$ is the
smaller mass). In the top left panel, we multiply the time by 50 to
check the consistency of the linear regime. Binaries 8 and 12 inspiral
in the same phase. Due to the small $q$ of run 13, its supersonic
phase is short, and later on it inspirals similarly to previous runs.
In the top right panel, the orbital eccentricities of runs 10 and
13 decay in the same fashion, but run 10 inspirals fast, the Mach
number is close to the sonic limit, and the eccentricity damping is
more efficient, while for run 13 the Mach numbers involved are subsonic.
In the bottom left panel, we multiplied the time by two for runs with
$Q=10^{-8}$ to account for the difference in $Q.$ Runs with $Q=2\cdot10^{-8}$
are presented with their real time (in Kyrs). 

In addition, note runs 3 and 12 in figures (\ref{fig:Results-for-runs})
and (\ref{fig:5}). They have the same parameters for the inspiral
time, except $q$, and indeed, the inspiral time of run 12 is twice
as fast.

The inward migration of the primary in runs with $q\ll1$ is effectively
similar to migration of single planetesimals. These planetesimals
migrate faster than their equal mass binary counterparts. This is
due to \emph{loss of gravitational focusing}, discussed in \citet{2014ApJ...794..167S}
and \citet{2011ApJ...726...28B}. The latter is true as long as the
other parameters are comparable. At some point, the relevant Mach
number for run 10 is $\mathcal{M}_{bin}$, and when it becomes supersonic,
it dominates also the evolution of the primary orbital elements. As
a consequence, the CM in run 10 continues to migrate inward at the
same phase, while the CM migration in run 13 is slowed down due to
the circularization of the primary and the low Mach numbers of the
binary revolution. On the top right panel, the picture is similar
to Fig. \ref{fig:fig4}, where the more massive binaries involve larger
Mach numbers. 

To conclude this section, there is a good consistency between the
predicted inspiral timescales and the simulations. The results are
consistent with different mass ranges $Q$ and $q$, and deviation
from the linear regime is well captured by eqn.   \ref{eq:deviation}.
The supersonic expansion rate is also consistent with the prediction
of eqn. {\ref{eq:tbreak}.

\subsection{Inclined binaries }
\label{sec:INCLINED}

\subsubsection{Equal mass binaries ($q=1$)}

On the left panel of fig. (\ref{inc1}) we plot the evolution of binary
elements of equal mass of $Q=2\cdot10^{-8}$ circular CM orbit.

\begin{figure*}
\begin{centering}
\includegraphics[height=5.5cm]{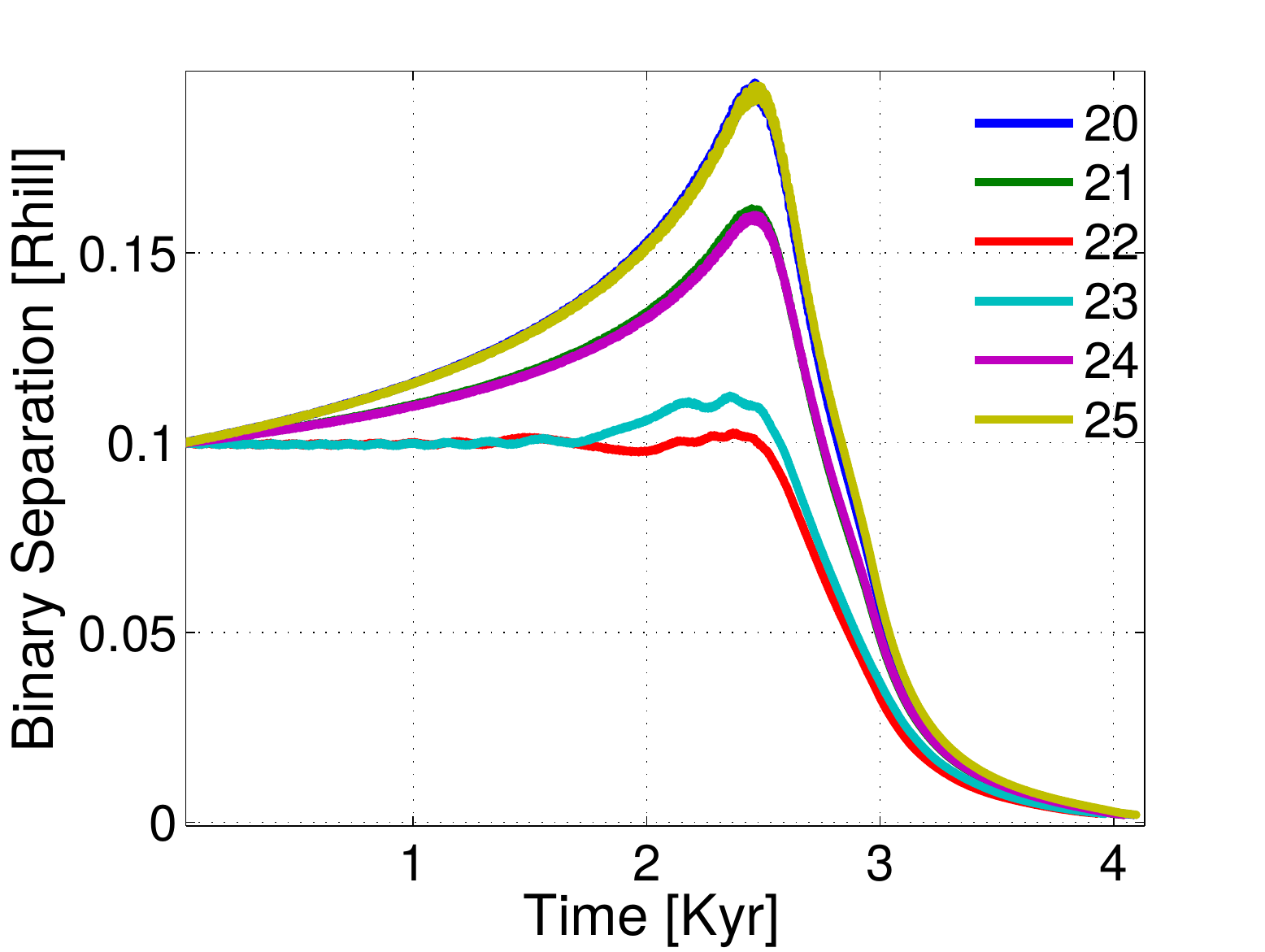}\includegraphics[height=5.5cm]{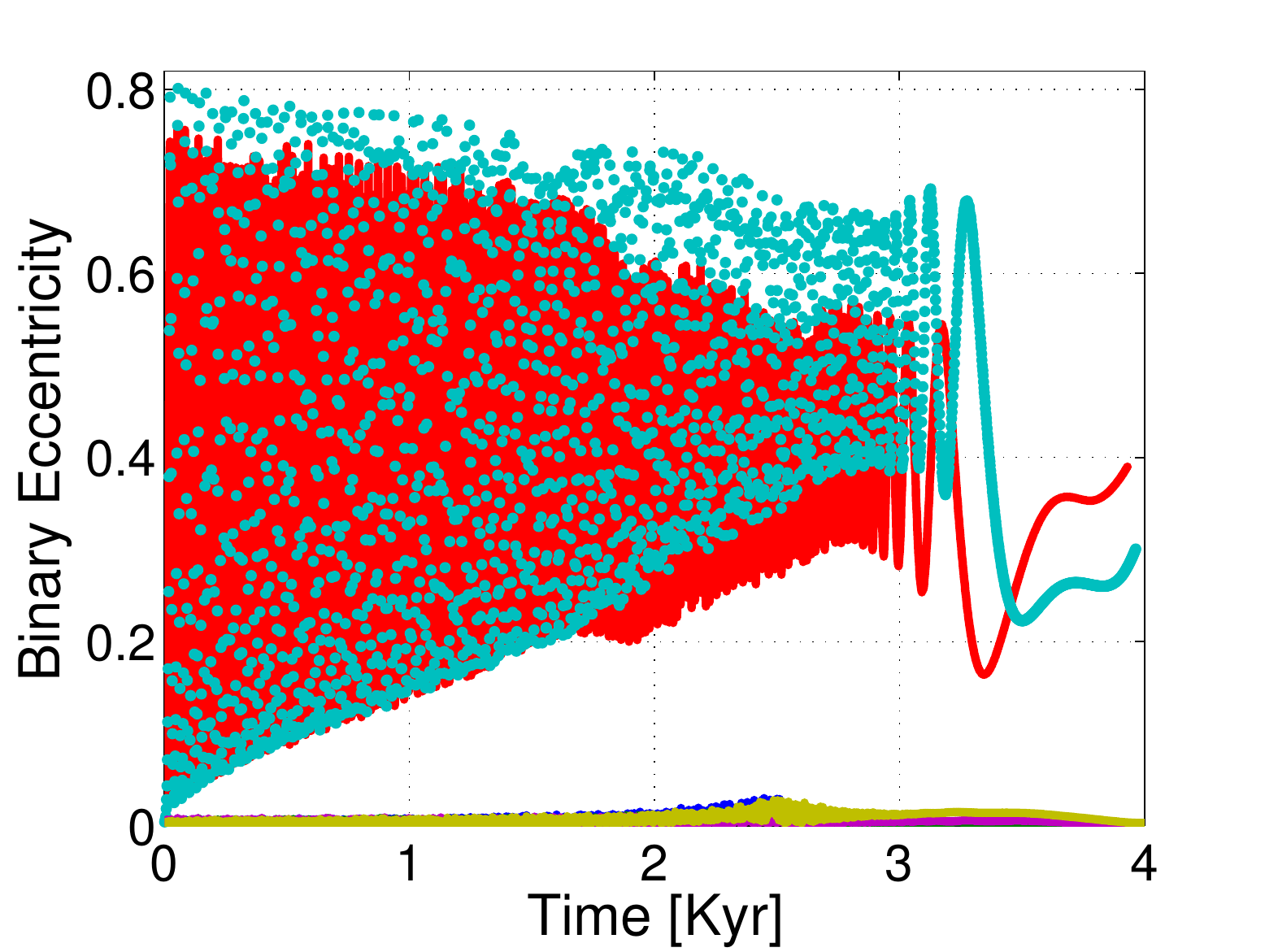}
\par\end{centering}

\begin{centering}
\includegraphics[height=5.5cm]{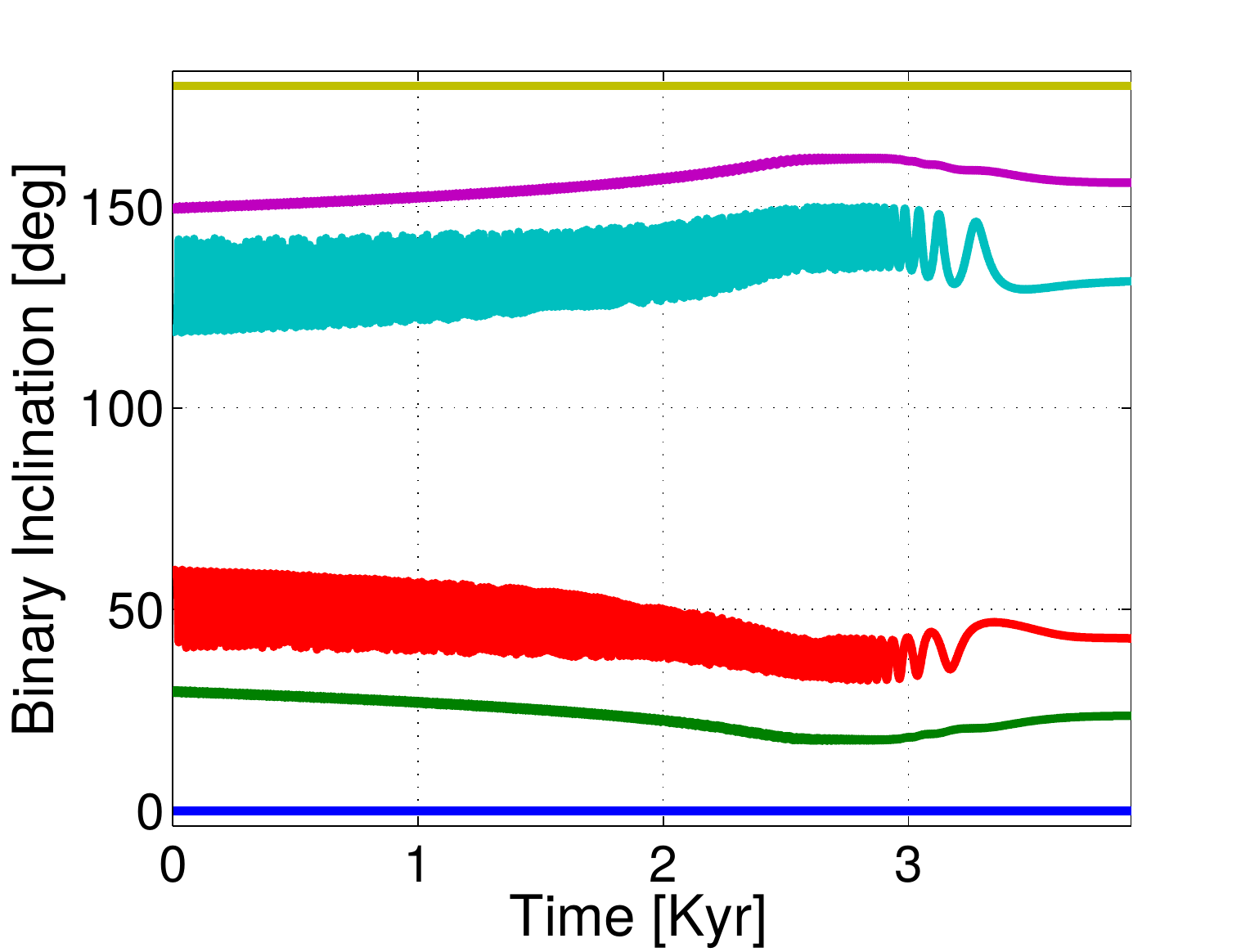}\includegraphics[height=5.5cm]{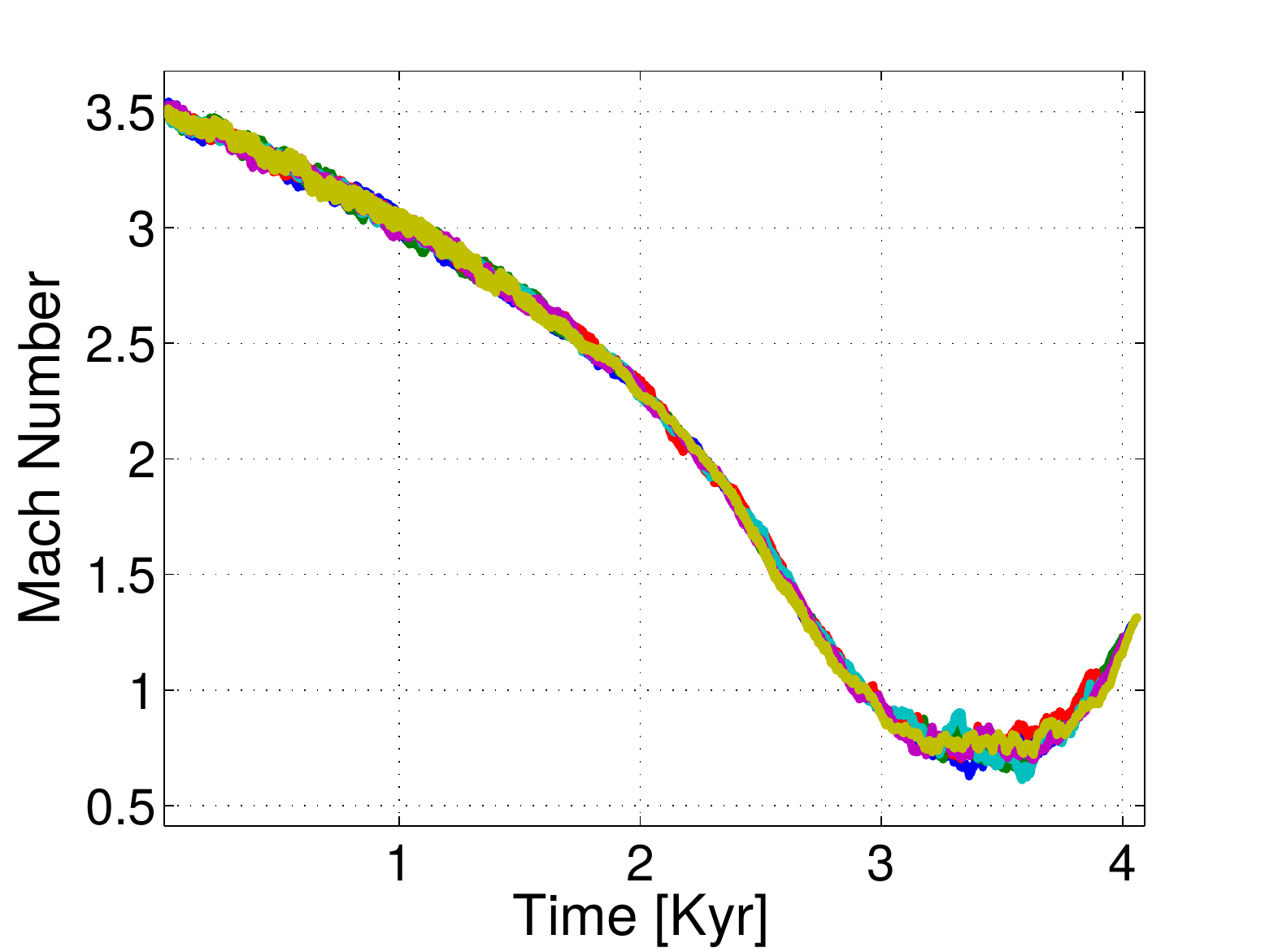}
\par\end{centering}

\caption{\label{inc_eccentricv}Same as Fig. \ref{inc1}, but with orbital
eccentricity $e_{out}=0.1$}
\end{figure*}

On the top left panel we see that the binary inspiral rate is indifferent
to binary inclination. The binary inspiral rate deviates from that
of co-planar binaries due to the increased periods of time that the
binary spends at higher eccentricities due to the effects of secular
evolution through Kozai-Lidov (KL) cycles \citet{1962AJ.....67..591K,1962P&SS....9..719L}, which can significantly affect BPs \citep{2009ApJ...699L..17P}.
For higher eccentricity, the Mach numbers involved are higher and
GDF is more efficient. Since the KL timescale scales as $T_{KL}\propto a_{bin}^{-3/2}$,
the time spent at higher eccentricities increases with cycle number.
The top right and bottom left panels show the evolution of binary
eccentricity and inclination, respectively. Again, the KL timescales
increase due to binary inspiral. On the bottom right panel we plot
the evolution of the Mach numbers. Generally, a highly inclined binary
contributes less to the relative velocity, and the average Mach number
is lower than the co-planar orbit case. The higher Mach numbers after
$\sim7$Kyr are due to the fact that the inclined binary is inspiraling
faster than the circular binary near the sonic limit where the Mach
numbers involved are close to unity.

On the left panels of fig. (\ref{inc2}) we plot the binary eccentricity
(top) and inclination (bottom) where we zoom-in on the first $600$yr
. The KL timescale is given by $T_{KL}\approx P_{out}^{2}(1-e_{out}^{2})^{3/2}/P_{bin}$
, where $P_{out}=1yr$ is the period of the primary, and $P_{bin}=f^{3/2}P_{out}$
is the binary period. For $e_{out}=0$, the KL timescale is $T_{KL}=f^{-3/2}yr$.
The initial timescale is $T_{KL}\approx32$yr for $f=0.1$, where
$T_{KL}$ increases as the binary inspiral. At $t=3Kyr$, $f=0.05$
and $T_{KL}\approx90$yr (middle panels), and at $t=5$Kyr, $f\approx0.027$
and $T_{KL}\approx230$yr (right panels).

We now turn to investigating inclined binaries with eccentric primary
orbit (initial eccentricity $e_{out}=0.1$). In Fig. \ref{primary_inc}
we plot the orbital semi-major axis (left) and the eccentricity (right).
We see that the decay of $a_{out}$ and $e_{out}$ is insensitive
to the binary orbital elements, as expected. The orbital semi-major
axis $a_{out}$ and eccentricity $e_{out}$ are insensitive to the
inner structure of the binary. The latter breaks down only when $\mathcal{M}_{bin}\gtrsim\mathcal{M}_{cm}$.

In Fig. \ref{inc_eccentricv} we plot the binary orbital elements.
On the top left panel we see the general feature of the supersonic
expansion and rapid inspiral. The difference is that for larger inclinations
the expansion rate is quenched, and for inclinations of $60$ and
$120$ deg there is no expansion. The reason is that the gained angular
momentum is now used to damp the binary inclination (bottom left panel). In the bottom right panel, we see that the Mach numbers are dominated
by $e_{out}$ for the first $3$Kyr, where $e_{p}\gtrsim2H_{0}$.
Later on, the Mach numbers are dominated by the binary inspiral, where
the differences between binary separations is small. 
\begin{figure*}
\begin{centering}
\includegraphics[height=5.45cm]{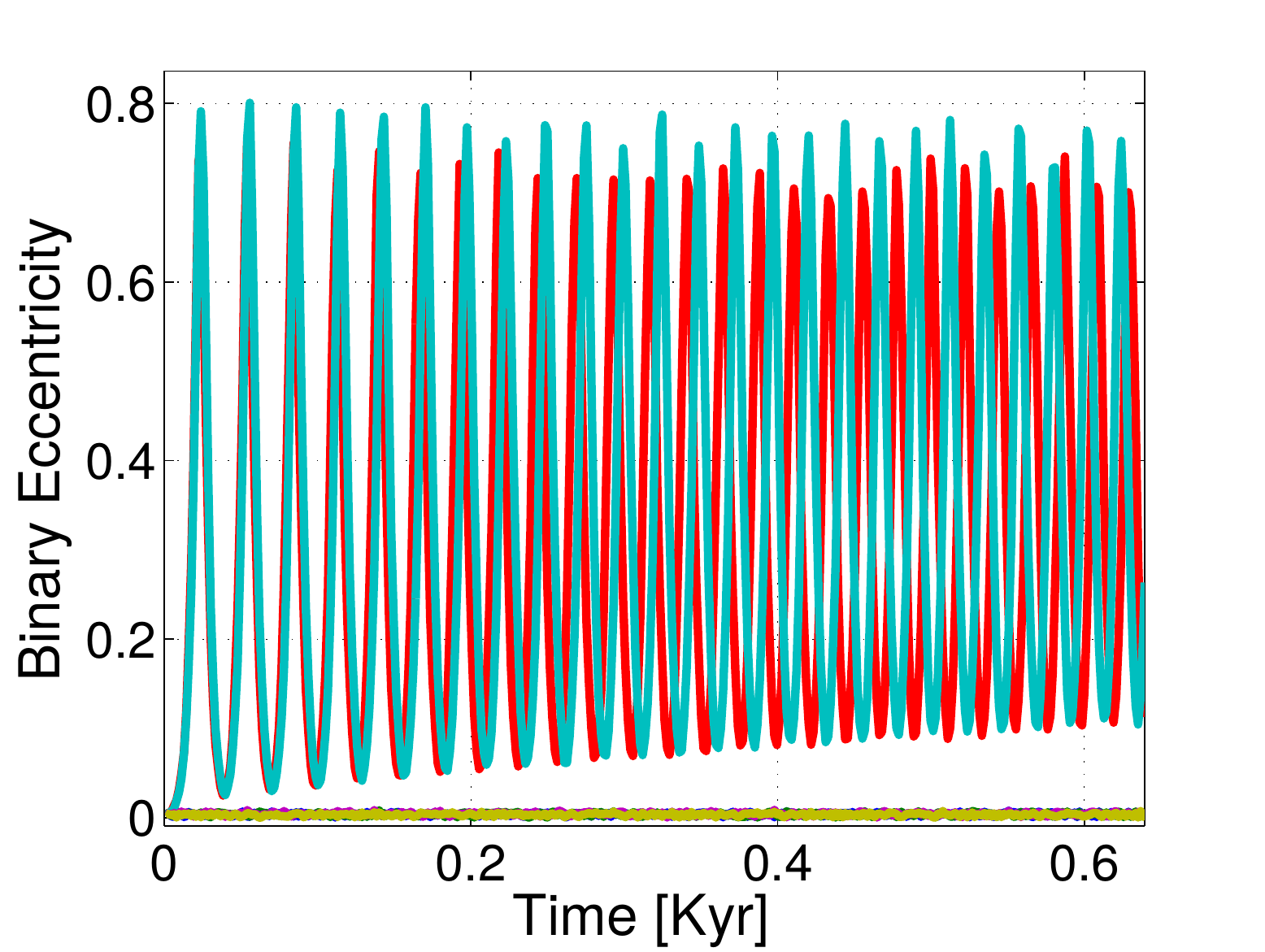}\includegraphics[height=5.45cm]{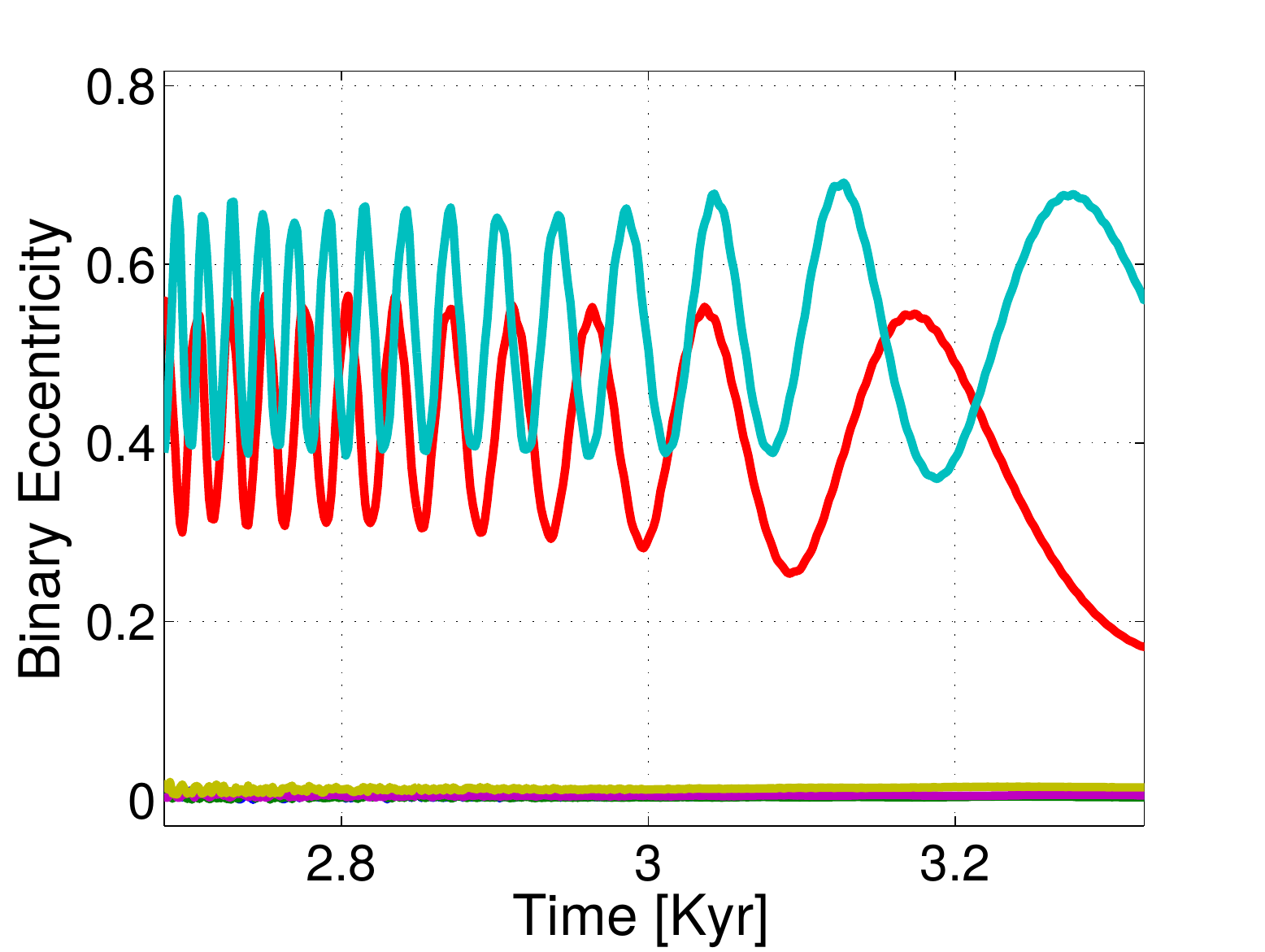}
\par\end{centering}

\begin{centering}
\includegraphics[height=5.45cm]{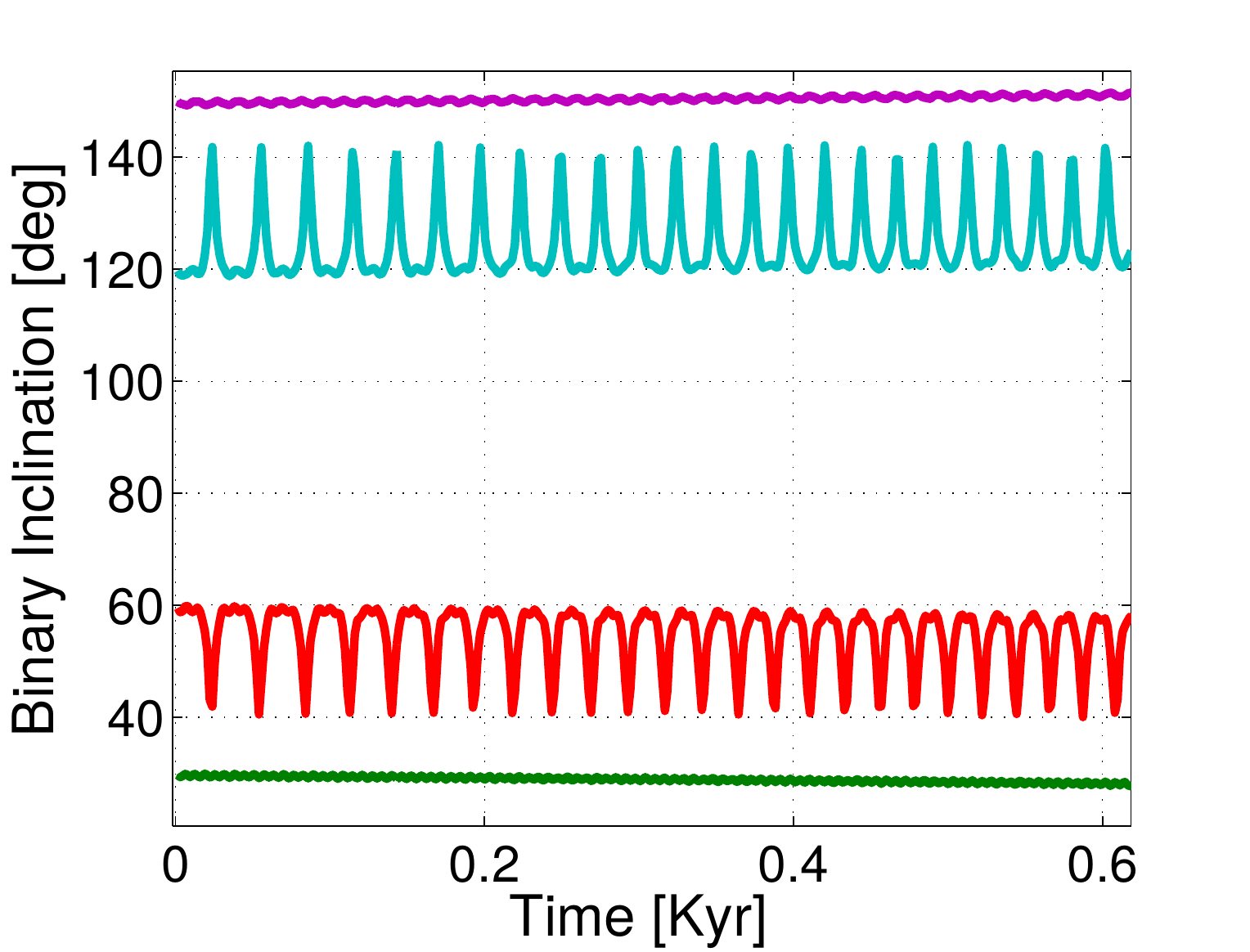}\includegraphics[height=5.45cm]{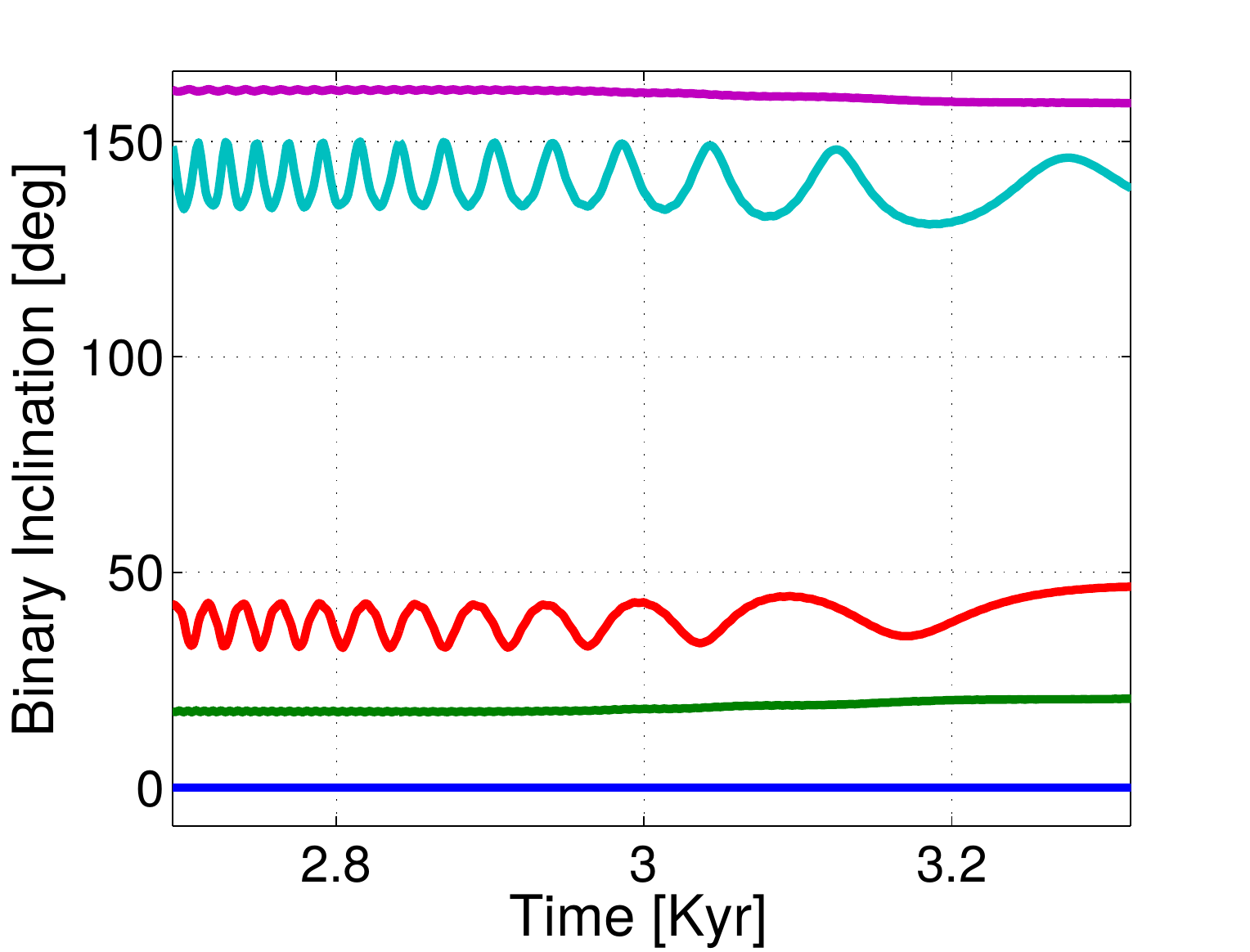}
\par\end{centering}

\caption{\label{inc3}Same as figs (\ref{inc2}) but with $e_{out}=0.1$}
\end{figure*}

In Fig. \ref{inc3} we plot the zoomed-in snapshots of the evolution
of the binary eccentricity (top) and the inclination (bottom). Contrary
to Fig. \ref{inc2}, the KL timescale $T_{KL}$ does not vary much,
since $a_{bin}$ is roughly constant. Only around $\sim2.8$Kyr does
the binary begin to inspiral rapidly and the KL timescale increases.
The binary eccentricity initially oscillates in the range of $\sim0-0.8$,
and in range of $\sim0.3-0.5$ for prograde orbit and $\sim0.4-0.65$
for retrograde orbit. It hints that the retrograde orbits are excited
to higher eccentricities and are less stable for some configurations,
as will be shown below. The inclinations are reduced by $\sim30\%$
near $t\sim2.8$Myr.

\subsubsection{Low mass ratio binaries ($q\ll1$)}

We now turn to binaries with small mass ratio $q\ll1$. 

\begin{figure*}
\begin{centering}
\includegraphics[height=5.45cm]{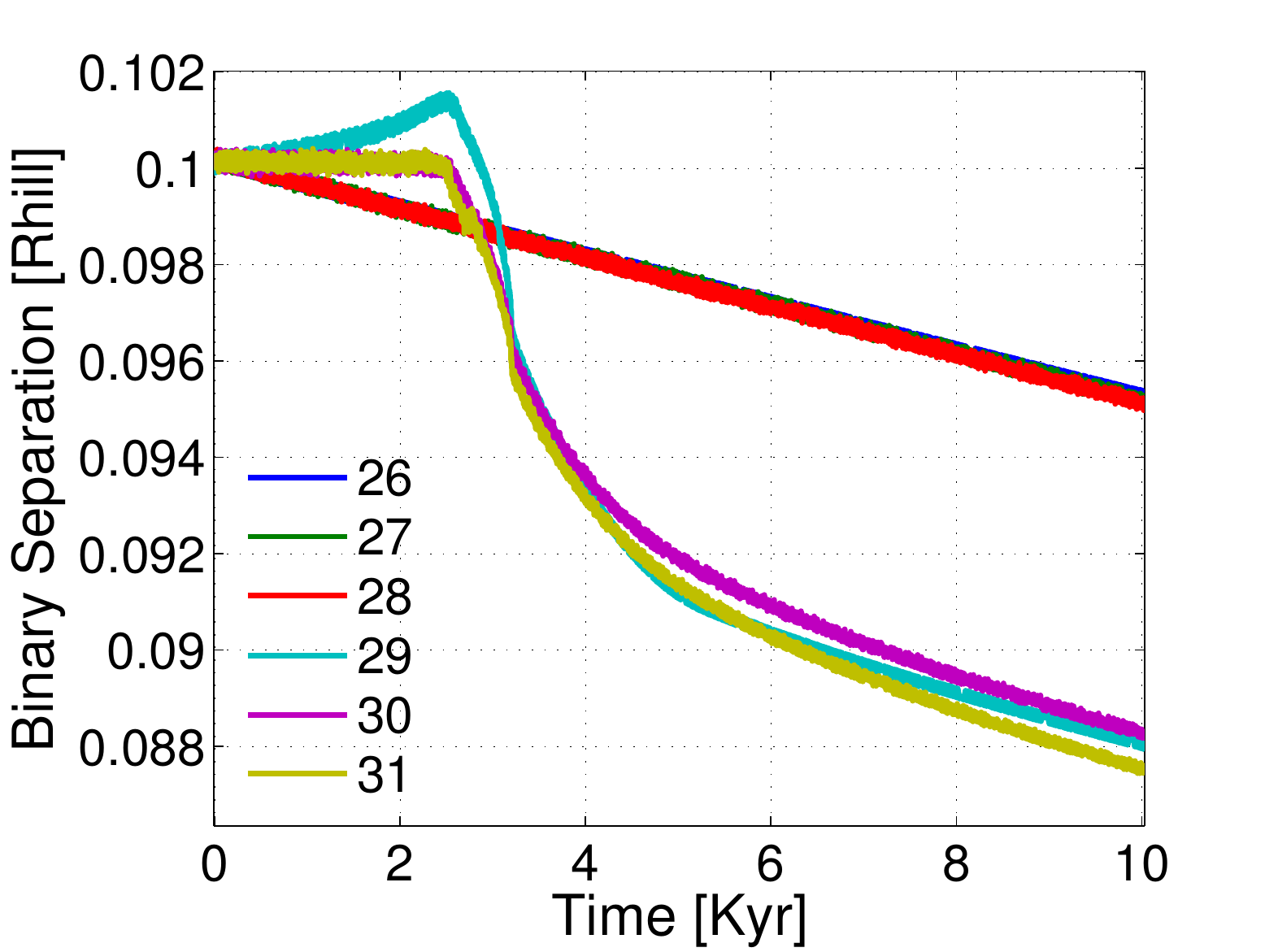}\includegraphics[height=5.45cm]{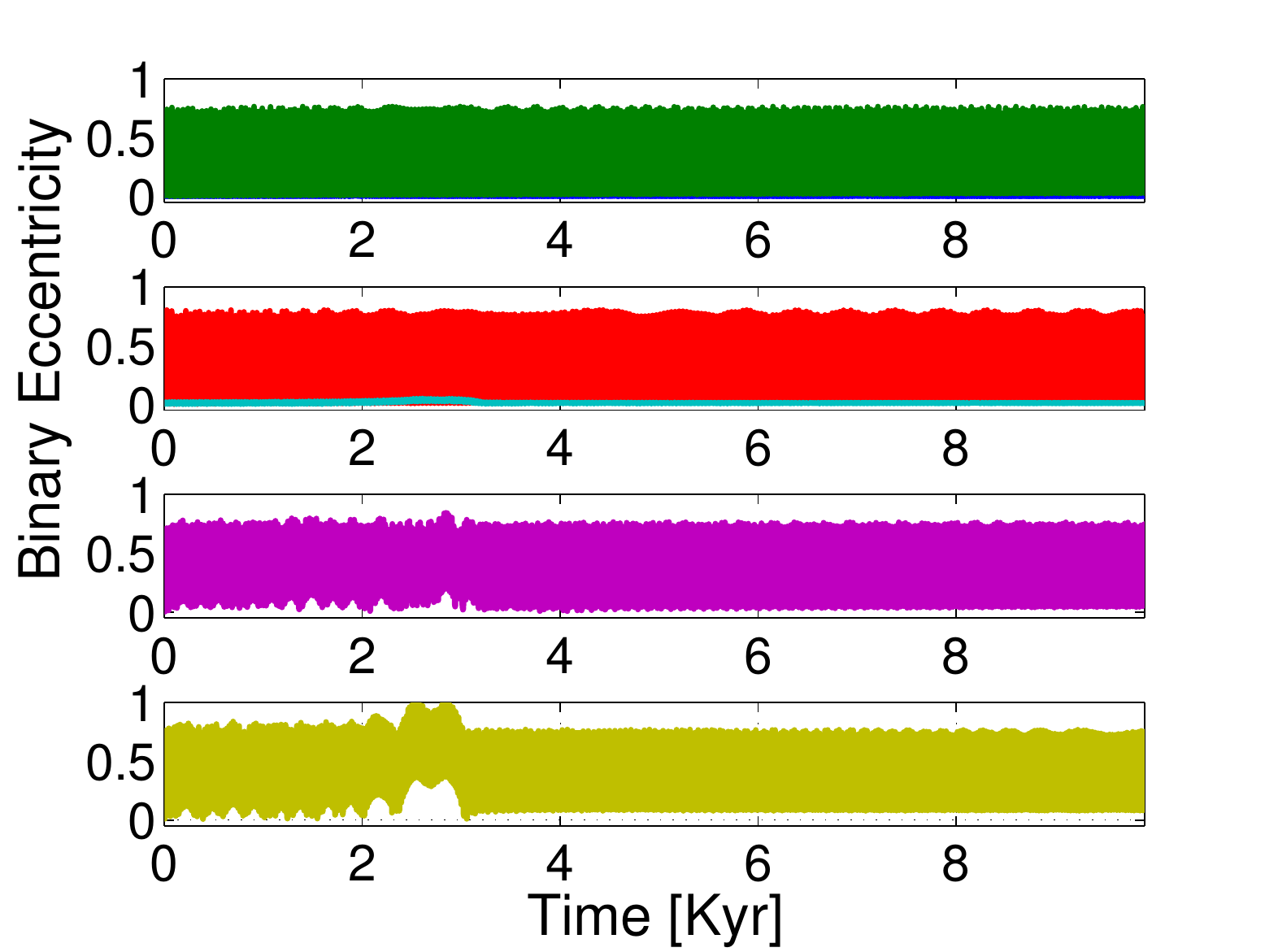}
\par\end{centering}

\begin{centering}
\includegraphics[height=5.45cm]{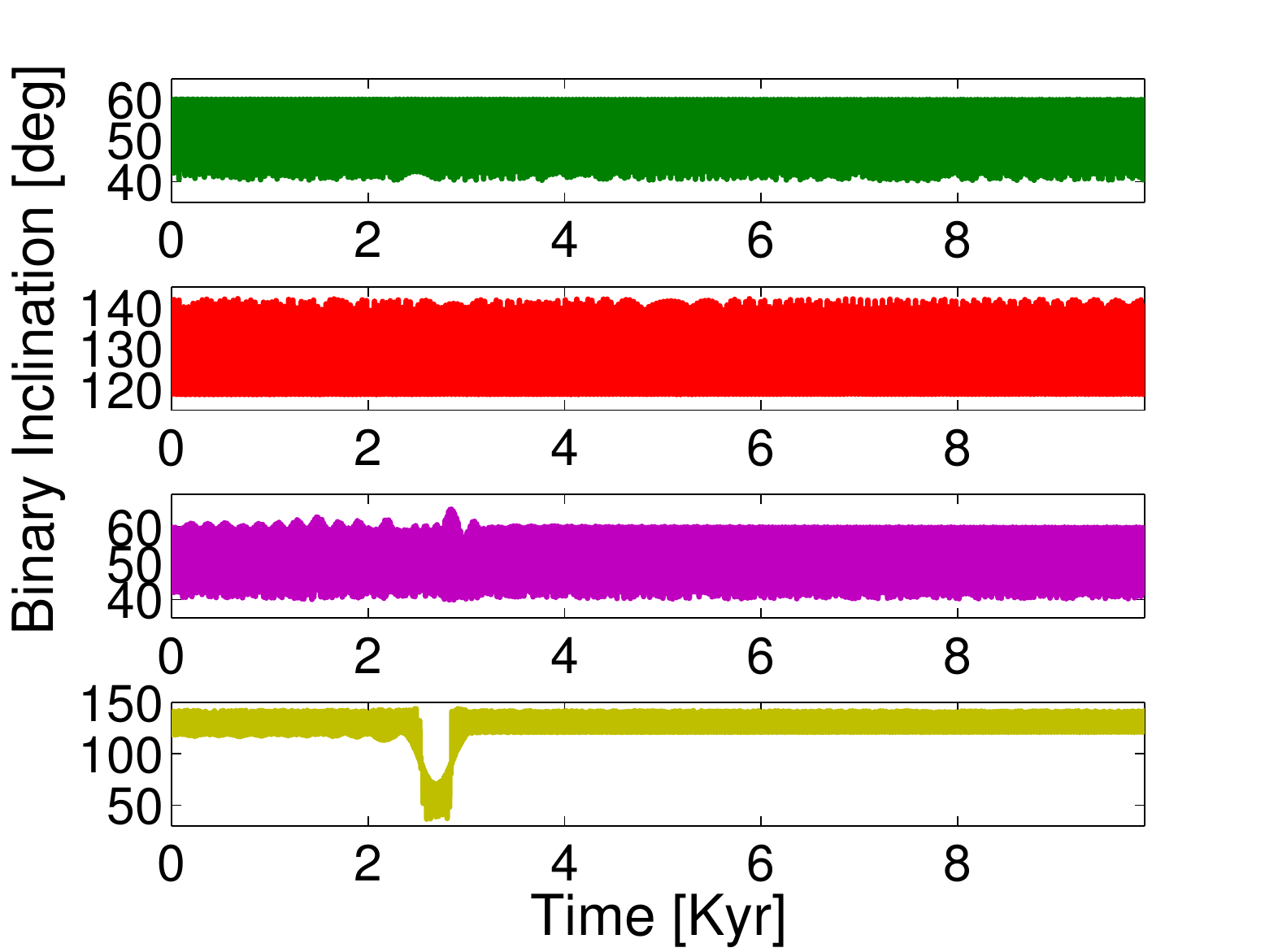}\includegraphics[height=5.45cm]{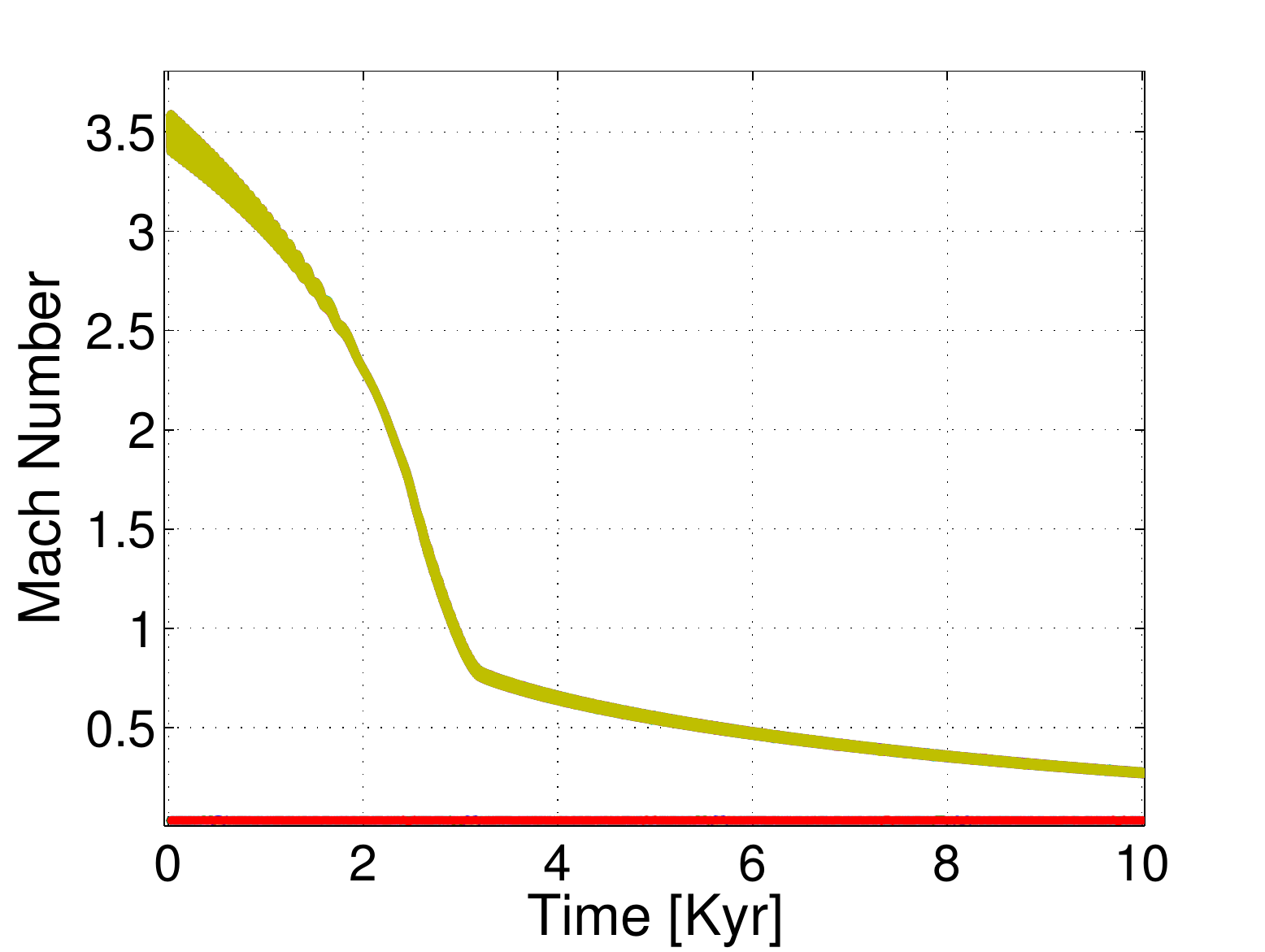}
\par\end{centering}

\caption{\label{hir1} Evolution of hierarchical binaries with $Q=10^{-8}$
and $q=10^{-2}$ with various binary inclinations and CM outer eccentricity
(see Table \ref{tab:incline} for list of initial conditions). Top
left: Evolution of the binary separation. Top right: Evolution of
the binary eccentricity. Bottom left: Evolution of the binary inclination.
Bottom right: Evolution of the Mach number.}
\end{figure*}

In Fig. \ref{hir1} we plot the evolution of hierarchical binaries
with either circular ($e_{out}=0$) or eccentric $(e_{out}=0.1$)
outer orbits. The primary orbital elements ($a_{p}$,$e_{p}$) evolve
similarly to the co-planar case (runs 12 and 13, Fig \ref{fig:5}).
On the top left panel we plot the evolution of the binary separation
versus time. Due to the small mass ratio $q\ll1$, the maximal change
in separation for the supersonic regime is shorter by a factor of
$q$, since the supersonic regime timescale is dominated by the outer
eccentricity decay, $\dot{e}_{out}$, which is proportional to $m_{b}.$
Hence, the supersonic regime timescale is shorter. For most of the
orbits, the binary inclination and eccentricity follow the KL evolution.
The exception is the retrograde orbit of run 31, which evolution is
erratic. As the orbital eccentricity decreases and the Mach number
approaches the sonic limit from above $(\mathcal{M}\to1^{+}$), the
binary is subjected to large jerks due to the steep slope of $d\mathcal{I}(\mathcal{M})/d\mathcal{M}$.
The differential GDF force excites the binary eccentricities and the
(sine of the) inclinations to larger values approaching unity. The
inclination flips to prograde at $t\approx2.7$Kyr. The orbit stays
prograde for $\approx200$yr and then it goes back to retrograde.
Later on the evolution returns to regular KL evolution with slightly
higher inclination.

The dramatic GDF-induced dynamical effects bear no signature on the
Mach number (bottom right) and the orbital evolution of the primary.

\begin{figure*}
\begin{centering}
\includegraphics[ height = 7.5cm]{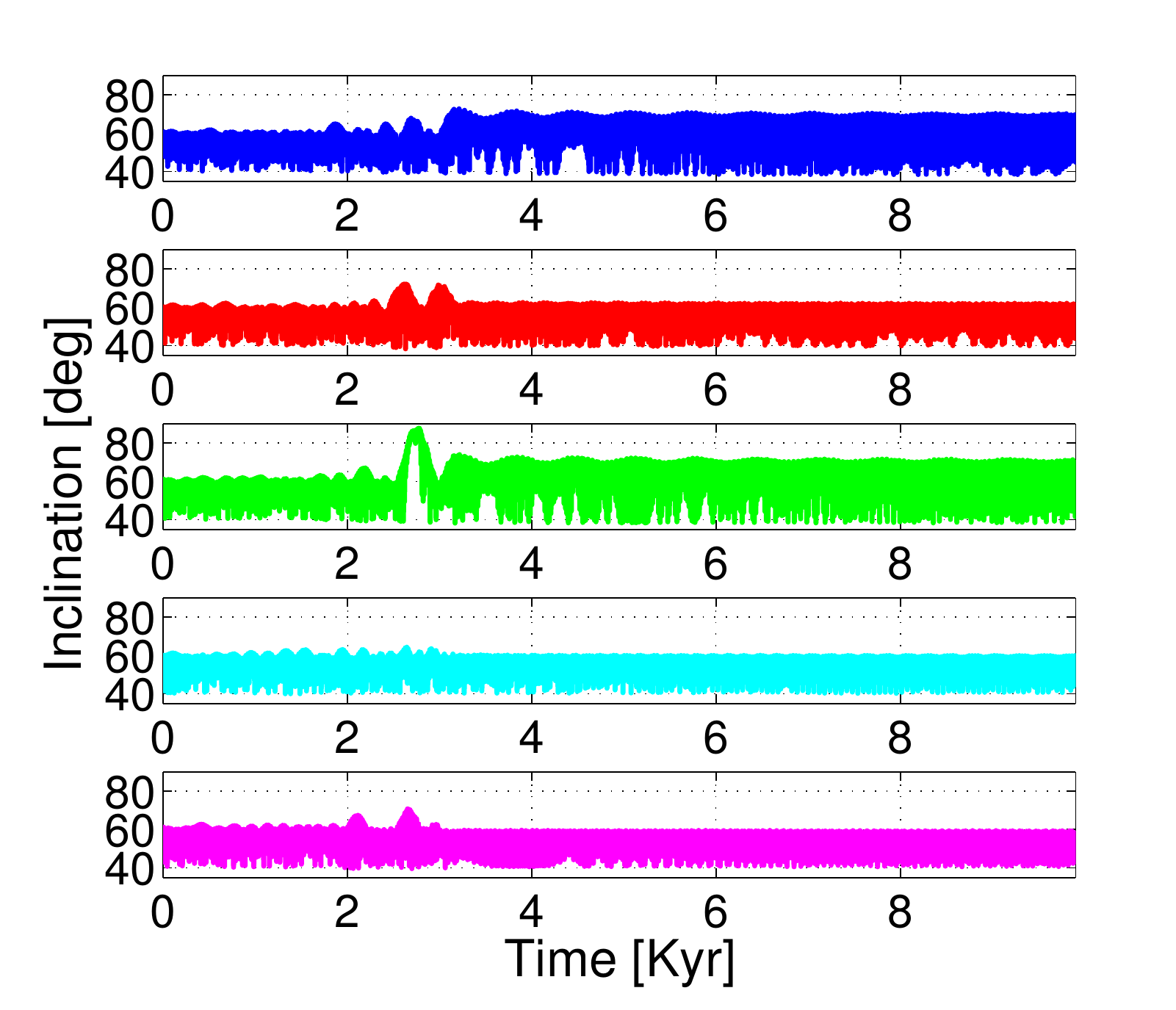}\includegraphics[ height=7.5cm]{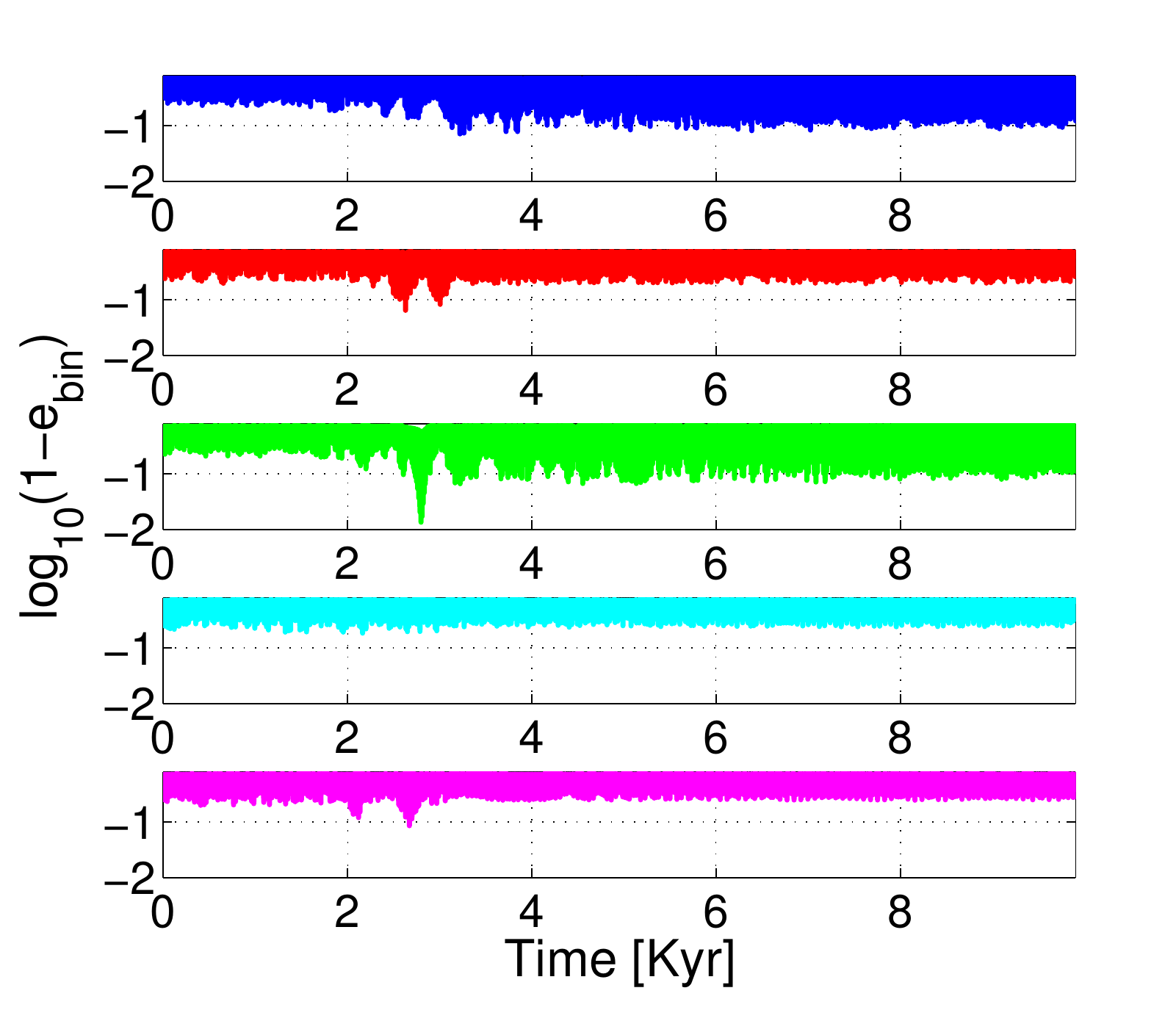}
\par\end{centering}

\begin{centering}
\includegraphics[ height = 7.5cm]{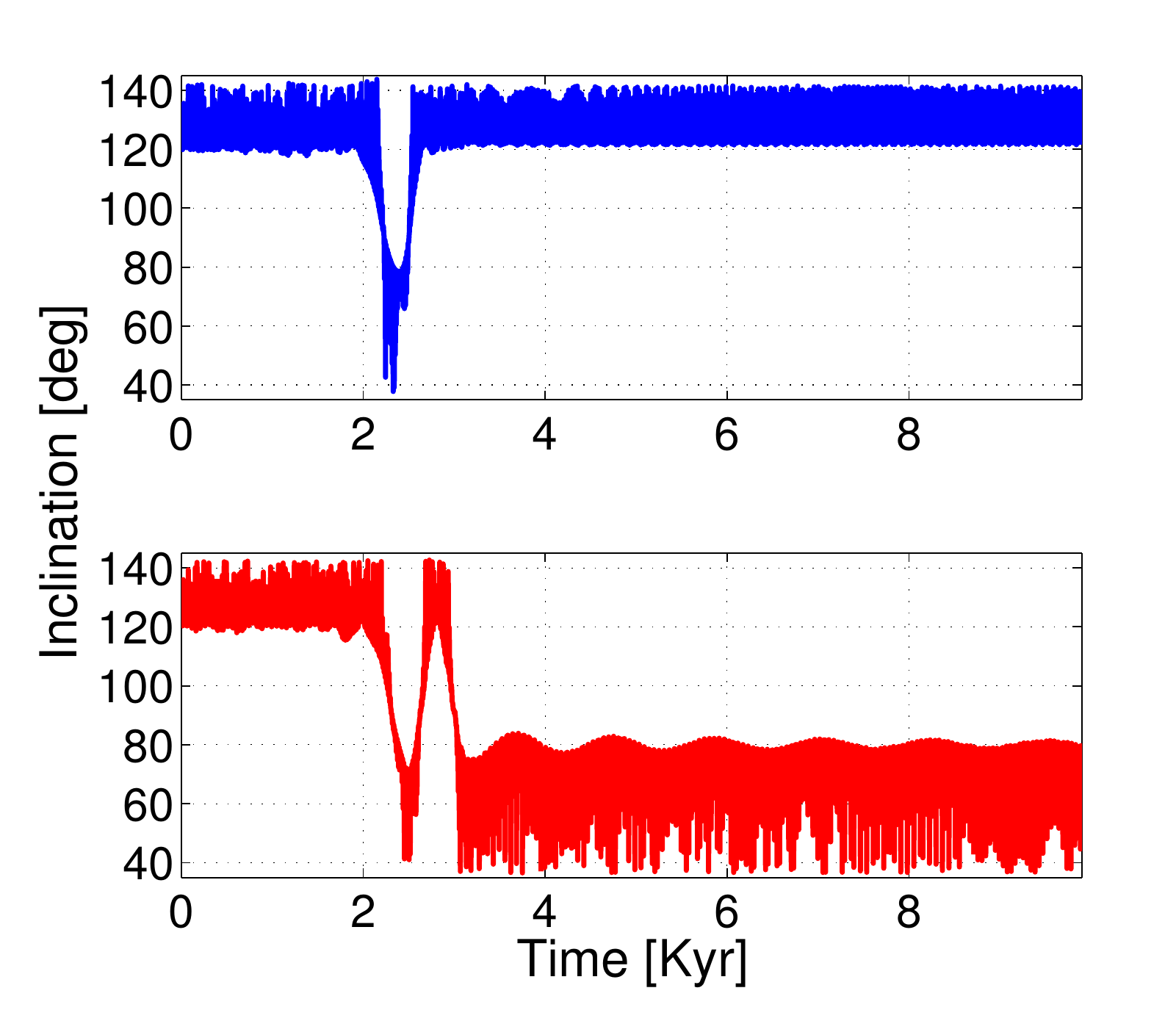}\includegraphics[ height=7.5cm]{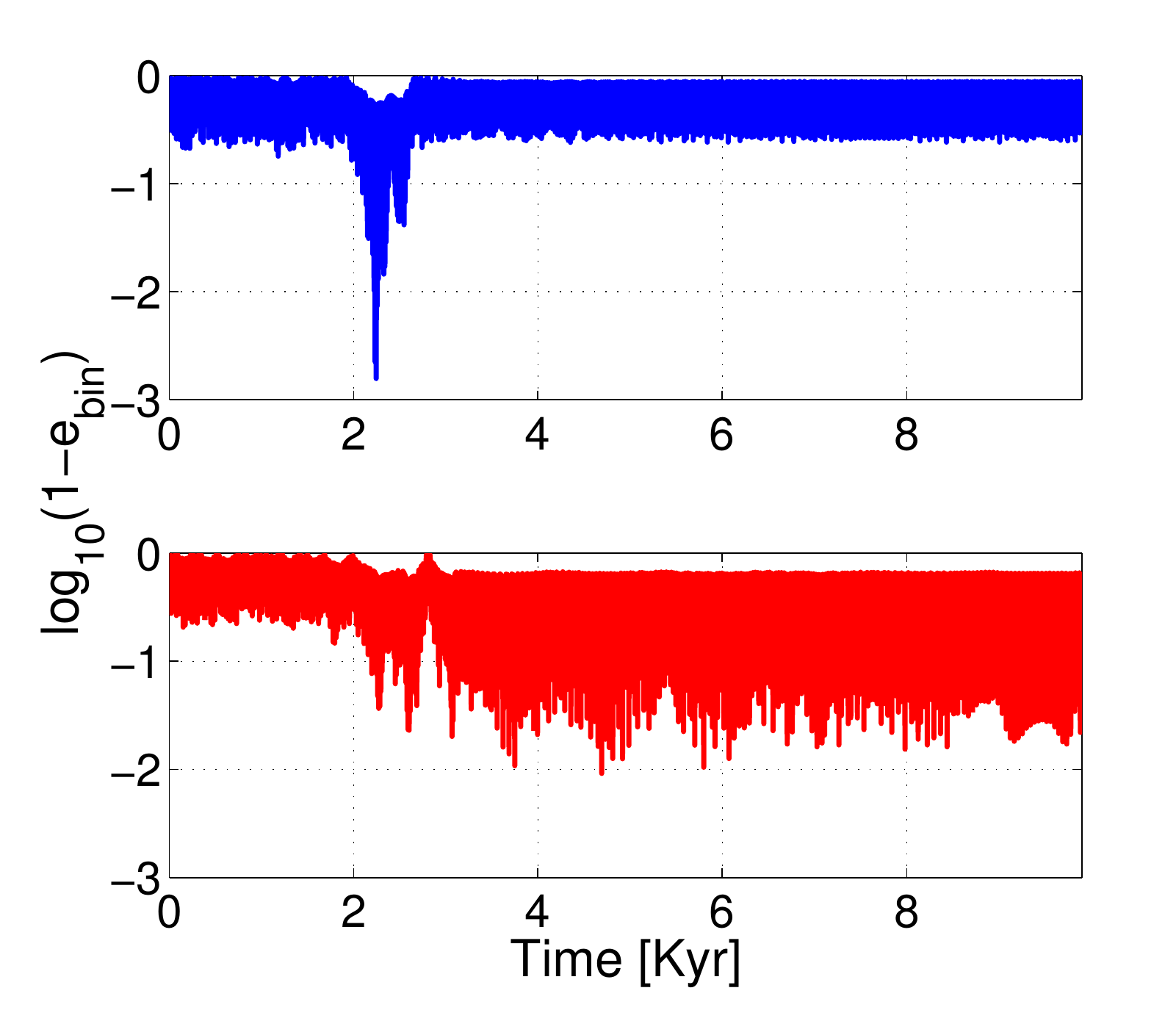}
\par\end{centering}

\caption{\label{fig:hir2}Evolution of binary orbital elements with random
phase (i.e. true anomaly $\nu_{bin}$ is chosen randomly), $q=10^{-2}$,
$Q=10^{-8},$ and outer eccentricity $e_{out}=0.1$. Top panels: $I=60$ deg. Bottom panels: $I=120$ deg. Left: Binary inclinations. Right: Binary eccentricity. $2/5$ of the retrograde binaries
have survived and are presented here. }
\end{figure*}

\begin{figure*}
\begin{centering}
\includegraphics[ height = 7.5cm]{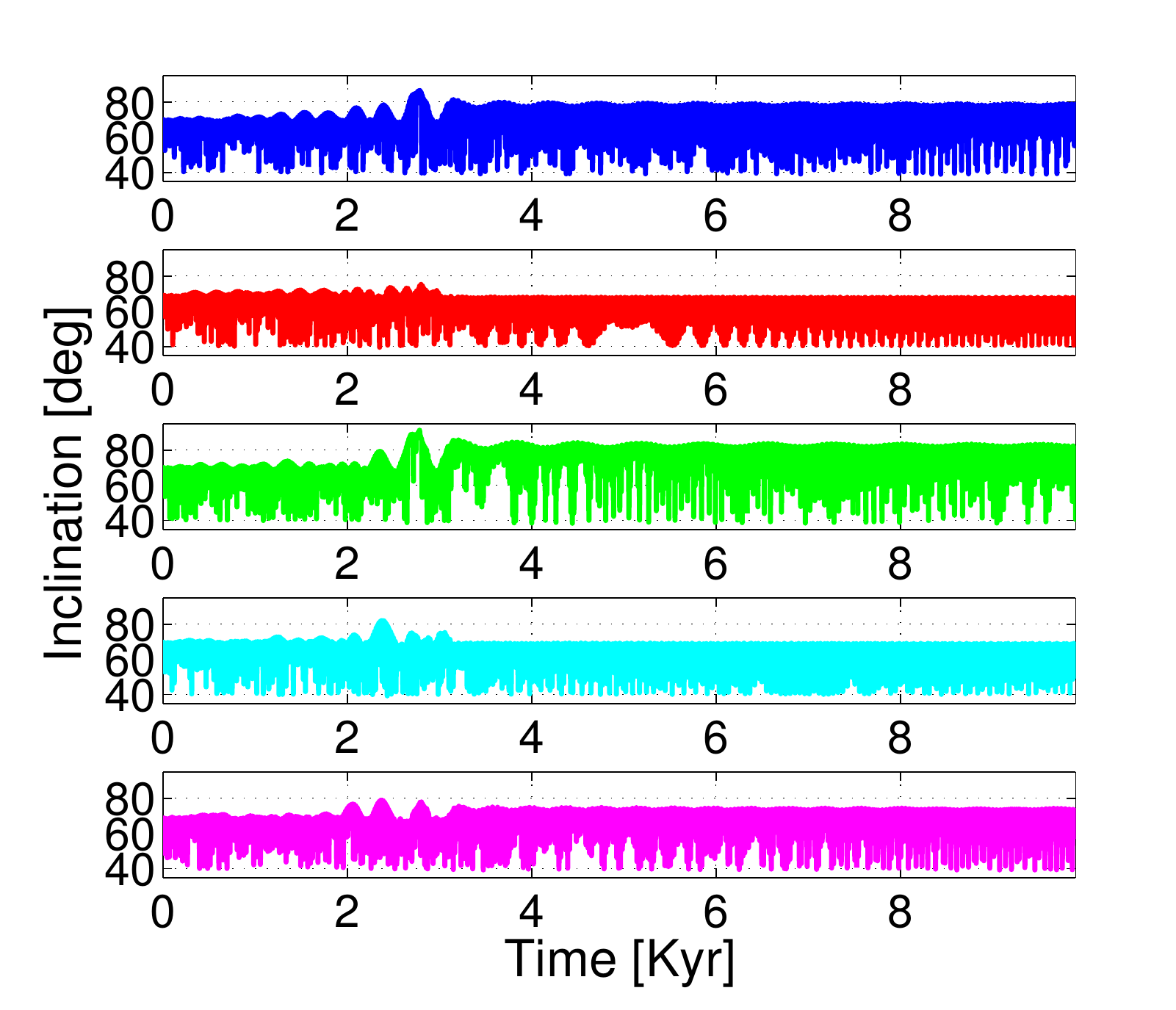}\includegraphics[ height=7.5cm]{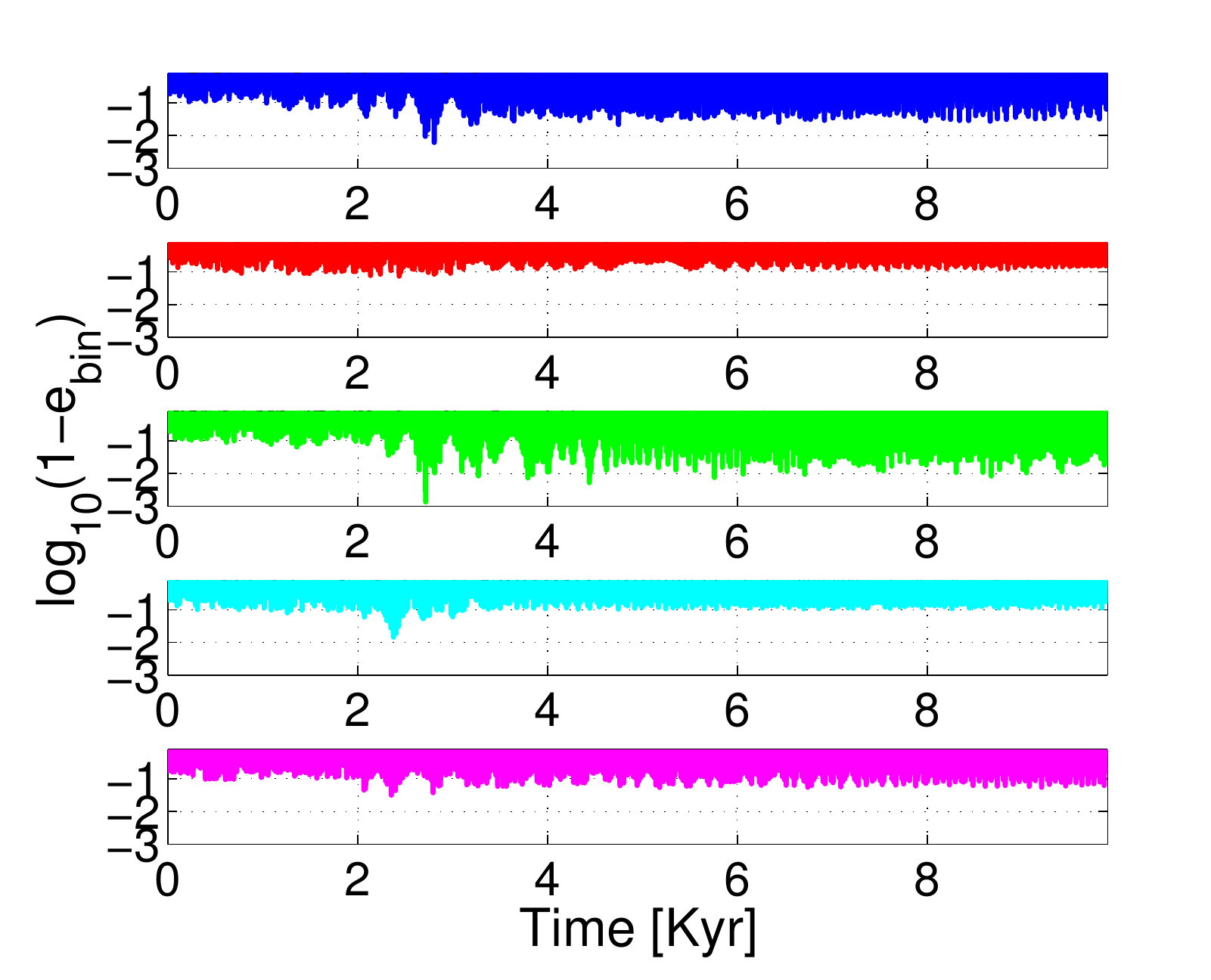}
\par\end{centering}

\begin{centering}
\includegraphics[ height = 7.5cm]{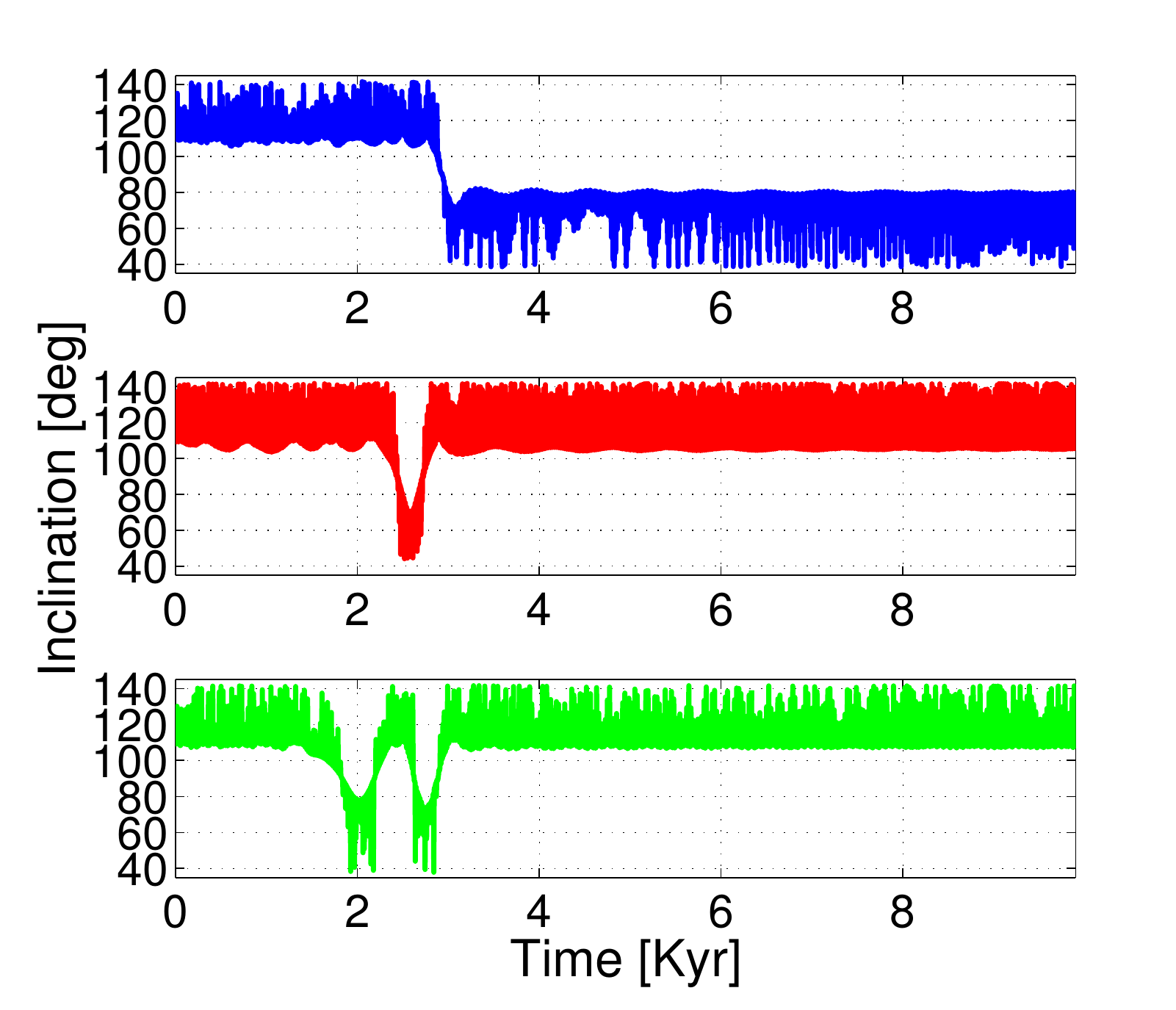}\includegraphics[ height=7.5cm]{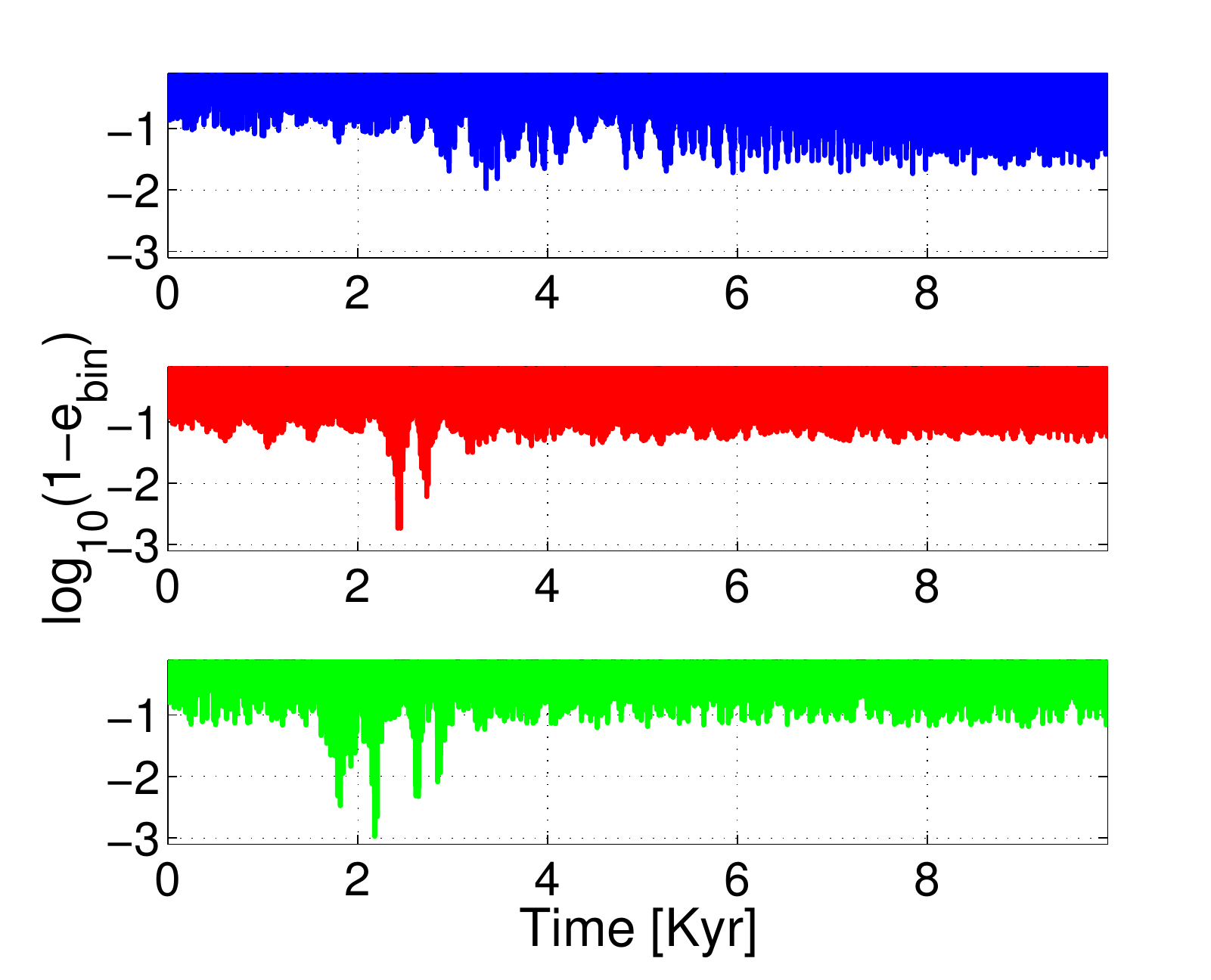}
\par\end{centering}

\caption{\label{fig:hir3}Evolution of binary orbital elements with random
phase (i.e. true anomaly $\nu_{bin}$ is chosen randomly), $q=10^{-2}$,
$Q=10^{-8},$ and outer eccentricity $e_{out}=0.1$. Top panels: $I=70$ deg. Bottom panels: $I=110$ deg. Left: Binary inclinations. Right: Binary eccentricity. $3/5$ of the retrograde binaries
have survived and are presented here.}
\end{figure*}

The combination of high inclination, small $q$ and large GDF jerks
near the sonic limit $\mathcal{M}\gtrsim1$ could lead to erratic
behavior of the binary. In principle, the interaction of the lower
mass body with the gas and the other two bodies depends on the binary
phase. Although the long term evolution does not depend on the phase,
the binary phase can be a crucial parameter when short term transient
interactions are taking place, and large instantaneous GDF jerks are
involved. In order to demonstrate that, we simulate the evolution of binary
inclinations and eccentricities of five binaries with random initial phases.

 Fig. \ref{fig:hir2} shows the evolution of binary inclinations and eccentricities for initial inclinations of  $60^{\circ}$ (top panel) and $ 120^{\circ}$ (bottom panel). Similarly,  Fig. \ref{fig:hir3} shows the evolution of binary inclinations and eccentricities for initial inclinations of  $70^{\circ}$ (top panel) and $ 110^{\circ}$ (bottom panel). In both figures prograde orbits are more stable; they all survive and
end up with somewhat higher inclinations and eccentricities. For the
orbits with the largest inclinations attained, post-interaction oscillations
on longer, secular timescales are evident, that decay over
time. Retrograde orbits are prone to ``torque kicks'' that change
the energy of the binary. Retrograde orbits are far less stable. Three
of the five orbits for $120^{\circ}$ and two of the five orbits for
$110^{\circ}$ have been destroyed. The rest have flipped their orientation
at least once. Some of them remained prograde. The maximal eccentricities
attained are in the range of $\log_{10}(1-e_{bin})\sim(-2)-(-3)$.
Given the above results, retrograde orbits are less stable and prone
to spin flips and destruction of the BPs. 

To conclude this section, the dynamical behavior of the primary orbital
elements for inclined binaries is the same as for co-planar binaries.
For low inclinations, the evolution of the binary elements is also
similar. For higher inclinations, the binaries are subjected to rapid
KL oscillations. If the outer eccentricity $e_{out}=0$, then the
behavior of the binary orbital elements is similar to the co-planar
case. If on the other hand, the initial eccentricity is $e_{out}=0.1$
(or any other large enough value), then the highly inclined orbits
quench the inspiral, and the loss (or gain) of angular momentum leads
to changes in $I_{bin}$ and $e_{bin}$. For low mass binary ratio
$q\ll1$, and particular binary orientation, the binary is excited
to high inclinations and eccentricities near the trans-sonic limit
$\mathcal{M}\gtrsim1$. Prograde BPs survive and remain close to their
initial conditions and evolve regularly with KL cycles. Retrograde
BPs flip their angular momentum vector, and some of them collide and/or
destroyed.

\section{Discussion}
\label{sec:DISCUSSION}

We have explored the evolution of intermediate mass BPs due to GDF
by mimicking the effects of the gaseous disk as a fiducial external
force in N-body simulations. The GDF force was derived following \citeauthor{1999ApJ...513..252O}'s
analytical theory. Naturally, it has limitations and does not take
into account other physical processes (e.g. companion wake, non-linear
effects, an inhomogenous medium, disk instabilities, turbulence etc.).
Further studies of the evolution of a bound binary embedded in gaseous
halo are required to model these additional effects. Detailed hydrodynamical
simulations could lead to a better understanding of these processes
and obtain the merger or break up timescale more rigorously.

The dynamical evolution of BPs can have important consequences for
planet formation and for the current structure of single merger-formed
planetesimals (asteroids/Kuiper-belt objects) and the configuration
of BPs \citep{2009ApJ...699L..17P,2011ApJ...727L...3P}.
As we found in this study, evolution of BPs due to interactions with
the gas (in particular due to GDF) could alter their initial distribution
in the planetesimal disk, and determine the initial conditions for
the later evolution once the gas has dispersed. We find that BPs on
circular CM orbits have inspiraled or merged due to GDF, depending
on the planetesimal-to-sun and inner binary mass ratios $Q$ and $q$,
respectively. Generally, GDF is more efficient in the inner regions
of the disk due to the density gradient. At 1AU $\sim10^{23}g$ mass
BPs inspiral and merge within 1Myr. If we choose the density scaling
$\rho_{g}\propto a^{-16/7}$ (see paper I for details), the inspiral
time increases by $\sim200$ at $10$AU, and only significantly affects
BPs with masses $\sim10^{25}g$. For Trans Neptunian and Kuiper Belt
objects, the inspiral time is longer by a factor of a few$\times10^{3}$,
hence it can only significantly affect masses of the order of $\sim10^{27}g$,
which is much more massive than Pluto, the largest known KBO member.
In addition, GDF approach does not apply to such large masses because
of accretion and non-linearity of the density wake (see paper I for
details). As the distance from the star increases, we expect that
BPs will be more common, with larger separations and larger masses.
In addition, the binary inspiral extracts internal energy and serves
as an additional heat source of the gaseous and planetesimal disks.

If the external CM orbit is eccentric, the binary expands. Starting
with initial separation of $0.1r_{H}$, none of the binaries has broken
up. However it is plausible that starting from larger separations,
initially gravitationally bound binaries will expand, reach the limit of dynamical stability, and  break. Hence, in this
case a more restricted stability criteria is required. The simulated
binaries have expanded roughly two times their initial separation,
hence the GDF eccentric stability radius should be reduced by a factor
of two. BPs with large separations have larger cross section for encounters
with other planetesimals, hence they are important for the evolution
of the planetesimal disk. If there is a mechanism that pumps the external
eccentricity above a certain threshold (e.g. scattering by larger
planetesimals and planets), the expansion could persist as long as
the gas is present. In this case determining the break up timescale
is critical to determine the fate of the BPs. 

Highly inclined BPs could be excited to large eccentricities and inclinations
due to KL oscillations as well as their coupling to GDF interactions.
Some of them can merge, break up or flip their orientations. It is
plausible that some of the highly inclined BPs survived the final
stages of planet formation.

\section{Summary }
\label{sec:Summary}

In this study we extended the study of the interactions of single
intermediate size planetesimals interaction with gas through gas dynamical
friction (paper I) and considered the effects on binary planetesimals
(BPs) in the same mass range of $10^{21}-10^{25}g$. At this mass
range aerodynamic gas drag is negligible, but planetesimals have sufficiently
low mass, so linear perturbation in the gas density is applicable.
In addition, for such binaries the Hill radius is much smaller than
the disk scale height, hence the approximation of spherical gaseous
halo is more easily attained than in the single planetesimal case.

Equal mass BPs of mass $\sim10^{23}g$ at $1$AU and initial low orbital
eccentricity inspiral in less than a Myr. This result is consistent
with \citet{2011ApJ...733...56P}'s merger timescale for planetesimals
of sizes of a few$\times100$km BPs, close to the critical size where
GDF and gas drag are comparable, and provides additional channel for
mergers and inspirals for larger planetesimals. Larger planetesimals
merge faster, with a general trend of $\tau_{merge}\propto m_{p}^{-1}$,
that breaks down near the sonic limit where $\mathcal{M}\sim1$.

BPs with external eccentricity $e_{out}\gtrsim2(h/a)$, experience
supersonic flow that applies a positive net torque. Such BPs gain
angular momentum and expand. As their outer orbit circularizes, the
torque direction is reversed, and the binary shrinks after the flow
has become subsonic. Higher initial separations or persistent external
eccentricity pumping could eventually lead to binary break up.

Inclined BPs are subjected to Kozai-Lidov oscillations that could
affect the binary evolution. While the inspiral timescale is the same
as in the co-planar case, the break up timescale is longer since the
expansion is quenched. Small mass binary ratios could result in a
stochastic interaction that could lead to merger, break up or inclination
flip.

The interactions of BPs with gas can significantly change their evolution
and thereby affect the planetesimal disk and affect the distribution
of BPs and their separations, as well as induce binary merger-formed
single planetesimals which could have a highly oblate structure. We
conclude that GDF can play an important role in the evolution of BPs,
and should be accounted for in the study of the early stages of planet
formation, as well as for the origin and early evolution of current
asteroids, Kuiper belt object and their binary components.

\section*{Acknowledgements}
We thank Erez Michaely for stimulating discussions and the referee, Keiji Ohtsuki, for helpful comments that lead to improvement of the manuscript. HBP acknowledges support from Israel-US bi-national science foundation, BSF grant number 2012384, European union career integration grant "GRAND", the the Minerva center for life under extreme planetary conditions and the Israel science foundation excellence center  I-CORE grant 1829.

\bibliographystyle{apj} 

\appendix{}

\section{A. Binary stability}
\label{sec:binary stability}

The Hill radius is defined as the distance at which the gravitational
tidal acceleration equals the gravitational acceleration of the binary.
Similarly to the definition of Hill radius, the GDF shearing radius
is defined as the the distance at which the gravitational tidal acceleration
equals the magnitude of the differential acceleration $\Delta a_{GDF}$.
Equating of the gravitational acceleration $a_{grav}=G(m_{b}+m_{s})/a_{bin}^{2}$
to the shearing acceleration yields the GDF shearing radius 
\[
R_{GDF}=\sqrt{\frac{G(m_{b}+m_{s})}{\Delta a_{GDF}}}.
\]

Taking $\mathcal{A}_{cm}$ and $\mathcal{A}_{bin}$ from Eqn. (\ref{eq:da_sep}),
the magnitude of the differential acceleration is $|\boldsymbol{\Delta a_{GDF}}|=(\mathcal{A}_{cm}^{2}+\mathcal{A}_{bin}^{2}+2\mathcal{A}_{cm}\mathcal{A}_{bin}\boldsymbol{\hat{\varphi}}\cdot\boldsymbol{\hat{\varphi}_{bin}})^{1/2}\le2\mathcal{A}$,
where $\mathcal{A}\equiv\max\{\mathcal{A}_{cm},\mathcal{A}_{bin}\}$.
The GDF shearing radius is bounded from below by 
\[
R_{GDF}\ge\sqrt{\frac{Gm_{bin}}{2\mathcal{A}}}.
\]
For each regime, the linear and the supersonic, there are two distinct
cases: equal and very small mass ratio binaries. 

For Linear regime and equal mass binaries, $q=1$, $m_{s}=m_{b}=m$
and $\mathcal{A}_{cm}=0$ , at $1$AU $c_{s,1AU}=6.7\cdot10^{4}cm/s$
\begin{eqnarray}
R_{GDF,linear}^{equal} & = & \sqrt{\frac{Gm_{bin}}{\mathcal{A}_{bin}}}=\sqrt{\frac{3c_{s}^{3}}{2\pi G\rho v_{bin}}}\nonumber \\
 & = & \sqrt{\frac{3H_{0}}{2\pi}}f^{1/4}Q^{-1/6}\frac{c_{s}}{\sqrt{G\rho}}\nonumber \\
 & = & 0.84\left(\frac{H_{0}}{0.022}\right)^{1/2}\left(\frac{f}{0.1}\right)^{1/4}\left(\frac{Q}{10^{-10}}\right)^{-1/6}\left(\frac{c_{s}}{c_{s,1AU}}\right)AU.\label{eq:Rgdf1}
\end{eqnarray}

For small mass binary, $q\ll1,$ $\mathcal{A}_{cm}\gg\mathcal{A}_{bin}$
we have 
\begin{eqnarray}
R_{GDF,linear}^{small} & \approx & \sqrt{\frac{Gm_{bin}}{\mathcal{A}_{cm}}}=\sqrt{\frac{1+q}{4\pi H_{0}(1-q)}}\frac{c_{s}}{\sqrt{G\rho}}\nonumber \\
 & \approx & 0.59\left(\frac{H_{0}}{0.022}\right)\left(\frac{c_{s}}{c_{s,1AU}}\right)AU.\label{eq:rgdf2}
\end{eqnarray}
The correction for Eqn (\ref{eq:rgdf2}) is of order $\mathcal{O}(q)$.
In this case, the shearing radius is smaller by $\sim70$\% compared
to the equal mass case. We expect that in the general case where $\mathcal{A}_{cm}$
and $\mathcal{A}_{bin}$ are comparable, the result will be somewhere
between these two extreme values. 

For the supersonic regime and small $q$, the first term in Eqn (\ref{eq:dasuper})
is the dominant one, and the GDF shearing radius is 
\begin{eqnarray}
R_{GDF,super}^{small} & = & \sqrt{\frac{(1+q)Ge_{p}^{2}v_{K}^{2}}{(1-q)C_{super}}}=\sqrt{\frac{(1+q)e_{p}^{2}v_{K}^{2}}{(1-q)4\pi G\rho_{g}\ln\Lambda}}\nonumber \\
 & \approx & 1.07e_{p}R_{GDF,linear}^{small}.\label{eq:rsuper1}
\end{eqnarray}
The latter result is valid only when the supersonic regime applies,
i.e. $e\gtrsim2H_{0}\approx0.044$, so the minimal GDF shearing radius
is $\sim20$ times smaller, but still much larger than the Hill radius.

For equal mass binaries, the second and third term is in Eqn (\ref{eq:dasuper})
are comparable, and the GDF shearing radius is 
\begin{equation}
R_{GDF,super}^{equal}\approx\frac{1}{2}\sqrt{\frac{2Ge_{p}^{3}v_{K}^{3}}{C_{super}3v_{bin}}}=\frac{1}{2}\sqrt{\frac{e_{p}^{2}v_{K}^{2}}{6\pi G\rho_{g}\ln\Lambda\beta}}\approx R_{GDF,super}^{small}\sqrt{\frac{1}{6\beta}}.\label{eq:rsuper2}
\end{equation}
Since $\beta$ is small, the GDF shearing radius is larger.

In summary, in all the cases we considered, the GDF shearing radius
is always much larger than the Hill radius, therefore the binaries
are stable to perturbations from GDF.

\section{B. Derivation of equation 17}
\label{sec:Derivation}

Starting from eqn \ref{eq:dasuper2}, expanding to first order in
$\beta$, rearranging terms and using the definition of the reduced
mass $\mu$, as well as replacing $\beta$ back to $v_{bin}$ leads
to

\begin{eqnarray*}
\frac{\boldsymbol{\Delta a}_{GDF}}{C_{super}} & = & \frac{1}{e_{out}^{2}v_{K}^{2}}\left[(m_{b}\boldsymbol{\hat{\varphi}}-m_{b}\frac{m_{s}}{m_{bin}}\beta\boldsymbol{\hat{\varphi}_{bin}})(1+3\frac{m_{s}}{m_{bin}}\beta)-(m_{s}\boldsymbol{\hat{\varphi}}+m_{s}\frac{m_{b}}{m_{bin}}\beta)\boldsymbol{\hat{\varphi}_{bin}}(1-3\frac{m_{s}}{m_{bin}}\beta)\right]\\
 & = & \frac{m_{b}(e_{out}v_{K}(1-q)+3qv_{bin})\boldsymbol{\hat{\varphi}}-2\mu v_{bin}\boldsymbol{\hat{\varphi}_{bin}}}{e_{out}^{3}v_{K}^{3}}
\end{eqnarray*}

\end{document}